\documentclass[11pt]{article}
\usepackage[utf8]{inputenc}
\usepackage{amsmath,amssymb,amsfonts}
\usepackage{dsfont}   
\usepackage{bm}
\usepackage[margin=1in]{geometry}
\usepackage{booktabs}
\usepackage{array}
\usepackage{multirow}
\usepackage{pifont}
\usepackage{graphicx}
\usepackage[numbers,sort&compress]{natbib}
\usepackage{authblk}
\usepackage{xcolor}
\usepackage[plainpages=false,pdfpagelabels]{hyperref}
\hypersetup{colorlinks=true,linkcolor=blue,citecolor=blue,urlcolor=blue}
\usepackage{enumitem}
\usepackage{placeins}   
\usepackage{textcomp}   
\usepackage{listings}
\DeclareTextSymbol{\textquotedbl}{OT1}{34}   
\usepackage{rotating}   
\newcommand{\xv}{\bm{x}}
\newcommand{\bet}{\bm{\beta}}
\newcommand{\Sbar}{\bar{S}}
\newcommand{\Pbar}{\bar{P}}
\newcommand{\E}{\mathbb{E}}
\newcommand{\ipd}{\mathrm{IPD}}
\newcommand{\agd}{\mathrm{AgD}}

\title{\bfseries Multilevel network meta-regression\\ for multistate models:\\
Population-adjusted joint synthesis of progression and survival from individual and aggregate data}

\author[1,2,3,4]{Jeroen P.\ Jansen\thanks{Correspondence: \texttt{jeroen.jansen@ucsf.edu}. ORCID: \texttt{0000-0003-2686-9217}}}
\affil[1]{Department of Clinical Pharmacy, School of Pharmacy, University of California, San Francisco}
\affil[2]{Associate Member in Cancer Control, UCSF Helen Diller Family Comprehensive Cancer Center, University of California, San Francisco}
\affil[3]{The Philip R.\ Lee Institute for Health Policy Studies, University of California, San Francisco}
\affil[4]{Precision AQ, Health Economics \& Outcomes Research, Precision Medicine Group}
\date{\today}

\begin{document}
\maketitle

\begin{abstract}
\noindent
In network meta-analysis (NMA) of oncology time-to-event outcomes, there is growing recognition that
progression-free survival (PFS) and overall survival (OS) are better synthesized jointly than in
separate analyses. The multistate NMA of Jansen et al.\ addresses this with a tri-state (stable, progressed, dead) Markov model fitted to aggregate survival curves. It cannot, however, adjust for imbalance in effect modifiers across trials, so its relative effects may be biased or not apply to the target population. Multilevel network meta-regression (ML-NMR) resolves covariate imbalance by integrating an individual-level model over each aggregate study's covariate distribution, coherently
combining individual participant data (IPD) and aggregate data (AgD); however, for time-to-event outcomes it has been developed only for a single endpoint. This paper introduces \emph{multilevel network meta-regression for
multistate models} (ML-NMR-MS), which embeds the ML-NMR integrated IPD-plus-AgD likelihood inside
an illness--death multistate model. For IPD studies the full multistate likelihood over
linked transitions is used; for AgD studies covariate-marginal state-occupancy probabilities
are computed by quasi-Monte Carlo integration of the occupancies implied by the transition
intensities over the reconstructed covariate distribution, and entered through a conditional-survival
likelihood. The resulting
population-adjusted joint PFS/OS treatment effects can directly parameterize a state-transition economic model. On an
illustrative network, ML-NMR-MS resolves a treatment's joint effect into its component transitions, separating
an effect on progression from an effect on post-progression survival, and produces conditional and marginal effects. Under correct specification, a simulation study with known truth shows negligible bias and near-nominal coverage for the population-adjusted treatment effects.
\par\vspace{0.5em}
\end{abstract}

\section{Introduction}\label{sec:intro}

Health technology assessment (HTA) in oncology routinely requires estimates of the relative
treatment effects of several treatments on both progression-free survival (PFS) and overall survival
(OS) in a target population. Standard network meta-analysis (NMA) synthesizes time-to-event
outcomes through a single univariate effect measure (a hazard ratio (HR), a survival probability at a
landmark, or a median), which discards the time course of the treatment effect; a constant HR is moreover
biased when the proportional-hazards (PH) assumption fails \citep{ouwens2010,jansen2011fp}. Several responses to this exist: synthesis
of parametric survival models on the exact likelihood \citep{campbell2025parametric}, fractional
polynomials \citep{jansen2011fp}, and flexible M-splines \citep{phillippo2026mspline}, each of which
retains the time course of the treatment effect. In addition, PFS and OS are not independent: they are
two functionals of the same underlying disease process, and analyzing them separately can produce
mutually incoherent extrapolations (e.g., crossing curves). Responses here include bivariate
meta-analysis of the two log-hazard-ratios \citep{bujkiewicz2019bivariate} and non-parametric joint
synthesis of the areas under the two curves \citep{daly2022nonparam}; both acknowledge the
dependence without modeling the process that generates it.

Jansen et al.~\citep{jansen2023multistate} addressed both issues with a multistate NMA. Patients occupy
one of three states, namely stable (pre-progression), progressed, or dead, and the model synthesizes the
three transition intensities (stable$\to$progression, stable$\to$death, progression$\to$death)
with time-varying (fractional-polynomial) baseline hazards and treatment effects. Because published
Kaplan--Meier curves
for PFS and OS do not reveal which patient's progression time corresponds to which death time, the
method works with the state-occupancy probabilities implied by the transition intensities, fitted through a conditional-survival binomial likelihood. This delivers a coherent joint PFS/OS synthesis from aggregate data (AgD) and a ready-made
parameterization of a state-transition cost-effectiveness model.

What the multistate NMA cannot do is adjust for imbalance in (patient-related) effect-modifying covariates across trials. When treatment benefit depends on patient characteristics (e.g., disease stage, a biomarker) and those characteristics are distributed differently across trials, a NMA that
ignores them is biased. When there are differences between these study population characteristics and the target population of interest, the estimated relative treatment effects are not relevant (i.e. externally biased), even when there are no differences between studies. 
One could fit a conventional meta-regression on the covariate
summaries that trials report (e.g., mean age or the proportion at a given disease stage); however, regressing aggregate effects on aggregate covariates is prone to aggregation
(ecological) bias (the across-study association between a trial's mean covariate and its estimated
relative effect need not equal the individual-level effect-modifier relationship) and, with only a handful of
trials, such a regression is imprecise and cannot separate within-trial from across-trial effect
modification \citep{phillippo2016tsd18}. With individual participant data (IPD) for every trial, one
could fit a ``gold-standard'' IPD network meta-regression, which estimates the effect-modifier
interactions directly at the individual level and free of aggregation bias; but IPD are rarely
available for more than a subset of studies. Multilevel network meta-regression (ML-NMR) \citep{phillippo2020mlnmr,phillippo2024general} solves
exactly this: it specifies an individual-level regression and, for the studies for which only summary
information is available (AgD), integrates the individual-level model, or in its
general-likelihood form the individual likelihood, over the study's covariate distribution, thereby avoiding aggregation bias, coherently combining IPD and AgD, and producing population-adjusted estimates in
any target population. ML-NMR has been extended to general likelihoods, including
single-endpoint survival, \citep{phillippo2024general}, but not to multistate or joint-endpoint
settings.

The two methods are complementary and, to date, disjoint: multistate NMA gives the joint PFS/OS
structure but only unadjusted, aggregate-only synthesis; ML-NMR gives population adjustment and
the combination of IPD and AgD but stops at a single time-to-event endpoint. This paper closes the gap by introducing
\emph{multilevel network meta-regression for multistate models} (ML-NMR-MS), and makes three
contributions. First, it embeds the ML-NMR integrated IPD-plus-AgD likelihood inside an
illness--death multistate model: IPD studies contribute the full multistate likelihood over their
linked transitions, while aggregate studies contribute covariate-marginal state-occupancy
probabilities obtained by quasi-Monte Carlo integration of the occupancies implied by the transition
intensities over the reconstructed covariate distribution, entered through a conditional-survival likelihood. ML-NMR-MS is an instance of the general-likelihood ML-NMR of Phillippo et al.~\citep{phillippo2024general}.
The original ML-NMR relates to it too: for binary outcomes the general form recovers its one-parameter binomial
likelihood, and its other aggregate likelihoods approximate the general one in closed form. What is new with ML-NMR-MS is that the individual likelihood contribution
integrated over the covariate distribution is the state occupancy of an illness--death process, which
links progression to death and makes the PFS and OS contributions dependent. Second, under conditional constancy of relative effects, the method
yields population-adjusted conditional transition effects for a decision-relevant target population.
Given a baseline hazard for that population (Section~\ref{sec:estimands}), it also yields conditional
endpoint and marginal PFS/OS effects and the absolute state-occupancy quantities, which together can
directly parameterize a state-transition economic model. Third, it
illustrates ML-NMR-MS on a network of multiple myeloma trials and examines its bias and coverage against a full-IPD gold standard in a limited simulation study with fully known truth. 

\section{Multilevel network meta-regression for multistate models}\label{sec:methods}

\subsection{Setting and notation}\label{sec:setting}
Consider a connected network of $j=1,\dots,J$ randomized trials comparing treatments
$\mathcal{K}_j\subseteq\{1,\dots,K\}$, with treatment $1$ the network reference. An arm is the pair
$(j,k)$, $k\in\mathcal{K}_j$, so within a study arms are identified by the treatment they receive.
The outcome is a multistate time-to-event process \citep{putter2007} on a finite set of health
states connected by a set of transitions $r\in\mathcal{R}$. The model is presented for a
general process of this kind in Section~\ref{sec:model} and then developed in full for the tri-state
illness--death model that recurs in oncology: at time $u$ (from
randomization) a patient is stable ($S$), progressed ($P$), or dead ($D$), with transitions
$\mathcal{R}=\{SP,SD,PD\}$.
PFS is then the probability of remaining stable and OS the
probability of being alive (stable or progressed), so both endpoints are functions of the same
underlying process. Two data modalities coexist: IPD
studies ($j\in\mathcal{J}_\ipd$) provide, for each individual $i=1,\dots,N_{jk}$ in arm $k$ of study $j$, covariates $\xv_{ijk}\in\mathfrak{X}\subseteq\mathbb{R}^Q$ and event histories sufficient to
reconstruct the realized transitions and their times, where $Q$ is the number of covariates and
$\mathfrak{X}$ is the covariate support; AgD studies
($j\in\mathcal{J}_\agd$) provide published PFS and OS Kaplan--Meier curves together with baseline
covariate summaries, but neither individual covariates nor the linkage between a patient's
progression and death times.

\subsection{Model}\label{sec:model}
For a general multistate process, each transition $r\in\mathcal{R}$ has a time-inhomogeneous
intensity, modeled on the log scale by a regression in which a flexible function of time
carries the study baseline and the treatment effect, while the covariates carry the prognostic and
effect-modifier terms:
\begin{equation}
\log h^r_{jk}(u\mid\xv) \;=\; \bm\varphi^r(u)^{\!\top}\big(\bm\mu^r_{j} + \bm\gamma^r_k\big)
 \;+\; \xv^{\!\top}\!\big(\bet^r_1 + \bet^r_{2,k}\big),
\label{eq:lp}
\end{equation}
where $\bm\varphi^r(u)=\big(1,\varphi^r_1(u),\dots,\varphi^r_{F_r}(u)\big)^{\!\top}$ is a time basis
of an intercept and $F_r$ time-varying terms, and $\bm\mu^r_j$ are study-specific baseline scale/shape coefficients. The treatment
coefficients $\bm\gamma^r_k$ ($\bm\gamma^r_1=\bm 0$, the reference treatment) act on those baseline
coefficients, and $\bet^r_{2,k}$ ($\bet^r_{2,1}=\bm 0$) is the effect-modifier interaction, both
evaluated at the treatment $k$ received in that arm; $\bet^r_1$ are prognostic effects. Here $\xv$ is an individual patient's covariate vector, so
\eqref{eq:lp} is an individual-level model shared by both data types: it supplies the transition
intensities at each patient's observed $\xv_{ijk}$ in IPD studies (Section~\ref{sec:ipd}) and at each
point of the reconstructed covariate distribution in aggregate studies (Section~\ref{sec:agd}), where
the implied state occupancies are then averaged over that distribution.

Time $u$ runs forward from randomization for every transition (including progression$\to$death, whose intensity depends on
time since randomization, not time since progression), so the process is a \emph{clock-forward},
time-inhomogeneous Markov process. The baseline is chosen for flexibility: (i) fractional polynomials of first or
second order with powers $p_1$ (and $p_2$ at second order) and the Box--Tidwell convention $u^{(0)}=\log u$ (continuity with Jansen et al.~\citep{jansen2023multistate};
first order recovers Weibull and Gompertz, and a scalar intercept $\bm\varphi^r\equiv1$ the
constant-hazard case), or (ii) cubic M-splines on the baseline hazard \citep{phillippo2026mspline,phillippo2024general}, with a prior on the simplex of basis weights; other flexible parametric baselines such as Royston--Parmar splines \citep{royston2002} could be
substituted. (See Appendix~\ref{app:baselines} for a discussion of the choice between M-splines and
fractional polynomials.) The baseline need not share the treatment's time basis: it may carry more terms, or, as in (ii),
replace $\bm\varphi^r(u)^{\!\top}\bm\mu^r_j$ in \eqref{eq:lp} altogether with a general function of time, $c^r_j+\log\big(\sum_{l=1}^{L} w^r_{jl} M_l(u)\big)$ for an
M-spline, where $c^r_j$ is a study-specific log-scale, $M_1,\dots,M_L$ are the basis functions and the weights $w^r_{j1},\dots,w^r_{jL}$ lie on the simplex. The weights may differ between studies or be shared by all of them, which gives the baselines a common shape. The treatment effect retains its
own basis $\bm\varphi^r(u)^{\!\top}\bm\gamma^r_k$ throughout (Appendix~\ref{app:baselines}). 

Writing a single set of treatment effects $\{\bm\gamma^r_k\}_{k=1}^{K}$ versus the network reference
treatment ($\bm\gamma^r_1=\bm 0$) imposes the usual NMA consistency assumption. The relative effect of treatment $b$ versus $a$ on transition $r$ is then
$\bm\varphi^r(u)^{\!\top}(\bm\gamma^r_b-\bm\gamma^r_a)+\xv^{\!\top}(\bet^r_{2,b}-\bet^r_{2,a})$.
It varies with the covariate value $\xv$ at which it is evaluated, but at any given $\xv$ it is the
same in every study, so direct and indirect evidence are coherent. The transition log-hazard-ratio of treatment $a$ versus the reference is
$\bm\varphi^r(u)^{\!\top}\bm\gamma^r_a + \xv^{\!\top}\bet^r_{2,a}$: time-varying in general, and constant (proportional hazards) exactly when the non-intercept
components of $\bm\gamma^r_a$ are zero, which includes the case of no treatment effect at all.

These treatment effects may be treated as fixed or random across studies. Writing $\bm\delta^r_{jk}$ for the study-specific counterpart of $\bm\gamma^r_k$ (a vector over the
basis coefficients) on transition $r$, a random-effects model replaces $\bm\gamma^r_k$ in
\eqref{eq:lp} by $\bm\delta^r_{jk}$ and, stacking across transitions, takes a multivariate Normal (MVN)
$\bm\delta_{jk}\sim\mathrm{MVN}(\bm\gamma_k,\bm\Sigma)$: the means are the network-reference
treatment effects, $\bm\Sigma$ is the between-study heterogeneity covariance (a common heterogeneity
variance and correlation are the usual simplifications), and the network reference is held fixed,
$\bm\delta^r_{j1}=\bm\gamma^r_1=\bm 0$. Because the means are anchored to the network reference,
the study-specific coefficient contrast $\bm\delta^r_{jb}-\bm\delta^r_{ja}$ has mean
$\bm\gamma^r_b-\bm\gamma^r_a$, so consistency is untouched. Covariate imbalance between arms does not change
this, because the effect-modifier term is evaluated at a common $\xv$. The fixed-effect model is the special case $\bm\Sigma=\bm 0$, i.e.\
$\bm\delta^r_{jk}=\bm\gamma^r_k$; multi-arm trials are accommodated by the standard
correlated-effects construction \citep{achana2014}. Appendix~\ref{app:re} gives the Stan code for random
effects on stable$\to$progression.

Covariate adjustment adds \emph{conditional constancy of relative effects}: the effect-modifier and treatment coefficients carry
no study index ($\bet^r_{2,j,k}\equiv\bet^r_{2,k}$ and, under fixed effects, $\bm\delta^r_{jk}\equiv\bm\gamma^r_k$; under random effects $\bm\delta^r_{jk}$ varies about $\bm\gamma^r_k$ while $\bet^r_{2,k}$ stays common), so the
conditional relative effect is the same function of the covariates in every study and transports to a
target population, only the baseline $\bm\mu^r_j$ remaining study-specific. The effect-modifier
interactions may be treatment-specific ($\bet^r_{2,k}$, as in \eqref{eq:lp}), common within a treatment class $c$
($\bet^r_{2,k}\equiv\bet^r_{2,c}$), or common to all active treatments ($\bet^r_{2,k}\equiv\bet^r_2$). The last two form the
\emph{shared effect-modifier assumption} \citep{phillippo2020mlnmr}, which trades flexibility for identifiability and lets a
treatment studied only in aggregate borrow effect-modification information.

\subsection{Data and likelihood}\label{sec:datalik}
IPD and AgD enter a single likelihood through the shared parameter vector $\bm\xi$. The individual-level contribution (the full multistate likelihood) is given first, then the aggregate-level contribution, obtained by integrating the state occupancy implied by that same individual model over the covariate distribution, and finally their joint synthesis.

\subsubsection{Individual participant data}\label{sec:ipd}
For an IPD individual the realized illness--death path is observed. Writing
$H^r_{jk}(u_1,u_2\mid\xv)$ for the cumulative intensity $\int_{u_1}^{u_2} h^r_{jk}(u\mid\xv)\,du$, the individual
likelihood is the standard multistate/competing-risks product of exit intensities and sojourn
survival:
\begin{equation}
\begin{split}
L^{\mathrm{Con}}_{ijk}(\bm\xi;y_{ijk},\xv_{ijk}) ={}&
\underbrace{e^{-H^{SP}_{jk}(0,u_S\mid\xv_{ijk})-H^{SD}_{jk}(0,u_S\mid\xv_{ijk})}}_{\text{stable sojourn}}
\cdot \big[h^{SP}_{jk}(u_S\mid\xv_{ijk})\big]^{\mathds{1}(S\to P)}
\big[h^{SD}_{jk}(u_S\mid\xv_{ijk})\big]^{\mathds{1}(S\to D)}\\
&\cdot \underbrace{\Big(e^{-H^{PD}_{jk}(u_S,u_P\mid\xv_{ijk})}
\big[h^{PD}_{jk}(u_P\mid\xv_{ijk})\big]^{\mathds{1}(P\to D)}\Big)^{\mathds{1}(S\to P)}}_{\text{progressed sojourn}},
\end{split}
\label{eq:ipdlik}
\end{equation}
where the realized path $y_{ijk}$ of patient $i$ consists of $u_S$, the exit time from the stable state (to progression or
death, or the censoring time if still stable), $u_P$, the exit time from the progressed state (death or censoring), and
the indicators $\mathds{1}(\cdot)$ of which transitions were observed. The progressed state is entered at $u_S$, and the outer
$\mathds{1}(S\to P)$ retains the progressed-state factor only for patients who progressed; $\bm\xi$ is the full parameter set. The superscript ``Con'' marks this as the
covariate-conditional likelihood (the individual likelihood as a function of the outcome
$y_{ijk}$ and covariates $\xv_{ijk}$), whose covariate-marginal form $L^{\mathrm{Mar}}$ is derived for AgD in Section~\ref{sec:agd} by integrating over $\xv$
the individual likelihood of the coarsened outcome (PFS or OS alone), itself obtained from \eqref{eq:ipdlik}
by marginalizing over the unobserved progression status and time. This likelihood uses the linked progression$\to$death information
available in IPD (which patient's progression precedes which death). 

\subsubsection{Aggregate data}\label{sec:agd}
The defining feature of ML-NMR is that the aggregate-data model is derived from the individual-level
model rather than posited separately, so both data types share the parameters $\bm\xi$. Integrating the covariate-conditional likelihood over an arm's joint covariate distribution $f_{jk}(\xv)$ gives the individual marginal likelihood function
\citep{phillippo2024general},
\begin{equation}
L^{\mathrm{Mar}}_{ijk}(\bm\xi;y_{ijk})=\int_{\mathfrak{X}}
 L^{\mathrm{Con}}_{ijk}(\bm\xi;y_{ijk},\xv)\, f_{jk}(\xv)\,d\xv ,
\label{eq:mlnmr}
\end{equation}
which no longer depends on $\xv$. This is the likelihood contribution of each patient whose individual outcome is
observed but whose individual covariate vector is unknown; only the arm's covariate distribution $f_{jk}(\xv)$ is known. When a study also reports the outcomes only as
summaries $y_{\bullet jk}$, the aggregate marginal likelihood function of the arm can be derived, up to a
normalizing constant, from the product of these contributions over its $N_{jk}$ patients
\citep{phillippo2024general},
\begin{equation}
L^{\mathrm{Mar}}_{\bullet jk}(\bm{\xi};y_{\bullet jk})\;\propto\;\prod_{i=1}^{N_{jk}}L^{\mathrm{Mar}}_{ijk}(\bm{\xi};y_{ijk}),
\label{eq:aggmar}
\end{equation}
where $\bullet$ marks quantities aggregated over patients and $N_{jk}$ is the number of patients in arm
$k$ of study $j$. The product is written in the individual outcomes $y_{ijk}$, so it can be evaluated from the
reported data only when it can be written in terms of the summaries $y_{\bullet jk}$ alone. That holds for
discrete outcomes, but not in general \citep{phillippo2024general}.

For time-to-event data the aggregate quantities are the separately reported PFS and OS
curves. Pseudo-individual data can be reconstructed from each curve \citep{guyot2012}, but PFS and OS
are reconstructed separately, with no way to recover which patient's progression precedes which death.
Equation~\eqref{eq:mlnmr} therefore cannot be applied to each patient's full illness--death history,
whose covariate-conditional likelihood is \eqref{eq:ipdlik}. However, the state occupancy supplies the missing
link under the model's assumptions: its illness--death structure ties PFS to OS through the shared transition model. The aggregate
marginal likelihood \eqref{eq:aggmar} is then applied to each curve separately, with each patient's
covariate-conditional likelihood built from the occupancy.

The transition intensities
$h^r_{jk}(u\mid\xv)$ from the individual model \eqref{eq:lp} translate into state-occupancy probabilities
$S_{jk}(u\mid\xv),P_{jk}(u\mid\xv)$ through the illness--death Kolmogorov forward system, started from all
patients stable, $(S,P,D)(0)=(1,0,0)$. The progressed occupancy has no closed form for time-varying
hazards in general. The system is therefore solved on a grid of intervals $U_m=[u_m,u_{m+1})$, with the intensities
held constant at each interval's midpoint value $h^r_{jk,m}=h^r_{jk}\big((u_m+u_{m+1})/2\mid\xv\big)$. The grid
can be refined to approximate arbitrary hazards, while the free parameters remain the
low-dimensional regression coefficients. On each interval, the system has the closed-form solution~\citep{jansen2023multistate}
\begin{align}
S_{jk}(u\mid\xv) &= S_{jk}(u_m\mid\xv)\,e^{-(h^{SP}_{jk,m}+h^{SD}_{jk,m})(u-u_m)}, \label{eq:occS}\\
P_{jk}(u\mid\xv) &= P_{jk}(u_m\mid\xv)\,e^{-h^{PD}_{jk,m}(u-u_m)}
  + \frac{S_{jk}(u_m\mid\xv)h^{SP}_{jk,m}\big(e^{-(h^{SP}_{jk,m}+h^{SD}_{jk,m})(u-u_m)}-e^{-h^{PD}_{jk,m}(u-u_m)}\big)}
         {h^{PD}_{jk,m}-h^{SP}_{jk,m}-h^{SD}_{jk,m}}, \label{eq:occP}\\
D_{jk}(u\mid\xv) &= 1-S_{jk}(u\mid\xv)-P_{jk}(u\mid\xv), \label{eq:occD}
\end{align}
with the removable singularity at $h^{PD}_{jk,m}=h^{SP}_{jk,m}+h^{SD}_{jk,m}$ handled by a series
expansion near it (Appendix~\ref{app:ode}); occupancies are chained across intervals.

Integrating the occupancy over the arm's covariate distribution $f_{jk}$ realizes the individual marginal likelihood
\eqref{eq:mlnmr} concretely, since the stable and alive covariate-conditional likelihoods are the occupancies
$S_{jk}$ and $S_{jk}+P_{jk}$. The covariate-marginal occupancies, the arm's population-average
probabilities of being stable ($\Sbar_{jk}$) and progressed ($\Pbar_{jk}$), are
\begin{equation}
\begin{aligned}
\Sbar_{jk}(u)&=\int S_{jk}(u\mid\xv)\,f_{jk}(\xv)\,d\xv\approx\tilde N^{-1}\textstyle\sum_{l=1}^{\tilde N} S_{jk}\big(u\mid\tilde\xv^{(l)}_{jk}\big),\\
\Pbar_{jk}(u)&=\int P_{jk}(u\mid\xv)\,f_{jk}(\xv)\,d\xv\approx\tilde N^{-1}\textstyle\sum_{l=1}^{\tilde N} P_{jk}\big(u\mid\tilde\xv^{(l)}_{jk}\big),
\end{aligned}
\label{eq:marg}
\end{equation}
quasi-Monte Carlo averages over $\tilde N$ Sobol' points $\tilde\xv^{(l)}_{jk}$ drawn from $f_{jk}$; the alive
(OS) occupancy is $\Sbar_{jk}+\Pbar_{jk}$. The order of operations matters. The occupancy map is
nonlinear in $\xv$, so each point's trajectory is evaluated and only then averaged; evaluating at the
arm's mean covariate vector $\bar{\xv}_{jk}=\int\xv f_{jk}(\xv)\,d\xv$ instead gives
$S_{jk}(u\mid\bar{\xv}_{jk})\neq\Sbar_{jk}(u)$, the multistate analogue of the aggregation bias that
motivates ML-NMR.

Only summaries of the covariates are reported, so the baseline joint distribution $f_{jk}(\xv)$ is
reconstructed by matching each covariate's marginal to its summary and inducing dependence through a
Gaussian copula whose correlation is borrowed from the IPD studies~\citep{phillippo2020mlnmr}
(Appendix~\ref{app:qmc}). That this dependence transports across populations is an assumption probed
by simulation (Section~\ref{sec:sim}).

The marginal occupancies give the interval conditional-survival probabilities (covariate-marginal
quantities) for the life-table intervals of the published curves. A life-table interval $[u_m,u_{m'})$,
with $m'>m$, runs between two points of the occupancy grid and may span several of its intervals; it is
labeled by the index $m$ of its start. The probability is the chance of surviving
from $u_m$ to $u_{m'}$ given survival to $u_m$, for PFS (staying stable) and OS (staying alive):
\begin{equation}
p^{\mathrm{cPFS}}_{jk,m}=\frac{\Sbar_{jk}(u_{m'})}{\Sbar_{jk}(u_m)},\qquad
p^{\mathrm{cOS}}_{jk,m}=\frac{\Sbar_{jk}(u_{m'})+\Pbar_{jk}(u_{m'})}{\Sbar_{jk}(u_m)+\Pbar_{jk}(u_m)}.
\label{eq:cond}
\end{equation}
The aggregate curves
are summarized (Appendix~\ref{app:data}) into interval conditional-survival counts: for each interval $[u_m,u_{m'})$, the number at risk $n^c$ at its start $u_m$ and the number of event-free survivors $r^c$
implied by the curve at its end $u_{m'}$, separately for PFS and OS. The occupancy grid may be refined below the life-table intervals without changing the form of the
likelihood or the counts entering it; in the analyses here
the two coincide at monthly intervals ($m'=m+1$). Following Jansen et al.~\citep{jansen2023multistate}, each interval
contributes a binomial conditional-survival likelihood,
\begin{equation}
r^{\mathrm{cPFS}}_{jk,m}\sim\mathrm{Binomial}\big(n^{\mathrm{cPFS}}_{jk,m},\,p^{\mathrm{cPFS}}_{jk,m}\big),
\qquad
r^{\mathrm{cOS}}_{jk,m}\sim\mathrm{Binomial}\big(n^{\mathrm{cOS}}_{jk,m},\,p^{\mathrm{cOS}}_{jk,m}\big).
\label{eq:agdbinom}
\end{equation}
This is the form the aggregate marginal likelihood \eqref{eq:aggmar} takes for each endpoint. Grouping each
patient's PFS or OS time into the life-table intervals, with censoring at interval ends, makes the outcome
discrete, and the product then factorizes by interval into these binomials, because the marginal occupancy at $u_m$ ($\Sbar_{jk}$ for PFS,
$\Sbar_{jk}+\Pbar_{jk}$ for OS) is the product of the preceding conditional-survival probabilities.
The same factorization shows why each probability in \eqref{eq:cond} is a ratio of marginal occupancies
rather than the average of each patient's interval survival over $f_{jk}$ (for PFS,
$S_{jk}(u_{m'}\mid\xv)/S_{jk}(u_m\mid\xv)$): the patients at risk at $u_m$ are survivors, whose covariate
distribution is reweighted by survival, and averaging over that distribution gives the ratio exactly
(Appendix~\ref{app:data}).

The construction assumes that \eqref{eq:occS}--\eqref{eq:occD}, evaluated at the intensities
\eqref{eq:lp}, are the covariate-conditional occupancies. This rules out an unobserved frailty, since
a frailty-marginal process would not be Markov. The assumption concerns the individual-level model,
so it applies equally to the IPD likelihood \eqref{eq:ipdlik}. Aggregate curves cannot separate a
frailty from the rest of the model: the freely estimated study-specific baselines absorb its effect on the shape of the
hazards, and the treatment effects absorb its weakening of the between-arm contrast, as bias (Section~\ref{sec:sim}). No-frailty is therefore an
assumption the aggregate data cannot test; it is probed by simulation (Section~\ref{sec:sim}; listed in
Section~\ref{sec:sensitivity}).

\subsubsection{Composite versus multinomial aggregate likelihood} \label{sec:com_multinomial}
Each of the binomials is exact for its own endpoint. But PFS and OS are dependent functionals of the same trajectory (being stable
contributes to both), so treating them as two independent binomials forms a composite
(pseudo-)likelihood \citep{varin2011composite}: point estimates remain consistent, but the resulting
pseudo-posterior need not be calibrated: its credible intervals (CrIs) lack an exact coverage
interpretation. Information criteria can in principle be adapted to a composite likelihood \citep{varin2011composite}, but the composite log-likelihood is not the
log predictive density of the aggregate data. The deviance information criterion (DIC), the widely
applicable information criterion (WAIC) and leave-one-out cross-validation (LOO) computed from it
would treat the dependent PFS and OS contributions as independent, overstating the effective sample
size. Each endpoint's binomial likelihood, however, is exact for that endpoint, so a criterion computed
from the PFS terms alone, or from the OS terms alone, does not count the shared information twice. Models
can therefore be compared on three separate contributions, the IPD, the aggregate PFS and the aggregate OS,
overall and by study (Section~\ref{sec:computation}). The aggregate PFS and OS criteria must not be added to
each other, although either may be added to the IPD criterion, since the IPD studies contribute other
patients. A leave-one-out criterion for one endpoint is still computed under the composite
pseudo-posterior and remains conditional on the other endpoint, whose count for the left-out interval
shares patients with it and stays in the fit; leaving out both counts of each interval together removes
this overlap (Appendix~\ref{app:baselines}). The alternative treats the interval jointly: those alive at $u_m$ are
partitioned at $u_{m'}$ into the three occupancy states $(n_S,n_P,n_D)_{jk,m}$. Throughout, a single
subscript $m$ labels the \emph{interval} $[u_m,u_{m'})$, not the time at which a quantity is
evaluated, so counts observed at the interval end $u_{m'}$ still carry the interval's index $m$; the
evaluation time appears only inside the occupancy arguments. The covariate-marginal
probabilities of the three states follow from the marginal occupancies \eqref{eq:marg} as
$\bar q_{S,jk,m}=\Sbar_{jk}(u_{m'})/A_{jk,m}$, $\bar q_{P,jk,m}=\Pbar_{jk}(u_{m'})/A_{jk,m}$ and
$\bar q_{D,jk,m}=1-\bar q_{S,jk,m}-\bar q_{P,jk,m}$, with
$A_{jk,m}=\Sbar_{jk}(u_m)+\Pbar_{jk}(u_m)$ the marginal probability of being alive at $u_m$, giving a
single multinomial
\begin{equation}
(n_S,n_P,n_D)_{jk,m}\sim\mathrm{Multinomial}\big(n^{\mathrm{cOS}}_{jk,m},\,
(\bar q_{S,jk,m},\bar q_{P,jk,m},\bar q_{D,jk,m})\big),
\label{eq:multinom}
\end{equation}
a marginalized Welton--Ades aggregate-Markov likelihood \citep{weltonades2005} that respects the
PFS/OS dependence within each interval. It conditions each interval only on the number alive at $u_m$, however,
not on how many of them are stable and how many progressed, so the stable fraction observed in earlier intervals
re-enters every later one: the product over intervals is not the exact grouped-data likelihood, and its posterior is
overconfident even when the counts are genuine.

The choice, however, is not free, because it turns on what AgD are available. The
multinomial requires the joint state-occupancy counts $(n_S,n_P,n_D)$ for a common at-risk set. These
are directly available from IPD arms, from studies that report partitioned-survival or stacked
state-occupancy tables, and for a reference-treatment arm for a target population when it comes from
patient-level registry data (Section~\ref{sec:estimands}).
When a study reports only separate PFS and OS Kaplan--Meier curves (the common
case), the occupancy probabilities are still recovered exactly ($S=\text{PFS}$, $P=\text{OS}-\text{PFS}$,
$D=1-\text{OS}$), but turning them into joint counts requires a shared at-risk denominator and
so treats the two curves, with generally different censoring, as a single observed partition, an
extra assumption the two-binomial form does not make. The two-binomial thus uses only what such a
study directly reports, without a guarantee of calibration; the multinomial respects the dependence but
assumes joint counts that, absent occupancy reporting, must be reconstructed, and as specified in \eqref{eq:multinom} it
overstates the information they carry. The two-binomial is therefore the default (continuity with Jansen
et al.~\citep{jansen2023multistate}). Using genuine state-occupancy counts efficiently would need a multinomial that conditions
each interval on the stable and progressed counts at its start, which is not developed here. The simulation sub-study of
Appendix~\ref{app:calibration} shows the practical consequence: the multinomial is overconfident, whereas the
two-binomial stays close to nominal.

\subsubsection{Joint likelihood}\label{sec:joint}
The two data types enter a single synthesis through their contributions under
the shared parameter vector $\bm\xi$,
\begin{equation}
L(\bm\xi)\;=\;\prod_{j\in\mathcal{J}_\ipd}\prod_{k\in\mathcal{K}_j}\prod_{i=1}^{N_{jk}}
 L^{\mathrm{Con}}_{ijk}(\bm\xi;y_{ijk},\xv_{ijk})
\;\times\;\prod_{j\in\mathcal{J}_\agd}\prod_{k\in\mathcal{K}_j}\prod_{m}
 L^{\mathrm{Mar}}_{\bullet jk,m}(\bm\xi),
\label{eq:joint}
\end{equation}
where $L^{\mathrm{Mar}}_{\bullet jk,m}$ is the contribution of interval $m$ to the aggregate marginal likelihood \eqref{eq:aggmar}: the
two-binomial composite \eqref{eq:agdbinom} for curve-only evidence, or the multinomial \eqref{eq:multinom}
when joint occupancy counts are available. IPD studies enter through the individual
multistate likelihood \eqref{eq:ipdlik}. Both contributions are built from the same model of Section~\ref{sec:model}.
Because every study informs the same transition-level regression
parameters, the IPD studies identify the covariate--effect-modifier relationships that de-bias the
aggregate studies, while the aggregate studies extend the network's connectivity and precision, the
essence of multilevel network meta-regression. Weakly informative priors on $\bm\xi$ complete the
model; estimation and diagnostics follow in Section~\ref{sec:computation}.

\subsection{Producing estimates for a target population}\label{sec:estimands}
ML-NMR-MS provides \emph{conditional} and \emph{marginal} relative treatment effects, together with \emph{absolute} quantities, for a target population $\mathcal{P}$ with covariate distribution $f^{(\mathcal{P})}$. This target population does not need to be represented by one of the studies in the network, and in
practice is often best represented by a registry or cohort study \citep{phillippo2024general}; transporting relative effects to it
rests on the conditional constancy of relative effects of Section~\ref{sec:model}. The two kinds of relative effect do not coincide: HRs are non-collapsible, so averaging a contrast over $\mathcal{P}$ and contrasting population
averages generally give different answers, and which is wanted depends on the decision
\citep{phillippo2025effectmod,remiroazocar2025marginal}.

\paragraph{Conditional treatment effects.} A population-average conditional treatment effect is the effect of one treatment versus another for patients who share the same covariates, averaged over the covariate distribution of $\mathcal{P}$. Because the linear predictor \eqref{eq:lp} is linear in $\xv$, contrasts integrate to plug-in means, giving the population-average conditional transition log-hazard ratio
\begin{equation}
d^r_{ab}(u;\mathcal{P})=\bm\varphi^r(u)^{\!\top}\big(\bm\gamma^r_b-\bm\gamma^r_a\big)
 +\bar\xv_{(\mathcal{P})}^{\!\top}\big(\bet^r_{2,b}-\bet^r_{2,a}\big),
\label{eq:condHR}
\end{equation}
where $\bar\xv_{(\mathcal{P})}=\int\xv f^{(\mathcal{P})}(\xv)\,d\xv$ is the population-mean covariate vector. It depends only on the mean of the effect-modifying covariates in $\mathcal{P}$, not on the distribution of purely prognostic covariates or on the baseline hazard. It varies with time whenever $\bm\gamma^r_b$ and $\bm\gamma^r_a$ differ on the non-intercept components of the basis.

Decisions, however, are made on endpoints rather than on single transitions. Conditioning PFS and OS on the same mean covariate profile gives
\begin{equation}
\begin{aligned}
\mathrm{HR}^{\mathrm{PFS}}_{ab}(u;\mathcal{P})
 &=h^{\mathrm{PFS}}_b(u\mid\bar\xv_{(\mathcal{P})})\big/h^{\mathrm{PFS}}_a(u\mid\bar\xv_{(\mathcal{P})}),\\[2pt]
\mathrm{HR}^{\mathrm{OS}}_{ab}(u;\mathcal{P})
 &=h^{\mathrm{OS}}_b(u\mid\bar\xv_{(\mathcal{P})})\big/h^{\mathrm{OS}}_a(u\mid\bar\xv_{(\mathcal{P})}),
\end{aligned}
\label{eq:condEndpointHR}
\end{equation}
with both endpoint hazards formed from the occupancy \eqref{eq:occS}--\eqref{eq:occD} propagated at
the fixed covariate value $\bar\xv_{(\mathcal{P})}$ under the baseline specified for $\mathcal{P}$, so
that no study index appears: $h^{\mathrm{PFS}}_k(u\mid\xv)
=-\mathrm{d}\log S_k(u\mid\xv)/\mathrm{d}u$ and $h^{\mathrm{OS}}_k(u\mid\xv)
=-\mathrm{d}\log\big(S_k(u\mid\xv)+P_k(u\mid\xv)\big)/\mathrm{d}u$. These do not follow from
\eqref{eq:condHR}: $h^{\mathrm{PFS}}_k=h^{SP}_k+h^{SD}_k$ pools two intensities, and
$h^{\mathrm{OS}}_k=\big(S_k\,h^{SD}_k+P_k\,h^{PD}_k\big)\big/(S_k+P_k)$ weights the two death intensities by where the patient is, so in neither case does the baseline cancel from the ratio. An endpoint ratio is therefore time-varying in general, even when every transition carries a constant conditional HR. The PFS ratio is then constant exactly when stable$\to$progression and stable$\to$death carry the same HR, or when their intensities stay proportional to each other over time. The OS ratio also depends on the share of survivors who are still stable, $S_k/(S_k+P_k)$, which treatment shifts, so even a common HR on all three transitions leaves it time-varying unless $h^{SD}_k=h^{PD}_k$. The endpoint log-hazard ratios are also not linear in $\xv$, so, unlike \eqref{eq:condHR}, \eqref{eq:condEndpointHR} is the effect for a patient at the mean covariate profile rather than the average of the covariate-conditional endpoint effects over $\mathcal{P}$.

\paragraph{Marginal treatment effects.} A marginal effect is the effect of one treatment versus another for the target population as a whole: outcomes are averaged over $f^{(\mathcal{P})}$ first and contrasted second, rather than contrasted at fixed covariates and averaged after. The averaging is over covariate-conditional occupancies, by Monte Carlo integration over draws from $f^{(\mathcal{P})}$, giving the population-average occupancies
$\Sbar^{(\mathcal{P})}_k(u)=\int S_k(u\mid\xv)f^{(\mathcal{P})}(\xv)\,d\xv$ and
$\Pbar^{(\mathcal{P})}_k(u)$, the target-population form of \eqref{eq:marg}. These occupancies, and the survival and state-membership curves read off them, are the \emph{absolute} estimands. The population-average hazards follow: at the transition level as occupancy-weighted mean intensities, at the endpoint level as the hazards of the marginal survival curves,
\begin{equation}
\begin{gathered}
\bar h^{r,(\mathcal{P})}_k(u)=\frac{\int h^r_k(u\mid\xv)\,Z^r_k(u\mid\xv)\,f^{(\mathcal{P})}(\xv)\,d\xv}
 {\int Z^r_k(u\mid\xv)\,f^{(\mathcal{P})}(\xv)\,d\xv},\\[3pt]
\bar h^{\mathrm{PFS},(\mathcal{P})}_k(u)=-\frac{\mathrm{d}}{\mathrm{d}u}\log\Sbar^{(\mathcal{P})}_k(u),
\qquad
\bar h^{\mathrm{OS},(\mathcal{P})}_k(u)=-\frac{\mathrm{d}}{\mathrm{d}u}
 \log\big(\Sbar^{(\mathcal{P})}_k(u)+\Pbar^{(\mathcal{P})}_k(u)\big),
\end{gathered}
\label{eq:marghaz}
\end{equation}
where $Z^r_k$ is the occupancy of the state from which $r$ departs ($S_k$ for stable$\to$progression
and stable$\to$death, $P_k$ for progression$\to$death): each individual intensity is weighted by the
probability that the patient is still present to make that transition. With this weighting the
endpoint-level hazards keep the form of their conditional counterparts: dropping the superscript
$(\mathcal{P})$, $\bar h^{\mathrm{PFS}}_k=\bar h^{SP}_k+\bar h^{SD}_k$ and
$\bar h^{\mathrm{OS}}_k=\big(\Sbar_k\,\bar h^{SD}_k+\Pbar_k\,\bar h^{PD}_k\big)\big/\big(\Sbar_k+\Pbar_k\big)$.
They are evaluated numerically by backward differences on the occupancy grid. The marginal HR is then
\begin{equation}
\Delta^{\mathrm{HR},e}_{ab}(u;\mathcal{P})
 =\bar h^{e,(\mathcal{P})}_b(u)\big/\bar h^{e,(\mathcal{P})}_a(u),
\label{eq:margHR}
\end{equation}
where $e$ indexes the level, either a transition $r$ or an endpoint (PFS or OS); it is the marginal counterpart of the conditional transition HR $\exp\{d^r_{ab}(u;\mathcal{P})\}$ from \eqref{eq:condHR} and of the conditional endpoint HRs \eqref{eq:condEndpointHR}, formed from population-average rather than covariate-conditional hazards. Unlike the conditional effects, which use the covariates only through their mean in $\mathcal{P}$, the marginal ones depend on the whole distribution of every covariate, prognostic as well as effect-modifying. Like the conditional endpoint effects, they also depend on the baseline hazard. In particular, when the covariates affect the intensities the marginal HR is in general time-varying even under a conditional PH model, because the covariate composition of each arm's at-risk set (the weights $Z^r_a$ and
$Z^r_b$ in \eqref{eq:marghaz}) changes over follow-up, and differently in the two arms, so the marginal HR varies over time even when every conditional HR is constant. It stays constant only in the special case where treatment leaves the departing state's occupancy unchanged.

Absolute quantities, marginal effects and the conditional endpoint HRs \eqref{eq:condEndpointHR} also require a baseline hazard
for $\mathcal{P}$, which the network does not supply; only the conditional transition effects \eqref{eq:condHR} are free of this
requirement. The baseline can be observed in $\mathcal{P}$ itself, for example by entering a reference-treatment curve from a
representative registry as a single-arm study, and then no further assumption is needed. If it is instead borrowed from a study
in the network, a judgment is required that the source's covariate-conditional baseline applies in $\mathcal{P}$. That judgment
attaches to the baseline rather than to the relative effects: randomization protects the contrasts within a study but not the
absolute level of risk.

\subsection{Estimation}\label{sec:computation}
Estimation is performed in a Bayesian framework. The log-likelihood of Section~\ref{sec:datalik} is coded directly, with the
covariate integral \eqref{eq:marg} evaluated inside the model, and the posterior is sampled by
Hamiltonian Monte Carlo in Stan \citep{carpenter2017stan}; Weibull shapes are sampled on the log scale,
and a dense mass matrix improves mixing because the transition parameters are strongly correlated
\emph{a posteriori}. The number of quasi-Monte Carlo integration points $\tilde N$ governs the accuracy
of \eqref{eq:marg} and is set by the doubling rule of Phillippo et al.~\citep{phillippo2024general}:
half the chains are run at $\tilde N$ and half at $\lceil\tilde N/2\rceil$, and, once the $\hat R$ within
each half indicates convergence, $\tilde N$ is doubled until no $\hat R$ computed across all chains
exceeds $1.05$. Because $\hat R$ responds only weakly to a small shift between resolutions, the posterior
medians are worth comparing across resolutions as well (Appendix~\ref{app:qmc}).

Two further discretizations enter the aggregate likelihood. The first is the partition of follow-up
into the life-table intervals that carry the conditional-survival counts, chosen when the counts are
constructed from the digitized curves (Appendix~\ref{app:data}). Shorter intervals retain more of the
time course in the curve, bringing the grouped likelihood closer to that of the reconstructed event
times and reducing the error from ignoring censoring within an interval. They cannot, however, recover
detail beyond the resolution of the digitized curve, their counts become sensitive to digitization
error, and they raise the computational cost, because each life-table interval must be a union of
occupancy-grid intervals. Longer intervals are less sensitive to digitization error but blur the time course of the hazards. The
second discretization is the occupancy grid. It may be refined below the life-table intervals
without altering the counts, and a finer grid approximates the time-varying intensities, and hence the occupancies, more accurately
at proportionally higher cost; the error is largest where the hazard changes fastest relative to the
spacing, typically just after randomization. The grid can be checked by halving the spacing and confirming
that the posterior summaries move by only a small fraction of their posterior standard deviations. No
refit is needed: the grid enters only the aggregate likelihood, so the existing draws can be reweighted
to the finer grid by Pareto-smoothed importance sampling, whose diagnostic $\hat k$ shows whether the
reweighting is reliable \citep{vehtari2024psis}.

Fitting requires several specification choices, each made per transition. The baseline time function sets how the baseline hazard varies with time, ranging from a constant
(exponential)
hazard through Weibull, Gompertz, and fractional polynomials to M-splines (Section~\ref{sec:model});
transitions need not share a form. Which transitions carry a treatment effect is
itself a modeling choice, since a treatment may act on progression without acting on post-progression
mortality. Each such effect may be constant (proportional hazards) or time-varying, the latter obtained by giving $\bm\gamma^r_k$ non-zero components on the time basis rather than an intercept alone. Finally,
the effect-modifier interactions may be treatment-specific, shared within a treatment class, or common
to all active treatments (Section~\ref{sec:model}), trading flexibility against identifiability.

Prior choice depends on the evidence base rather than being fixed by the method. Where the network
carries substantial individual data, non-informative or weakly informative priors on all parameters
suffice and the posterior is dominated by the likelihood. Where evidence is sparse, or where a quantity
is only weakly identified from aggregate curves, more informative priors on the affected baselines or
coefficients stabilize estimation. The stable$\to$death intensity is the usual candidate: the
conditional-survival counts inform the stable and progressed occupancies directly, but where
pre-progression deaths are rare this intensity is a small residual of the total stable-state exit rate
once progressions are accounted for, and is correspondingly imprecise.

Competing specifications can be compared by leave-one-out cross-validation \citep{vehtari2017loo} on
the IPD contribution, whose pointwise terms are independent observations, and, as separate criteria, on
the aggregate PFS and the aggregate OS contributions, whose pointwise terms are the conditional-survival
binomials of their life-table intervals (Section~\ref{sec:com_multinomial}). Each criterion should be reported for each study as well as in total, since poor fit in
one study can be masked by good fit in others. Because late follow-up, where few patients remain at
risk, contributes little to such a criterion, the fitted PFS and OS curves of every arm should also
be compared with the observed Kaplan--Meier curves and, where IPD record the transitions, observed
with expected transition counts.

\subsection{Assumptions}\label{sec:sensitivity}
The assumptions divide into those in force for every specification and those incurred by particular
modeling choices.
\begin{itemize}[leftmargin=1.4em,itemsep=1pt,topsep=2pt]
\item \emph{Conditional constancy of relative effects.} At any given covariate value the relative
effect on each transition is the same in every study in which that comparison could have been made,
and so transports to a target population; absolute outcomes may still differ between study
populations. This is the central assumption on which the synthesis of relative effects rests, and it is the conditional form
of what the network meta-analysis literature calls transitivity or exchangeability: consistency of
the network is its consequence, not a separate requirement, so a single set of $\bm\gamma^r_k$
relative to the network reference reproduces both the direct and the indirect evidence, and a
violation surfaces as residual heterogeneity or inconsistency
\citep{phillippo2022validating,phillippo2020mlnmr}. It is weaker than the unconditional constancy
assumed by aggregate-data NMA, which requires the covariates themselves to be balanced. Equivalently,
the treatment and effect-modifier coefficients carry no study index: only the baseline does. Under the
random-effects specification of Section~\ref{sec:model} the treatment effects are instead exchangeable
about a common mean, while the effect-modifier coefficients remain common. Either way, every covariate
that modifies a relative effect must be measured and included, and each treatment's effect modification
is a single network-level quantity rather than a study-level one.
\item \emph{Clock-forward intensities.} Post-progression mortality depends on time since randomization
rather than on how long a patient has already spent in the progressed state, so the future of the
process is determined by the current state and current time alone and the occupancy solves
\eqref{eq:occS}--\eqref{eq:occD}.
\item \emph{No unobserved shared frailty.} Conditional on the covariates, no residual patient-level
trait links the transitions: patients who progress early are not, for that reason, also more likely to
die early.
\item \emph{Aggregate likelihood form.} With curve-only evidence the PFS and OS streams are taken to
carry independent information, the two-binomial composite scoring each against its own risk set; the
multinomial \eqref{eq:multinom} instead takes the reported counts to identify the joint
stable/progressed/dead occupancy.
\item \emph{Covariate-distribution reconstruction.} An aggregate study's joint covariate distribution
is the one implied by its reported covariate summaries, under assumed marginal forms, together with a
correlation structure taken from the IPD studies~\citep{phillippo2020mlnmr}
(Appendix~\ref{app:qmc}); the dependence between covariates is assumed not to differ across
populations.
\end{itemize}
The following arise from specification choices.
\begin{itemize}[leftmargin=1.4em,itemsep=1pt,topsep=2pt]
\item \emph{Baseline-hazard form.} The chosen baseline time function can represent the true shape
of each transition's baseline hazard over the follow-up modeled.
\item \emph{Treatment-effect placement and time course.} A transition given no treatment term is assumed
unaffected by treatment. Where a transition carries one, the time-basis components of $\bm\gamma^r_k$ can
represent the true time course of the effect; an intercept alone holds the hazard ratio constant over
time (proportional hazards).
\item \emph{Shared effect-modifier assumption.} The interactions $\bet^r_{2,k}$ are specified as treatment-specific, shared
within a treatment class, or common to all active treatments (Section~\ref{sec:model}). Where the
available data are insufficient to estimate independent interactions for every treatment, a shared
effect-modifier assumption is needed, and treatments pooled at the chosen level are modified
identically: a covariate modifying one treatment's effect on a transition modifies the others' by the
same amount.
\item \emph{Fixed versus random effects.} With $\bm\Sigma=\bm 0$ the covariate-adjusted relative
effects are identical across studies; the random-effects model instead takes them to be exchangeable
draws about a common mean (Section~\ref{sec:model}).
\end{itemize}

\section{Illustrative example}\label{sec:example}

\subsection{Evidence base}\label{sec:ndmm-network}
ML-NMR-MS is illustrated on a partly individual-level, partly aggregate network of
maintenance-therapy trials in newly diagnosed multiple myeloma (NDMM), taken from the \texttt{multinma} package \citep{phillippo2026multinma}. Its IPD are simulated to
resemble the original trials and its aggregate PFS is reconstructed from digitized curves. The package provides only PFS,
so an OS layer is simulated here for this example (Appendix~\ref{sec:ndmm-data}). The
example therefore demonstrates the method rather than reporting empirical findings about these
treatments.

Three treatments are compared: placebo/observation (Pbo, the network reference), lenalidomide (Len)
and thalidomide (Thal) (Figure~\ref{fig:network}). IPD are available
for three lenalidomide-versus-placebo trials (Attal2012~\citep{attal2012} ($n=614$), McCarthy2012~\citep{mccarthy2012}
($n=460$) and Palumbo2014~\citep{palumbo2014} ($n=251$)), and aggregate PFS/OS curves with baseline covariate
summaries for two more: Jackson2019~\citep{jackson2019} (lenalidomide vs placebo, $n=2001$) and Morgan2012~\citep{morgan2012}
(thalidomide vs placebo, $n=818$). The IPD trials supply individual PFS and OS with the progression indicator, together with
covariates; the aggregate trials supply PFS/OS curves and published covariate summaries. Baseline characteristics by trial and arm are given
in Table~\ref{tab:trials} and the trial-specific Kaplan--Meier curves in Figure~\ref{fig:km}
(Appendix~\ref{app:trials}); the structure of the analysis data is set out in
Appendix~\ref{app:datastruct}. Three covariates enter the model: age, International Staging System stage III (ISS-III), and achievement
of a complete or very-good-partial response after autologous stem cell transplant (ASCT). The populations are imbalanced: the IPD trials enrolled younger patients (mean age $\approx55$) and
the aggregate trials older ones ($\approx65$), with the measured covariates distributed differently
across trials (Table~\ref{tab:trials}).

\IfFileExists{figures/fig_ndmm_network.png}{%
\begin{figure}[htbp]\centering
\includegraphics[width=0.5\linewidth]{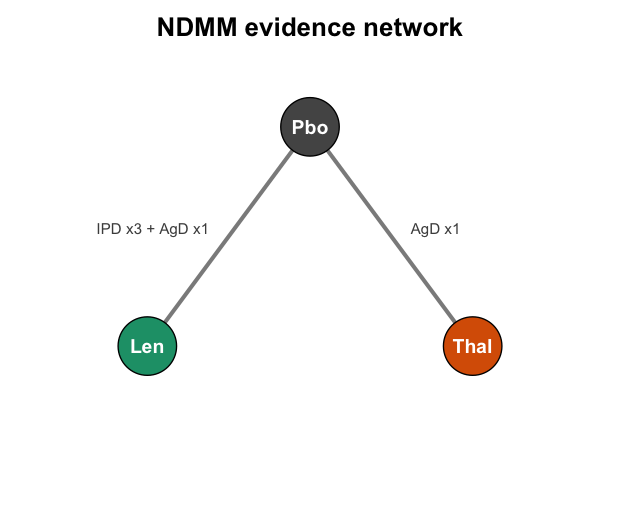}
\caption{NDMM evidence network. Nodes are treatments; edges are pairwise comparisons labeled by the
number of contributing IPD and AgD trials.}
\label{fig:network}
\end{figure}}{}

\subsection{Model and estimation}\label{sec:ex-model}
ML-NMR-MS is fitted by combining the full multistate likelihood \eqref{eq:ipdlik} for the three IPD
trials with the aggregate covariate-marginal state-occupancy likelihood \eqref{eq:marg}--\eqref{eq:agdbinom}
for the two AgD trials. The covariate vector is $\xv=(\text{age},\ \text{ISS-III},\ \text{ASCT response})$,
with age standardized as $(\text{age}-60)/10$; baselines are stratified by study $j$, and treatment
$1$ is placebo. The three transition log-intensities are
\begin{equation}
\begin{aligned}
\log h^{SP}_{jk}(u\mid\xv) &= c^{SP}_j + (v^{SP}_j-1)\log u + \xv^{\!\top}\bet^{SP}_1
   + \mathds{1}(k\neq1)\big[\gamma^{SP}_k + \gamma^{SPt}_k\log u + \xv^{\!\top}\bet^{SP}_2\big],\\
\log h^{SD}_{jk}(u\mid\xv) &= c^{SD}_j + (v^{SD}_j-1)\log u + \xv^{\!\top}\bet^{SD}_1,\\
\log h^{PD}_{jk}(u\mid\xv) &= c^{PD}_j + \xv^{\!\top}\bet^{PD}_1 + \mathds{1}(k\neq1)\,\gamma^{PD}_k .
\end{aligned}
\label{eq:ex-lp}
\end{equation}
Here, for transition $r\in\{SP,SD,PD\}$ and arm $k$ of study $j$: $c^r_j$ is the study-specific
log-scale intercept and $v^r_j$ the Weibull shape of the baseline; $\gamma^{SP}_k$ and $\gamma^{PD}_k$
are the treatment scale log-hazard-ratios on stable$\to$progression and progression$\to$death;
$\gamma^{SPt}_k$ is the stable$\to$progression time-variation (its effect on the log-time slope);
$\bet^r_1$ are the prognostic covariate effects on transition $r$; and $\bet^{SP}_2$ are the
effect-modifier (treatment-by-covariate) interactions on stable$\to$progression. All treatment terms
($\gamma$, $\bet^{SP}_2$) are zero for placebo (treatment $1$). Thus stable$\to$progression has a Weibull
baseline whose treatment effect acts on both the scale ($\gamma^{SP}_k$) and the log-time slope
($\gamma^{SPt}_k$), giving the non-proportional HR
$\mathrm{HR}^{SP}_k(u)=\exp(\gamma^{SP}_k+\xv^{\!\top}\bet^{SP}_2)\,u^{\gamma^{SPt}_k}$;
stable$\to$death has a Weibull baseline and no treatment effect; and progression$\to$death is
exponential (shape fixed at one, $v^{PD}_j\equiv1$) with a constant treatment HR
$\exp(\gamma^{PD}_k)$, a parsimonious choice given the immature overall survival. The Weibull
stable$\to$progression baseline is the specification presented here;
Appendix~\ref{app:baselines} refits it as a second-order fractional polynomial and as a degree-3
M-spline with a shape common to all studies, the fractional polynomial being preferred on leave-one-out cross-validation
of both the IPD and the aggregate PFS (Table~\ref{tab:baseline-loo}), and compares the results the three forms imply. Prognostic effects
$\bet^r_1$ enter every transition; the effect-modifier interactions $\bet^{SP}_2$ act on
stable$\to$progression and are shared across the active treatments (one set for lenalidomide and
thalidomide, with the placebo interaction fixed at zero); thalidomide appears in a single aggregate
study, so its own interactions are not separately identified.

Weakly informative independent priors were used, on the log hazard per month with standardized
covariates: $c^r_j\sim\mathrm N(-3,2^2)$ on the log-scale intercepts ($95\%$ of the mass on monthly
baseline hazards between $0.001$ and $2.5$), $\log v^r_j\sim\mathrm N(0,0.5^2)$ on the log-shapes
(Weibull shapes between $0.4$ and $2.7$), $\gamma^{SP}_k,\gamma^{PD}_k\sim\mathrm N(0,2^2)$ on the
treatment effects (hazard ratios between $0.02$ and $50$), $\gamma^{SPt}_k\sim\mathrm N(0,1^2)$ on the
time variation (a hazard ratio changing by a factor between $1/7$ and $7$ per unit of $\log u$), and
$\bet^r_1,\bet^{SP}_2\sim\mathrm N(0,2^2)$ on the regression coefficients (hazard ratios between
$0.02$ and $50$ per covariate unit). With time in years each intercept shifts by $v^r_j\log 12$
($\log 12\approx2.5$ for an exponential baseline), and $\gamma^{SP}_k$, the log-hazard ratio at $u=1$, by
$\gamma^{SPt}_k\log 12$, so the priors must be moved with them.
The posterior variance is at most $6.6\%$ of the prior variance for every parameter except the
stable$\to$death shapes, which the rare pre-progression deaths identify only weakly ($22\%$ in
Jackson2019), so that is where an informative prior matters most (Appendix~\ref{app:priorpost}). An
analyst reusing these priors should check the same contraction in their own data. For each AgD arm the
covariate-marginal occupancies \eqref{eq:marg} are evaluated by quasi-Monte Carlo integration at
$\tilde N=32$ Sobol' points drawn from a Gaussian copula matched to the study's covariate
summaries, with the correlation between age and ASCT response borrowed from the IPD (Appendix~\ref{app:qmc}). The general
form allows an arm-specific $f_{jk}$. Here the arms of an aggregate study share one reconstructed distribution, because
randomization balances covariates within a study. In Morgan2012, however, the thalidomide arm is older than the placebo
arm (Table~\ref{tab:trials}), so that sharing pushes thalidomide's estimated effect toward the null. The life-table intervals and the occupancy grid are both monthly, and halving the grid moves the
posterior medians of the treatment effects and effect modifiers by at most $0.02$ posterior standard deviations (Appendix~\ref{app:grid}). By the $\hat R$-doubling rule $\tilde N=32$
suffices; doubling it to $64$ moves the posterior medians of the treatment effects and effect modifiers
by at most $0.13$ posterior standard deviations (Appendix~\ref{app:qmc}).

The model, whose Stan code is given in Appendix~\ref{app:stan}, was fitted with four chains of $3000$ iterations ($1500$ warmup; target acceptance $0.9$,
maximum tree depth $11$, dense mass matrix), started at the posterior means of a preliminary fit with
small random perturbations and run in parallel, taking about five minutes per chain on a laptop
(Appendix~\ref{app:mcmc}).
All rank-normalized $\hat R$ were at most $1.003$, and the Monte Carlo standard error of every
posterior median at most $0.02$ posterior standard deviations (Appendix~\ref{app:mcmc}).

As a check of fit, the model-implied PFS and OS of every arm were compared with the observed
Kaplan--Meier curves and, in the IPD arms, observed with expected transition counts
(Appendix~\ref{app:fit}). PFS is reproduced in every arm, and OS in every study except McCarthy2012 and
Palumbo2014, where the model understates the separation between lenalidomide and placebo because,
unlike the semi-synthetic data, it has no treatment effect on stable$\to$death.
The complete set of fitted parameters (treatment effects, effect modifiers, prognostic effects, and
study-specific baselines) is reported in Table~\ref{tab:recovery}.

\begin{table}[htbp]\centering\footnotesize\setlength{\tabcolsep}{4.5pt}
\caption{Estimated model parameters, by transition (posterior median [95\% CrI]).
Treatment effects are versus placebo: $\gamma^{SP}_k,\gamma^{PD}_k$ are the log-hazard-ratios (scale) and $\gamma^{SPt}_k$ the
stable$\to$progression time-variation; progression$\to$death has a constant effect and
stable$\to$death none. $\bet_1$ are prognostic effects; $\bet_2$ effect modifiers on the treatment
effect, shared across the active treatments. A dash denotes a parameter not in the model.
Time in months; age centered at 60 years, per 10 years. $c^r_j$ and $\gamma^{SP}_k$ are the log baseline
hazard and log-hazard ratio at $u=1$ month and $\xv=\bm 0$.}
\label{tab:recovery}
\begin{tabular}{lccc}
\toprule
Parameter & $S\!\to\!P$ & $S\!\to\!D$ & $P\!\to\!D$\\
\midrule
\multicolumn{4}{l}{\textit{Treatment effect, lenalidomide vs placebo}}\\
\quad $\gamma^r_k$ (scale)      & $-1.34\,[-1.73,-0.94]$ & --- & $-0.08\,[-0.26,+0.10]$\\
\quad $\gamma^{SPt}_k$ (time-var.) & $+0.16\,[+0.06,+0.27]$ & --- & ---\\
\multicolumn{4}{l}{\textit{Treatment effect, thalidomide vs placebo}}\\
\quad $\gamma^r_k$ (scale)      & $-0.53\,[-1.08,0.00]$ & --- & $-0.22\,[-0.62,+0.18]$\\
\quad $\gamma^{SPt}_k$ (time-var.) & $+0.07\,[-0.10,+0.24]$ & --- & ---\\
\multicolumn{4}{l}{\textit{Effect modifiers on the treatment effect ($\bet_2$; shared across active treatments)}}\\
\quad $\beta^{SP}_{2}$, age & $-0.01\,[-0.20,+0.19]$ & --- & ---\\
\quad $\beta^{SP}_{2}$, ISS-III & $+0.24\,[-0.12,+0.61]$ & --- & ---\\
\quad $\beta^{SP}_{2}$, ASCT response & $+0.21\,[-0.09,+0.52]$ & --- & ---\\
\multicolumn{4}{l}{\textit{Prognostic effects ($\bet_1$)}}\\
\quad $\beta^r_{1}$, age & $+0.76\,[+0.60,+0.92]$ & $+1.24\,[+0.87,+1.63]$ & $+0.22\,[+0.02,+0.42]$\\
\quad $\beta^r_{1}$, ISS-III & $+0.43\,[+0.15,+0.68]$ & $+0.24\,[-0.26,+0.70]$ & $+0.37\,[+0.09,+0.62]$\\
\quad $\beta^r_{1}$, ASCT response & $-0.19\,[-0.40,+0.02]$ & $+0.24\,[-0.17,+0.65]$ & $+0.10\,[-0.15,+0.35]$\\
\multicolumn{4}{l}{\textit{Study baseline log-intercepts $c^r_j$}}\\
\quad Attal2012      & $-4.30\,[-4.74,-3.90]$ & $-5.86\,[-6.74,-5.12]$ & $-4.01\,[-4.28,-3.73]$\\
\quad McCarthy2012   & $-4.74\,[-5.27,-4.25]$ & $-6.33\,[-7.30,-5.55]$ & $-4.06\,[-4.35,-3.78]$\\
\quad Palumbo2014    & $-3.35\,[-3.81,-2.93]$ & $-5.66\,[-6.74,-4.79]$ & $-3.85\,[-4.15,-3.56]$\\
\quad Jackson2019    & $-4.02\,[-4.36,-3.70]$ & $-7.73\,[-8.66,-6.94]$ & $-3.84\,[-4.15,-3.51]$\\
\quad Morgan2012     & $-4.02\,[-4.45,-3.64]$ & $-7.15\,[-8.32,-6.29]$ & $-3.94\,[-4.38,-3.52]$\\
\multicolumn{4}{l}{\textit{Study baseline Weibull shapes $v^r_j$ ($P\!\to\!D$ fixed at 1)}}\\
\quad Attal2012      & $1.34\,[1.21,1.48]$ & $1.07\,[0.82,1.36]$ & $1$\\
\quad McCarthy2012   & $1.32\,[1.17,1.48]$ & $1.04\,[0.80,1.34]$ & $1$\\
\quad Palumbo2014    & $1.06\,[0.91,1.22]$ & $1.05\,[0.72,1.42]$ & $1$\\
\quad Jackson2019    & $0.99\,[0.91,1.08]$ & $0.90\,[0.50,1.25]$ & $1$\\
\quad Morgan2012     & $0.96\,[0.84,1.09]$ & $1.13\,[0.67,1.45]$ & $1$\\
\bottomrule
\end{tabular}
\end{table}

\FloatBarrier
\subsection{Relative treatment effects}\label{sec:ex-relative}
Treatment effects enter on two transitions only: stable$\to$progression and progression$\to$death.  On stable$\to$progression, lenalidomide has a scale coefficient $\gamma^{SP}_{\mathrm{Len}}=-1.34$ with a positive
time-variation coefficient $\gamma^{SPt}_{\mathrm{Len}}=+0.16$, and thalidomide
$\gamma^{SP}_{\mathrm{Thal}}=-0.53$ with $\gamma^{SPt}_{\mathrm{Thal}}=+0.07$. On progression$\to$death both effects are small and their
intervals span zero (lenalidomide $-0.08$, thalidomide $-0.22$), and
stable$\to$death carries no treatment effect by specification. The effect-modifier interactions
$\bet^{SP}_2$, shared across the active treatments, are modest: $+0.24$ for ISS-III (a
smaller progression benefit in those patients), $+0.21$ for ASCT response, and essentially
null for age.

These coefficients translate into conditional HRs, reported in three
contrasting target populations that form the columns of Figure~\ref{fig:hrpops} (solid): a
\emph{favorable} population (mean age $53$ years, $5\%$ ISS-III, $85\%$ achieving an ASCT response),
an \emph{intermediate} population, taken to be that of the McCarthy2012 trial ($58$, $23\%$,
$67\%$), and an \emph{unfavorable} one ($67$, $40\%$, $55\%$). The favorable and unfavorable
populations are each represented by $1200$ Monte Carlo draws, with age normal (standard deviation $6$
years) and ISS-III and ASCT response independent; the intermediate one by the recorded covariates of the
McCarthy2012 patients. Every result that depends on the baseline hazard uses that of McCarthy2012,
which is thereby judged to apply in all three populations once their covariates are accounted for
(Section~\ref{sec:estimands}), and every derived quantity is computed from all $6000$ posterior draws.
The conditional HRs separate an effect on progression from an effect on post-progression survival. On stable$\to$progression the lenalidomide HR is time-varying: it is strongest early and attenuates
over follow-up, the positive $\gamma^{SPt}_{\mathrm{Len}}$ driving the
attenuation; the corresponding thalidomide coefficient is not distinguishable from zero. On
progression$\to$death the HR is constant because the treatment effect there carries only a scale
component, and it is close to null. The OS benefit is therefore largely inherited from the progression delay rather than a direct
post-progression effect. This is the structure built into the semi-synthetic OS layer
(Appendix~\ref{sec:ndmm-data}). It illustrates a decomposition that a multistate model provides and separate PFS and OS
analyses cannot; it is not a finding about these treatments. In that layer, part of the benefit also
runs through fewer pre-progression deaths, a route the example model omits (Appendix~\ref{app:fit}).
Across target populations these transition HRs shift only modestly. All three effect-modifier interactions
have intervals covering zero, so no direction of effect modification is established; the ISS-III and ASCT-response point
estimates are both positive, and because ISS-III becomes more prevalent and ASCT response less
prevalent in the less favorable populations, their contributions to
$\bar\xv_{(\mathcal{P})}^{\!\top}\bet^{SP}_2$ move in opposite directions and largely offset, while the
progression$\to$death HR is invariant, having no effect modifier and so transporting unchanged.
Combining the transitions gives the conditional HRs for PFS and OS; because an endpoint pools
transitions through the occupancy, these do shift with the target population (for lenalidomide the
conditional $12$-month OS HR moves from $0.68$ to $0.75$ between the favorable and unfavorable
populations). 

The marginal HRs are shown alongside the conditional ones in Figure~\ref{fig:hrpops} (dashed). These are attenuated toward the null relative to the conditional HRs by non-collapsibility, and
shift across the three populations by a similar amount ($0.71$ to $0.79$ for the same comparison). The
exception is thalidomide late in follow-up in the unfavorable population, where the marginal HR is no closer to the null than the
conditional one (stable$\to$progression at $48$ months in the unfavorable population: $0.92$ against
$0.93$; Table~\ref{tab:summ-sphr}). Appendix~\ref{app:baselines} gives the corresponding figures under each of the three stable$\to$progression baseline forms and, in Tables~\ref{tab:summ-sphr}--\ref{tab:summ-abs}, numerical summaries for each target population.

\FloatBarrier

\IfFileExists{figures/fig_ndmm_hr_pops.png}{%
\begin{figure}[htbp]\centering
\includegraphics[width=0.92\linewidth]{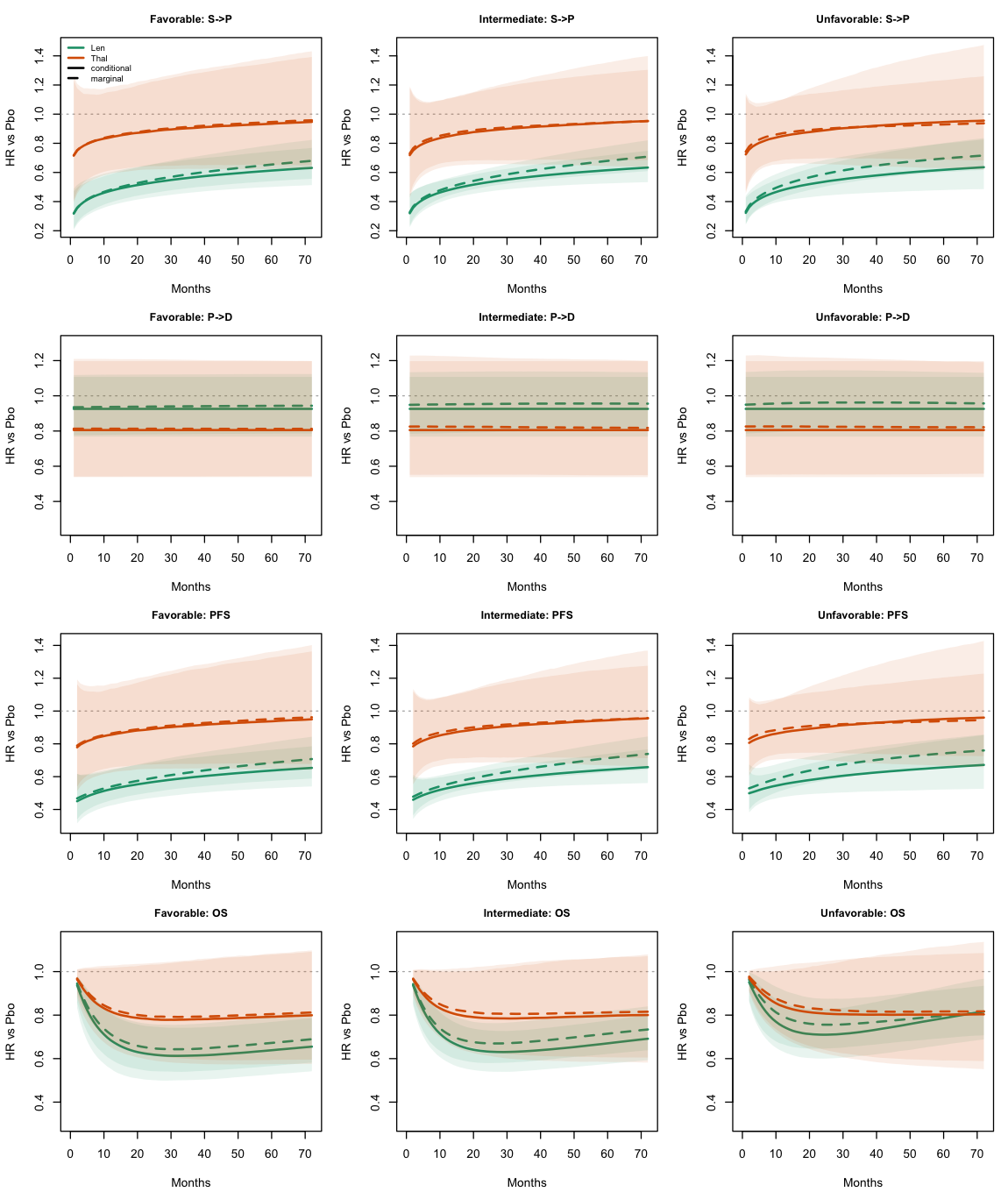}
\caption{Hazard ratios versus placebo in three target populations (columns) for lenalidomide (green)
and thalidomide (orange), each shown as conditional (solid, evaluated at the mean covariates of
the target population) and marginal (dashed, averaged over that population's covariate
distribution): the stable$\to$progression and progression$\to$death transition ratios (top two rows)
and the PFS and OS endpoint ratios (bottom two rows). The progression$\to$death conditional ratio is constant
across the target populations (it has no effect modifier, so it transports unchanged), whereas all other
ratios, conditional and marginal, shift with the target population; the conditional--marginal gap reflects
non-collapsibility and, for the endpoint ratios, evaluation at the mean covariate profile. Lines are posterior medians with
$95\%$ credible bands.}
\label{fig:hrpops}
\end{figure}}{}

\FloatBarrier

\subsection{Baseline hazards, state occupancy probabilities, and survival}\label{sec:ex-absolute}
\begin{sloppypar}
The estimated reference-treatment (placebo) transition hazard rates (an increasing
stable$\to$progression intensity, a low near-constant stable$\to$death intensity, and a constant
(exponential) progression$\to$death intensity) are shown across the same three populations in
Figure~\ref{fig:basepops}. They differ sharply between populations: the conditional stable$\to$progression intensity is about three and a half times higher in the
unfavorable than in the favorable population. The model holds that ratio constant over follow-up, because all
three populations share the McCarthy2012 baseline and the covariates act proportionally on it; the death intensities
are also higher in the unfavorable population.
\end{sloppypar}

\begin{sloppypar}
Each transition is shown both conditionally, at the population-mean covariates, and marginally, as the
occupancy-weighted population average (solid versus dashed). On both stable-state exits the marginal hazard starts above the conditional (by about $13\%$ for
progression and $30\%$ for death), the hazard being convex in the linear predictor, and then falls below it as the patients
at highest risk exit first and leave a progressively lower-risk stable population behind. On progression$\to$death the
marginal hazard also starts above, because early progressors are high-risk patients, and converges as new progressors
replenish the at-risk set. The gap that opens with follow-up is widest where the disease is
most aggressive.
\end{sloppypar}

Combining the conditional baseline hazards with the conditional relative effects gives each treatment's
occupancy trajectory at a given covariate value; averaging these trajectories over the target
population's covariate distribution gives the marginal state occupancy (the probability of being
stable, progressed or dead) over time (Figure~\ref{fig:occ}). Lenalidomide keeps a substantially larger fraction of
patients in the stable state and fewer dead; the progressed-state occupancy rises and then falls as patients move on to death, visibly so
within follow-up in the unfavorable population; and, like the baseline hazards, the whole occupancy shifts markedly across
populations, the disease running much faster in the unfavorable one.

Collapsing the occupancy to the two endpoints gives the marginal survival curves by treatment
(Figure~\ref{fig:survpops}). Across the populations the absolute survival differs
markedly, even though the treatment effects barely change: outcomes are driven mostly by the
baseline disease process rather than by the treatment effect.

\IfFileExists{figures/fig_ndmm_baseline_pops.png}{%
\begin{figure}[htbp]\centering
\includegraphics[width=\linewidth]{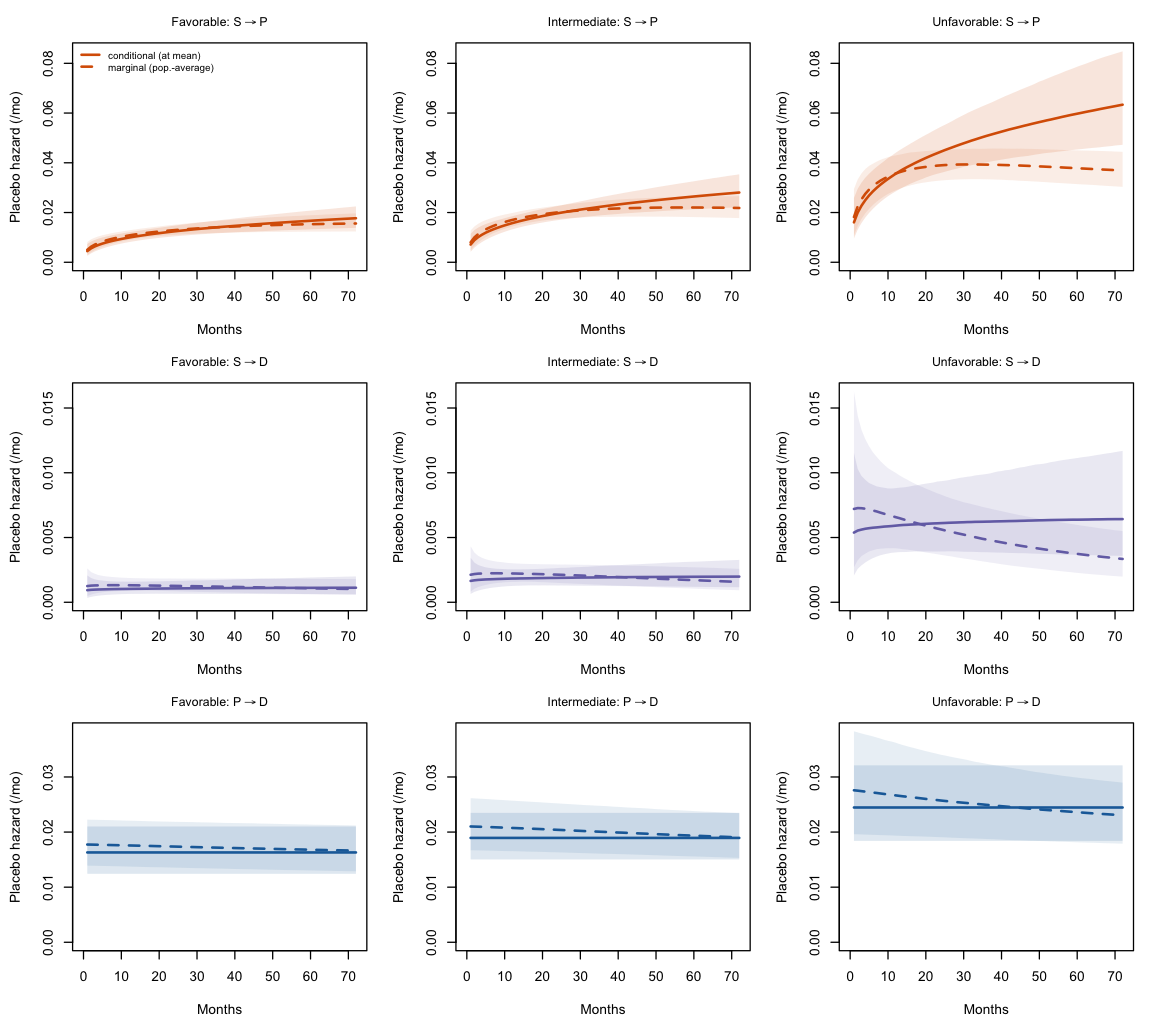}
\caption{Placebo baseline transition hazards over time (stable$\to$progression top, stable$\to$death
middle, progression$\to$death bottom) in three target populations (columns), each shown
conditionally (solid line: the hazard at the population-mean covariates) and
marginally (dashed line: the occupancy-weighted population-average hazard), both as posterior
median with 95\% CrI band.}
\label{fig:basepops}
\end{figure}}{}

\IfFileExists{figures/fig_ndmm_occupancy.png}{%
\begin{figure}[htbp]\centering
\includegraphics[width=\linewidth]{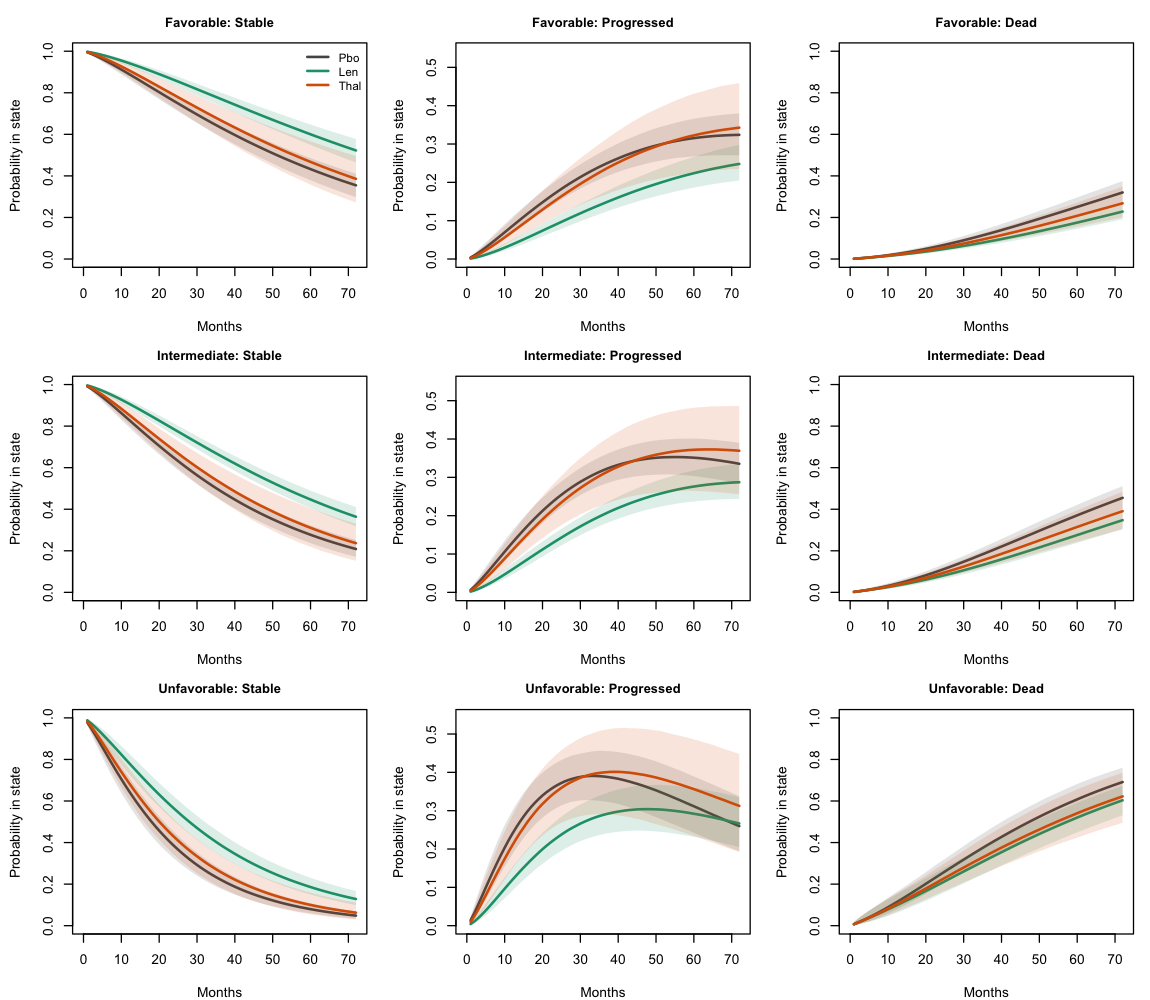}
\caption{Marginal state-occupancy probabilities over time (stable, progressed, and dead, columns) by
treatment (lines: placebo gray, lenalidomide green, thalidomide orange; posterior median with 95\%
CrI), in three target populations (rows). Progression-free survival is the stable probability and
overall survival the probability of being alive (stable or progressed).}
\label{fig:occ}
\end{figure}}{}

\IfFileExists{figures/fig_ndmm_surv_pops.png}{%
\begin{figure}[htbp]\centering
\includegraphics[width=\linewidth]{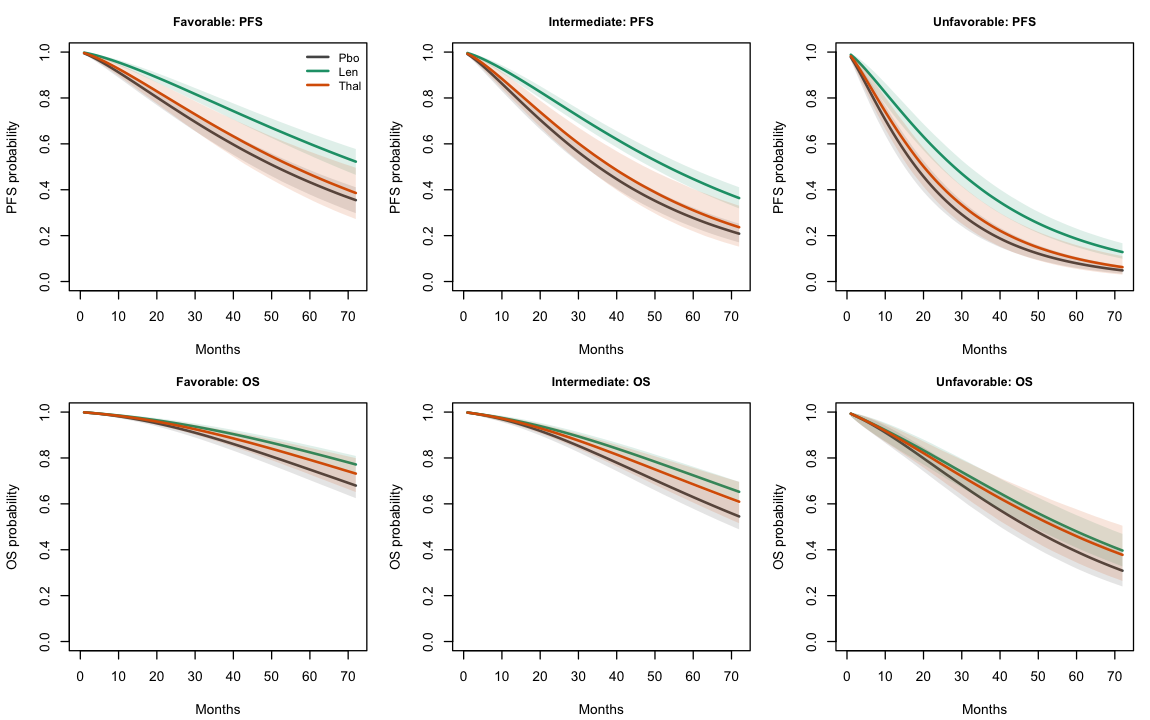}
\caption{Marginal PFS (top) and OS (bottom) by treatment in three target populations (columns), from
the same fitted model: posterior median with 95\% CrI. Only the target covariate distribution changes;
absolute survival shifts substantially.}
\label{fig:survpops}
\end{figure}}{}

\FloatBarrier
\section{Simulation study}\label{sec:sim}

ML-NMR-MS is assessed in a simulation study organized according to the ADEMP framework (aims, data-generating mechanisms, estimands, methods, performance measures)~\citep{morris2019adep}.

\subsection{Aims} The simulation has three aims: (i) to assess whether ML-NMR-MS can recover
population-adjusted transition effects governing joint PFS/OS in a target population; (ii) to quantify its precision and the information cost of replacing individual participant data by
aggregate survival curves, relative to a full-IPD comparator; and (iii) to assess the
robustness of ML-NMR-MS to violations of its assumptions (unobserved
heterogeneity, a misspecified covariate correlation, and non-proportional hazards).

\subsection{Data-generating mechanisms} IPD are simulated from the illness--death model with known transition intensities, covariate distributions, and treatment effects, so that every estimand has an exact value. Each transition carries a Weibull baseline on the log scale,
\begin{equation}
\log h^r_{k}(u\mid\xv)=c^r+(v^r-1)\log u+\xv^{\!\top}\bet^r_1
+\mathds{1}(k\neq A)\,\xv^{\!\top}\bet^r_2+\gamma^r_k,
\label{eq:dgm}
\end{equation}
over four treatments $k\in\{A,B,C,D\}$ with $A$ the reference and three covariates
$\xv=(x_1,x_2,x_3)$ ($x_1,x_3$ continuous, $x_2$ binary), standing in for age, ISS-III and ASCT
response as in Section~\ref{sec:example} and labeled accordingly in Table~\ref{tab:sim}. The true values are
$\bm c=(-2.30,-4.00,-2.60)$ and Weibull shapes $\bm v=(1.20,1.00,1.10)$ for $(SP,SD,PD)$; prognostic effects
$\bet_1$ of $(0.15,0.30,0.00)$, $(0.25,0.35,0.00)$ and $(0.20,0.30,0.00)$ on the three transitions;
effect-modifier interactions on stable$\to$progression only, $\bet^{SP}_2=(0.00,0.40,-0.30)$, shared
across the active treatments; and treatment effects $(\gamma^{SP}_B,\gamma^{SP}_C,\gamma^{SP}_D)=(-0.50,-0.70,-0.60)$ and
$(\gamma^{PD}_B,\gamma^{PD}_C,\gamma^{PD}_D)=(-0.10,-0.15,-0.10)$ for $B,C,D$ versus $A$, with no effect on stable$\to$death. Because
the covariate and treatment terms in \eqref{eq:dgm} do not depend on $u$, every transition HR
is constant over follow-up at the individual level: the base configuration satisfies proportional hazards. Scenario (e)
below departs from it directly, and the frailty of scenario (c) induces non-proportional hazards once it is integrated out.

The network comprises six two-arm studies of $200$ patients per arm ($2400$ in total) comparing
$A$--$B$, $A$--$C$, $B$--$C$, $A$--$D$, $C$--$D$ and $A$--$B$, with administrative censoring at
$24$--$36$ months and a small dropout rate. Study covariate means differ by design ($x_1$ from $-0.4$
to $0.5$, $\Pr(x_2=1)$ from $0.20$ to $0.70$), creating the cross-study imbalance that population
adjustment must resolve. The studies designated aggregate, four of the six in the base configuration, have their individual records
replaced by conditional-survival life-table counts (akin to Appendix~\ref{app:data}) together with covariate summaries, while the full-IPD comparator retains them. Both methods therefore see the same underlying replicate, so the difference between them reflects aggregation alone rather than sampling variation.

Starting from this base scenario, factors are varied one at a time, rather than fully
factorially, to isolate each mechanism: (a) effect-modifier strength; (b) the proportion
of studies contributing IPD, from an AgD-dominated network with a single IPD study, through the base
two-study mix, to four of the six studies (the all-IPD limit, at which ML-NMR-MS reduces to full-IPD
multistate network meta-regression, is represented by the full-IPD comparator); (c) an unobserved
shared frailty on the transition intensities, probing the no-frailty assumption of the individual-level
model; (d) the covariate-distribution reconstruction, a correct Gaussian copula versus a
misspecified correlation; and (e) non-proportional hazards, adding a time-variation coefficient $\gamma^{SPt}_k=0.15$ for every active treatment $k$ to the
stable$\to$progression log-time slope, so that
$\mathrm{HR}^{SP}_k(u\mid\xv)=\exp\big(\gamma^{SP}_k+\xv^{\!\top}\bet^{SP}_2\big)\,u^{\gamma^{SPt}_k}$ while the baseline and the other
transitions are unchanged. Holding the base scenario proportional isolates the aggregation and
assumption-violation mechanisms from time-variation. Table~\ref{tab:sim-dgm} gives the remaining
settings: each study's horizon and covariate means, the studies that contribute IPD, the target
populations, and what each scenario changes, with its true values.

\subsection{Estimands} The estimands are of two kinds. The first are the model's building-block
parameters: the prognostic effects $\bet_1$, the treatment effects $\gamma^{SP}$ and $\gamma^{PD}$ with
the time-variation coefficient $\gamma^{SPt}$, and the effect-modifier interactions $\bet^{SP}_2$. The
second is the population-adjusted contrast assembled from them: the conditional
stable$\to$progression log-hazard-ratio $d^{SP}_k(\mathcal{P})=\gamma^{SP}_k+\bar\xv_{(\mathcal{P})}^{\!\top}\bet^{SP}_2$ of \eqref{eq:condHR} for each treatment versus the reference, in three target populations $\mathcal{P}$ (favorable, intermediate, and unfavorable); the two extremes
differ from every study population, while the intermediate is a network-mean-like profile that
coincides with that of study~1. Only stable$\to$progression carries a population-specific effect, since the
mechanism places effect modifiers on that transition alone; the progression$\to$death effect is
population-invariant and equals $\gamma^{PD}_k$. Scoring both the parameters and the contrast is deliberate, since
calibrated recovery of the coefficients need not by itself imply a calibrated interval on their
covariate-weighted combination. In the non-proportional-hazards scenario the estimand is the same
quantity at a landmark time, $\gamma^{SP}_k+\gamma^{SPt}_k\log u+\bar\xv_{(\mathcal{P})}^{\!\top}\bet^{SP}_2$,
scored at $u=6$, $12$ and $24$ months. All scored quantities are conditional; the marginal effects they induce are not evaluated here. 

\subsection{Methods} The study evaluates ML-NMR-MS (with the two-binomial composite aggregate likelihood
\eqref{eq:agdbinom} for AgD) across the evidence scenarios above, rather than ranking it against
competing synthesis methods. The one reference is the full-IPD multistate network meta-regression, which
uses IPD for every study and so bounds the attainable precision; it is fitted alongside ML-NMR-MS in
every scenario except the non-proportional-hazards one, and its precision gap to ML-NMR-MS quantifies the
information cost of aggregation (aim (ii)). The non-proportional-hazards scenario instead contrasts the
correctly specified time-varying-coefficient model with a misspecified proportional-hazards fit,
isolating the cost of model misspecification from that of aggregation. The fitted models have study-specific Weibull baselines on all
three transitions and the covariate and treatment terms of \eqref{eq:dgm}; the time-varying-coefficient model adds log-time
slopes $\gamma^{SPt}_k$ and $\gamma^{PDt}_k$ on stable$\to$progression and progression$\to$death, each with a $\mathrm N(0,1)$ prior. All models are fitted in Stan with the priors of Section~\ref{sec:example} and, for aggregate studies, its Gaussian-copula Sobol' integration, using $\tilde N=24$ integration points (fixed in advance rather than set by the doubling rule)
and four chains of $1000$ iterations ($500$ warmup), which give $2000$ posterior draws per fit. Chains start
from random values in a narrow range of plausible values that does not depend on the truth: log-scale
intercepts $\mathrm N(-3,0.3^2)$, log-shapes $\mathrm N(0,0.2^2)$, treatment and regression coefficients
$\mathrm N(0,0.2^2)$ and time slopes $\mathrm N(0,0.1^2)$, since Stan's default random starting values can
fail to initialize the model. Each replicate contributes the posterior mean
as its point estimate, the posterior standard deviation as its model-based standard error, and the
$2.5$--$97.5\%$ posterior quantiles as the interval whose coverage is scored. A fit is classed as non-converged, and excluded from that method's summaries, when its largest split-$\hat R$ (as computed by rstan) exceeds $1.05$. Two hundred replicates were simulated per scenario, and all $2800$ fits converged (largest $\hat R$ of $1.04$), so none was excluded.
Twenty-four fits ($0.9\%$) had at least one divergent transition after warmup (at most $14$ in a fit), nine of
them in the frailty scenario, in which the fitted model is misspecified; they were retained. Effective sample
sizes were recorded for every scored estimate. Their medians were about $2000$ for the bulk and $1600$ for the
tails of the posterior, and only $4$ of the $83{,}400$ scored estimates fell below the $400$ recommended for
four chains~\citep{vehtari2021rhat}.

\subsection{Performance measures} For a scalar estimand $\theta$ with estimates $\hat\theta_s$ and $95\%$ interval limits over the $s=1,\dots,n_\mathrm{sim}$ converged replicates ($n_\mathrm{sim}=200$ in every scenario and for every method), the following are reported, after
Morris et al.~\citep{morris2019adep}: bias $\widehat{\mathrm{Bias}}=\bar{\hat\theta}-\theta$; the
empirical standard error
$\mathrm{empSE}=\{(n_\mathrm{sim}-1)^{-1}\textstyle\sum_s(\hat\theta_s-\bar{\hat\theta})^2\}^{1/2}$, whose ratio to the full-IPD
comparator's measures the cost of aggregation (aim (ii));
the root mean squared error
$\mathrm{RMSE}=\{n_\mathrm{sim}^{-1}\textstyle\sum_s(\hat\theta_s-\theta)^2\}^{1/2}$;
the average model-based standard error (ModSE; posterior standard deviation), whose comparison with empSE
diagnoses interval calibration; and $95\%$ coverage
$\widehat{\mathrm{Cov}}=n_\mathrm{sim}^{-1}\textstyle\sum_s\mathds{1}\{\theta\in\mathrm{CrI}_s\}$,
with $\mathrm{CrI}_s$ the $2.5$--$97.5\%$ posterior interval from replicate $s$. Bias, empirical standard error and coverage are
reported with their Monte Carlo standard errors (MCSE):
$\mathrm{MCSE}(\widehat{\mathrm{Bias}})=\mathrm{empSE}/\sqrt{n_\mathrm{sim}}$,
$\mathrm{MCSE}(\mathrm{empSE})=\mathrm{empSE}/\sqrt{2(n_\mathrm{sim}-1)}$, and
$\mathrm{MCSE}(\widehat{\mathrm{Cov}})=\{\widehat{\mathrm{Cov}}(1-\widehat{\mathrm{Cov}})/n_\mathrm{sim}\}^{1/2}$.
Coverage, together with the agreement between empSE and the model-based standard error, is the primary
measure of interval calibration; bias and precision benchmark the recovery of the population-adjusted
effects against the full-IPD comparator. This replicate count trades Monte Carlo precision against
the substantial per-fit cost of the Bayesian multistate models: it puts the MCSE of coverage near the
nominal $95\%$ at $\sqrt{0.95\times0.05/200}\approx0.015$ (about one and a half percentage points) and,
with an empSE of about $0.10$ on the log-hazard-ratio scale, the MCSE of bias at about
$0.10/\sqrt{200}\approx0.007$. Per-fit Monte Carlo error adds little to these errors: the Monte Carlo
standard error of the interval limits (the larger of the two in each fit) has a median of $0.08$
posterior standard deviations ($99$th percentile $0.13$), against an interval about four posterior
standard deviations wide, and that of the posterior mean a median of $0.02$.

\subsection{Results} 
\paragraph{Aim (i): recovery in a target population.} Table~\ref{tab:sim} reports recovery of the data-generating treatment, prognostic and
effect-modifier parameters; its non-PH column is the correctly specified time-varying-coefficient fit,
the misspecified proportional-hazards fit being taken up under aim (iii). Bias and coverage are close to those of the
full-IPD comparator: across the four correctly specified proportional-hazards scenarios, mean absolute bias is $0.006$--$0.010$, against
$0.005$--$0.007$ with IPD throughout, at $95\%$ coverage close to nominal. The prognostic
effects ($\bet_1$) are well recovered, the largest residual bias falling on the pre-progression-death
($S\to D$) coefficients, weakly identified from the aggregate curves; the treatment effects ($\gamma^{SP}$ and $\gamma^{PD}$) and the effect-modifier interactions
($\bet^{SP}_2$) recover to within a few hundredths on the log-hazard-ratio scale in the
correctly specified scenarios, as does the time variation $\gamma^{SPt}$, which is estimated only in
the non-proportional-hazards scenario. The exception is the progression$\to$death effect in that scenario. There the fitted time-varying-coefficient
model also gives progression$\to$death its own log-time slope $\gamma^{PDt}_k$, whose true value is
zero, so the reported $\gamma^{PD}_k$ is the log-hazard-ratio at $u=1$ month. Under the
data-generating mechanism this equals the constant effect it is scored against, but few patients have
progressed by then, so the estimate rests on extrapolating the fitted slope back to that time. It shows a
bias of up to $-0.17$, $2.8$ to $6.1$ Monte Carlo standard errors from zero, and a posterior nearly
four times as wide as the proportional-hazards fit's, although coverage stays at $0.95$ to $0.99$. Refitting the same replicates with the proportional-hazards model, which carries neither time slope
(the misspecified fit of Table~\ref{tab:sim-nonph}), recovers the progression$\to$death effects
essentially exactly ($-0.10$, $-0.16$ and $-0.11$ against truths $-0.10$, $-0.15$ and $-0.10$), so the bias reflects the redundant time slope
rather than any loss of information about the transition. For the
transportability estimand itself (the population-adjusted conditional $S\to P$ log-hazard-ratio), the
lower rows of Table~\ref{tab:sim} report recovery for B, C and D versus A in the three target
populations. Its true value moves with the target: for B versus A, from $-0.57$ in the favorable
population to $-0.11$ in the unfavorable one. ML-NMR-MS recovers each population-specific effect with comparably small bias and near-nominal
coverage. Averaged over populations and contrasts its bias is negligible, with
coverage of $0.95$ in every correctly specified scenario (Table~\ref{tab:sim-robust}).

\paragraph{Aim (ii): precision and the cost of aggregate data.} The precision penalty for replacing
individual data by aggregate curves grows as the share of individual data falls
(Table~\ref{tab:sim-robust}): the ML-NMR-MS empirical standard error is $1.09$ times the full-IPD
comparator's when four of the six studies contribute IPD, $1.27$ at the base two-IPD mix, and $1.44$
when only one study does. Bias and coverage stay near their targets throughout, so the cost of
aggregation is paid in precision rather than in validity, and it is modest once two of the six studies report individual data (an empirical standard error $27\%$ larger,
$9\%$ with four).

\paragraph{Aim (iii): robustness to assumption violation.} Misspecifying the correlation borrowed for the reconstructed covariates leaves recovery
essentially unchanged (mean absolute bias $0.013$, coverage $0.95$, against $0.009$ and $0.95$ at base), so that
channel is not a practical vulnerability in this design; misspecification of the covariate marginals
themselves, and dependence involving the binary effect modifier, are not exercised here and remain
uncharacterized. An unobserved shared frailty, by contrast, biases both
ML-NMR-MS and the full-IPD analysis and lowers their coverage of the transportability estimand (to $0.88$ and $0.80$;
Table~\ref{tab:sim-robust}). The selected risk set attenuates the estimated effects toward the null (ML-NMR-MS bias $+0.08$ to
$+0.11$ against truths of $-0.50$ to $-0.70$; Table~\ref{tab:sim}, $\gamma^{SP}$ rows). Individual data alone therefore do not
remove the problem, which is consequential but not specific to the aggregate construction. Whether a
frailty-augmented multistate model fitted to the linked individual transition times could recover the
effects was not assessed here; from aggregate curves alone it could not, the frailty being absorbed
into the baselines and, as bias, into the treatment effects (Section~\ref{sec:agd}). Under non-proportional hazards, the correctly specified time-varying-coefficient model tracks the
stable$\to$progression HR across follow-up, holding near-nominal coverage at the $6$-, $12$-
and $24$-month landmarks (Table~\ref{tab:sim-nonph}) and recovering the time-variation
coefficient $\gamma^{SPt}$; a proportional-hazards fit, forced to a
constant ratio, grows biased as follow-up accrues. Overall, under correct specification this simulation shows recovery within a few hundredths on the
log-hazard-ratio scale, with coverage close to nominal under the two-binomial composite aggregate likelihood; the largest
remaining biases fall on the weakly identified stable$\to$death coefficients (up to $-0.06$). The one material exception, discussed above, is the progression$\to$death effect of the
time-varying-coefficient fit.
Among the factors varied, only violations of the fitted individual-level model degrade recovery, and they would affect
any analysis based on that model: unobserved frailty biases the full-IPD analysis at least as much, and a time-varying
treatment effect is recovered once the model allows for it.

\begin{sidewaystable}\centering\scriptsize\setlength{\tabcolsep}{3.5pt}
\caption{Recovery of the data-generating model parameters across all scenarios: posterior-mean bias
and, in parentheses, $95\%$ coverage for ML-NMR-MS, over the $200$ replicates of each scenario (all converged;
Monte Carlo SE of coverage $\approx0.015$ near nominal coverage, at most $0.03$). Bias is on the log-hazard-ratio scale. The Non-PH column is the time-varying-coefficient (TVC) fit. The Truth column gives the base values (for $\gamma^{SPt}$, the non-PH value). The strong effect-modifier (EM) scenario doubles the effect modifiers, which changes their truths and those of the population-adjusted effects (Table~\ref{tab:sim-dgm}); bias and coverage are computed against each scenario's own truth. A dash marks a parameter not in that scenario's model ($\gamma^{SPt}$) or an effect reported in Table~\ref{tab:sim-nonph} (the population-adjusted effects under non-PH).}
\label{tab:sim}
\begin{tabular}{ll r ccccccc}\toprule
 & & & \multicolumn{7}{c}{Scenario: bias\,(coverage)}\\
\cmidrule(lr){4-10}
Parameter & & Truth & Base & Strong EM & Non-PH & More IPD & Agg.-dom. & Frailty & Copula mis.\\\midrule
\multirow{9}{*}{Prognostic effects ($\beta^r_1$)} & age, S$\to$P & $+0.15$ & $+0.00$\,(0.97) & $-0.00$\,(0.95) & $-0.00$\,(0.94) & $-0.00$\,(0.96) & $+0.01$\,(0.94) & $-0.03$\,(0.90) & $+0.00$\,(0.95)\\
 & ISS-III, S$\to$P & $+0.30$ & $+0.01$\,(0.95) & $+0.02$\,(0.93) & $+0.01$\,(0.95) & $+0.01$\,(0.96) & $+0.01$\,(0.93) & $-0.05$\,(0.90) & $+0.01$\,(0.95)\\
 & resp., S$\to$P & $+0.00$ & $-0.00$\,(0.96) & $-0.00$\,(0.95) & $+0.00$\,(0.94) & $+0.00$\,(0.96) & $-0.01$\,(0.95) & $-0.00$\,(0.91) & $-0.00$\,(0.97)\\
 & age, S$\to$D & $+0.25$ & $+0.00$\,(0.91) & $+0.01$\,(0.94) & $-0.00$\,(0.92) & $+0.00$\,(0.92) & $+0.01$\,(0.94) & $-0.01$\,(0.93) & $+0.00$\,(0.91)\\
 & ISS-III, S$\to$D & $+0.35$ & $-0.04$\,(0.95) & $-0.02$\,(0.96) & $-0.03$\,(0.92) & $-0.03$\,(0.96) & $-0.06$\,(0.94) & $-0.10$\,(0.93) & $-0.05$\,(0.96)\\
 & resp., S$\to$D & $+0.00$ & $-0.01$\,(0.91) & $-0.02$\,(0.96) & $-0.00$\,(0.97) & $-0.00$\,(0.95) & $-0.01$\,(0.94) & $+0.03$\,(0.91) & $-0.00$\,(0.92)\\
 & age, P$\to$D & $+0.20$ & $-0.00$\,(0.95) & $+0.00$\,(0.94) & $+0.01$\,(0.95) & $-0.00$\,(0.96) & $-0.00$\,(0.95) & $-0.06$\,(0.81) & $-0.00$\,(0.94)\\
 & ISS-III, P$\to$D & $+0.30$ & $-0.01$\,(0.93) & $-0.00$\,(0.94) & $+0.01$\,(0.97) & $-0.01$\,(0.94) & $-0.01$\,(0.93) & $-0.12$\,(0.78) & $-0.01$\,(0.94)\\
 & resp., P$\to$D & $+0.00$ & $+0.01$\,(0.97) & $+0.00$\,(0.90) & $-0.00$\,(0.94) & $+0.00$\,(0.95) & $+0.01$\,(0.94) & $+0.02$\,(0.94) & $+0.01$\,(0.95)\\
\midrule
\multirow{3}{*}{Treatment effect, S$\to$P ($\gamma^{SP}_k$)} & B vs A & $-0.50$ & $+0.00$\,(0.94) & $+0.01$\,(0.94) & $-0.00$\,(0.94) & $+0.01$\,(0.95) & $+0.00$\,(0.94) & $+0.08$\,(0.84) & $+0.02$\,(0.94)\\
 & C vs A & $-0.70$ & $+0.00$\,(0.94) & $+0.02$\,(0.94) & $-0.02$\,(0.98) & $+0.01$\,(0.94) & $-0.01$\,(0.94) & $+0.11$\,(0.84) & $+0.01$\,(0.94)\\
 & D vs A & $-0.60$ & $+0.01$\,(0.96) & $+0.02$\,(0.95) & $-0.01$\,(0.97) & $+0.02$\,(0.96) & $+0.00$\,(0.97) & $+0.08$\,(0.92) & $+0.03$\,(0.96)\\
\midrule
\multirow{3}{*}{Time variation, S$\to$P ($\gamma^{SPt}$)} & B vs A & $+0.15$ & --- & --- & $+0.01$\,(0.94) & --- & --- & --- & ---\\
 & C vs A & $+0.15$ & --- & --- & $+0.02$\,(0.97) & --- & --- & --- & ---\\
 & D vs A & $+0.15$ & --- & --- & $+0.02$\,(0.97) & --- & --- & --- & ---\\
\midrule
\multirow{3}{*}{Treatment effect, P$\to$D ($\gamma^{PD}_k$)} & B vs A & $-0.10$ & $-0.00$\,(0.94) & $-0.00$\,(0.96) & $-0.06$\,(0.95) & $+0.00$\,(0.94) & $-0.01$\,(0.97) & $+0.07$\,(0.92) & $-0.00$\,(0.95)\\
 & C vs A & $-0.15$ & $-0.01$\,(0.94) & $-0.00$\,(0.94) & $-0.11$\,(0.99) & $-0.01$\,(0.95) & $-0.01$\,(0.94) & $+0.12$\,(0.85) & $-0.01$\,(0.95)\\
 & D vs A & $-0.10$ & $-0.00$\,(0.98) & $-0.00$\,(0.98) & $-0.17$\,(0.98) & $-0.00$\,(0.95) & $-0.00$\,(0.98) & $+0.08$\,(0.99) & $-0.00$\,(0.98)\\
\midrule
\multirow{3}{*}{Effect modifiers, S$\to$P ($\beta^{SP}_2$)} & age & $+0.00$ & $+0.00$\,(0.95) & $+0.01$\,(0.94) & $+0.01$\,(0.94) & $+0.00$\,(0.94) & $+0.01$\,(0.93) & $+0.02$\,(0.92) & $-0.00$\,(0.93)\\
 & ISS-III & $+0.40$ & $+0.00$\,(0.95) & $-0.01$\,(0.94) & $-0.00$\,(0.95) & $-0.01$\,(0.94) & $+0.01$\,(0.93) & $-0.03$\,(0.89) & $-0.01$\,(0.96)\\
 & response & $-0.30$ & $-0.01$\,(0.94) & $-0.01$\,(0.95) & $-0.01$\,(0.95) & $-0.01$\,(0.93) & $-0.00$\,(0.94) & $+0.04$\,(0.94) & $+0.00$\,(0.94)\\
\midrule
\multirow{3}{*}{Pop.-adj.\ S$\to$P, favorable ($d^{SP}_k$)} & B vs A & $-0.57$ & $-0.00$\,(0.97) & $+0.00$\,(0.97) & --- & $+0.00$\,(0.97) & $+0.00$\,(0.96) & $+0.08$\,(0.86) & $+0.02$\,(0.97)\\
 & C vs A & $-0.77$ & $-0.00$\,(0.94) & $+0.00$\,(0.96) & --- & $+0.00$\,(0.93) & $-0.01$\,(0.94) & $+0.12$\,(0.82) & $+0.01$\,(0.94)\\
 & D vs A & $-0.67$ & $+0.01$\,(0.94) & $+0.01$\,(0.96) & --- & $+0.01$\,(0.95) & $-0.00$\,(0.94) & $+0.08$\,(0.93) & $+0.03$\,(0.94)\\
\midrule
\multirow{3}{*}{Pop.-adj.\ S$\to$P, intermediate ($d^{SP}_k$)} & B vs A & $-0.34$ & $+0.01$\,(0.94) & $+0.01$\,(0.94) & --- & $+0.00$\,(0.95) & $+0.01$\,(0.94) & $+0.06$\,(0.85) & $+0.01$\,(0.95)\\
 & C vs A & $-0.54$ & $+0.00$\,(0.96) & $+0.01$\,(0.94) & --- & $+0.01$\,(0.96) & $-0.00$\,(0.96) & $+0.10$\,(0.80) & $+0.01$\,(0.96)\\
 & D vs A & $-0.44$ & $+0.01$\,(0.96) & $+0.02$\,(0.96) & --- & $+0.02$\,(0.94) & $+0.01$\,(0.97) & $+0.07$\,(0.90) & $+0.02$\,(0.96)\\
\midrule
\multirow{3}{*}{Pop.-adj.\ S$\to$P, unfavorable ($d^{SP}_k$)} & B vs A & $-0.11$ & $+0.01$\,(0.94) & $+0.02$\,(0.93) & --- & $+0.01$\,(0.94) & $+0.02$\,(0.94) & $+0.05$\,(0.90) & $+0.00$\,(0.94)\\
 & C vs A & $-0.31$ & $+0.01$\,(0.94) & $+0.02$\,(0.93) & --- & $+0.01$\,(0.95) & $+0.01$\,(0.95) & $+0.08$\,(0.90) & $+0.00$\,(0.93)\\
 & D vs A & $-0.21$ & $+0.02$\,(0.95) & $+0.02$\,(0.95) & --- & $+0.02$\,(0.96) & $+0.02$\,(0.95) & $+0.05$\,(0.94) & $+0.02$\,(0.94)\\
\bottomrule\end{tabular}\end{sidewaystable}
\begin{sidewaystable}\centering\footnotesize\setlength{\tabcolsep}{5pt}
\caption{Performance measures for the population-adjusted conditional S$\to$P log-hazard-ratio (the
transportability estimand), averaged over the three target populations and three treatment contrasts,
by scenario and method: mean absolute bias, empirical standard error (empSE), average model-based
standard error (ModSE, the mean posterior standard deviation), root mean squared error (RMSE), and
$95\%$ coverage, each with its Monte Carlo standard error (MCSE, the mean of the nine per-cell MCSEs) in
parentheses where defined. The full-IPD rows coincide across the base, more-IPD, aggregate-dominated and
copula scenarios, which share each replicate's simulated data and differ only in what ML-NMR-MS receives.}
\label{tab:sim-robust}
\begin{tabular}{ll ccccc}\toprule
Scenario & Method & $|$bias$|$\,(MCSE) & empSE\,(MCSE) & ModSE & RMSE & coverage\,(MCSE)\\\midrule
\multirow{2}{*}{Base} & ML-NMR-MS & $0.009$\,(0.008) & $0.109$\,(0.005) & $0.111$ & $0.110$ & $0.95$\,(0.02)\\
 & Full-IPD & $0.010$\,(0.006) & $0.086$\,(0.004) & $0.088$ & $0.087$ & $0.95$\,(0.02)\\
\cmidrule(l){2-7}
\multirow{2}{*}{Strong EM} & ML-NMR-MS & $0.013$\,(0.008) & $0.112$\,(0.006) & $0.113$ & $0.113$ & $0.95$\,(0.02)\\
 & Full-IPD & $0.011$\,(0.006) & $0.082$\,(0.004) & $0.088$ & $0.083$ & $0.96$\,(0.01)\\
\cmidrule(l){2-7}
\multirow{2}{*}{More IPD} & ML-NMR-MS & $0.009$\,(0.007) & $0.094$\,(0.005) & $0.096$ & $0.095$ & $0.95$\,(0.02)\\
 & Full-IPD & $0.010$\,(0.006) & $0.086$\,(0.004) & $0.088$ & $0.087$ & $0.95$\,(0.02)\\
\cmidrule(l){2-7}
\multirow{2}{*}{Agg.-dom.} & ML-NMR-MS & $0.008$\,(0.009) & $0.125$\,(0.006) & $0.127$ & $0.125$ & $0.95$\,(0.02)\\
 & Full-IPD & $0.010$\,(0.006) & $0.086$\,(0.004) & $0.088$ & $0.087$ & $0.95$\,(0.02)\\
\cmidrule(l){2-7}
\multirow{2}{*}{Frailty} & ML-NMR-MS & $0.078$\,(0.009) & $0.122$\,(0.006) & $0.118$ & $0.146$ & $0.88$\,(0.02)\\
 & Full-IPD & $0.094$\,(0.007) & $0.093$\,(0.005) & $0.092$ & $0.136$ & $0.80$\,(0.03)\\
\cmidrule(l){2-7}
\multirow{2}{*}{Copula mis.} & ML-NMR-MS & $0.013$\,(0.008) & $0.110$\,(0.006) & $0.112$ & $0.111$ & $0.95$\,(0.02)\\
 & Full-IPD & $0.010$\,(0.006) & $0.086$\,(0.004) & $0.088$ & $0.087$ & $0.95$\,(0.02)\\
\bottomrule\end{tabular}\end{sidewaystable}
\begin{table}[htbp]\centering\footnotesize\setlength{\tabcolsep}{5pt}
\caption{Non-proportional-hazards scenario: recovery of the time-varying stable$\to$progression
hazard ratio at landmark times $u$, for the correctly specified time-varying-coefficient (TVC)
model and the misspecified proportional-hazards (PH) model: mean bias (log-hazard-ratio
scale), empirical standard error (empSE), model-based standard error (ModSE), root mean squared error (RMSE), and $95\%$ coverage
(Monte Carlo SEs in parentheses, the mean of the per-cell SEs), averaged over contrasts and target populations.}
\label{tab:sim-nonph}
\begin{tabular}{ll ccccc}\toprule
Model & $u$ (mo) & bias\,(MCSE) & empSE & ModSE & RMSE & coverage\,(MCSE)\\\midrule
\multirow{3}{*}{TVC (correct)} & 6 & $+0.013$\,(0.008) & $0.115$ & $0.115$ & $0.116$ & $0.94$\,(0.02)\\
 & 12 & $+0.022$\,(0.010) & $0.142$ & $0.146$ & $0.144$ & $0.95$\,(0.02)\\
 & 24 & $+0.032$\,(0.013) & $0.183$ & $0.192$ & $0.186$ & $0.95$\,(0.02)\\
\cmidrule(l){2-7}
\multirow{3}{*}{PH (misspec.)} & 6 & $-0.047$\,(0.007) & $0.106$ & $0.107$ & $0.121$ & $0.91$\,(0.02)\\
 & 12 & $-0.151$\,(0.007) & $0.106$ & $0.107$ & $0.186$ & $0.66$\,(0.03)\\
 & 24 & $-0.255$\,(0.007) & $0.106$ & $0.107$ & $0.277$ & $0.34$\,(0.03)\\
\bottomrule\end{tabular}\end{table}

\FloatBarrier
\section{Discussion}\label{sec:discussion}

ML-NMR-MS unifies the joint PFS/OS multistate NMA of Jansen et al.~\citep{jansen2023multistate} with the
population-adjusted, IPD-plus-AgD ML-NMR of Phillippo et al.~\citep{phillippo2020mlnmr,phillippo2024general}. Its hallmark, integrating the multistate transition model over the aggregate covariate distributions to
form marginal state occupancies, addresses two problems at once: it removes aggregation bias and, under the
model's assumptions, supplies the progression$\to$death linkage missing from separately reported curves. The
resulting relative treatment effects are coherent with a clock-forward state-transition (i.e., Markov)
simulation model. ML-NMR-MS produces conditional transition effects in any target
population and, where a baseline hazard for it can be specified, conditional endpoint effects, marginal effects and absolute
quantities (Section~\ref{sec:estimands}).

The aggregate contribution enters through the conditional-survival likelihood on reconstructed
life-table counts rather than by treating pseudo-individual records, reconstructed from the published
curves \citep{guyot2012}, as if they were IPD. This is a deliberate choice. What is needed from each study is information about the transition intensities of the illness--death model, whose individual likelihood is built from transition-level events: who
progressed, who died, and when. Pseudo-individual reconstruction, carried out endpoint by endpoint, does
not deliver this: because progression-free and overall survival are reconstructed separately, the
individual link between progression and death is severed, so the post-progression sojourn that identifies
the progression$\to$death intensity cannot be recovered and the paired records need not respect the elementary constraint that PFS cannot exceed OS.
Reconstruction moreover returns
event times but no covariates, so any covariate assignment to link
PFS and OS indirectly would invent a covariate-outcome association the data do not contain, biasing the prognostic
and effect-modifier estimates the regression targets. The reconstructed records therefore add no information about the
transition intensities beyond the separate curves they came from, while presenting as individual data.
Pahuta et al.~\citep{pahuta2019transitions} take a different route, approximating transition rates
from published PFS and OS curves by resolving the ties between them, at the cost of strong
assumptions about how the two curves correspond at the individual level.
The conditional-survival likelihood instead uses only what is reported, the survival curves and the
covariate summaries, and matches them to the model's population-average occupancy, itself a
function of the transition intensities: the integral of the individual model over the
reconstructed covariate distribution (not a plug-in at the mean covariates), evaluated by quasi-Monte Carlo. The multinomial form additionally respects the
stable/progressed/dead occupancy constraint, but as specified here it is overconfident (Appendix~\ref{app:calibration}), so the
two-binomial composite is the default. ML-NMR-MS is an instance of the general-likelihood ML-NMR rather than an alternative to it
(Appendix~\ref{app:mlnmr-compare}, Table~\ref{tab:mlnmr-compare}). The general-likelihood
formulation \citep{phillippo2024general} integrates the individual likelihood \eqref{eq:mlnmr}; the
original ML-NMR \citep{phillippo2020mlnmr} integrates the individual-level model, that is the
covariate-conditional mean outcome $g^{-1}(\eta(\xv))$, and its one-parameter binomial likelihood is recovered from the general case
for binary outcomes; and ML-NMR-MS integrates the state occupancy, which for a
survival outcome is again an individual likelihood contribution. The contribution of the present
paper therefore lies not in the integration itself but in the integrand: a coupled illness--death
process that models the progression$\to$death linkage absent from separately reported curves,
and a pair of dependent endpoint streams whose joint treatment
(Section~\ref{sec:com_multinomial}) has no counterpart in single-endpoint synthesis. When the model is collapsed to a single transition, its individual-data contribution and linear predictor coincide with those of survival ML-NMR, while the aggregate contribution matches grouped state-occupancy counts rather
than reconstructed event times, the two coinciding in the limit of life-table intervals containing at most one event, which published
curves do not deliver. 

Like the multistate NMA of Jansen et al.~\citep{jansen2023multistate}, the present model is \emph{clock-forward}: every transition intensity, including progression$\to$death, is a function of time since randomization, so the process is a
(time-inhomogeneous) Markov process. A \emph{clock-reset} (i.e., semi-Markov) alternative would instead make
the progression$\to$death intensity depend on time since progression, which is often clinically
more natural, because post-progression mortality typically accrues from the moment of progression
rather than from randomization. The two formulations coincide for the stable$\to$progression and
stable$\to$death transitions, and can differ only for progression$\to$death. Even there they coincide
whenever that intensity is constant, as in the example. A genuine difference arises only when the progression$\to$death intensity is time-varying. For individual data the change is routine: the progression$\to$death
contribution is simply written in the sojourn time $u-u_S$. The obstacle is the aggregate
contribution: the progressed-state occupancy no longer follows the
closed-form step \eqref{eq:occP} but a renewal (convolution) equation that
integrates the sojourn-time survival against the distribution of progression times, itself integrated
over the covariate distribution, while the stable-state survival \eqref{eq:occS} is unchanged. This is implementable (one propagates the entry-time distribution into the progressed state along the occupancy grid and convolves) at appreciably higher computational cost,
but it is a worthwhile extension to develop.

The multistate NMA of Jansen et al.~\citep{jansen2023multistate} forms its aggregate likelihood from
conditional-survival counts at three evaluation points within each interval, all sharing the
interval-start risk set, a device that stabilizes estimation when events are sparse; there the
transition intensities are already smooth, their interval levels tied across time through a fractional
polynomial. ML-NMR-MS reduces the same reported curves to conditional-survival counts but constructs them
differently: it records one count per (monthly) interval, each with its own interval-start risk
set, so that for each endpoint consecutive intervals are disjoint and the counts form a proper grouped-survival likelihood
rather than nested evaluation points sharing a denominator. This single count per interval suffices in the mixed network because the covariate and
effect-modifier structure is shared with, and largely identified by, the individual-data studies, so
no aggregate interval need by itself stabilize that structure, whereas an aggregate-only fit must
extract enough from each interval to stabilize the whole model. The two constructions are not compared
directly here.

Because ML-NMR-MS models the joint progression--death process rather than PFS and OS separately, it is
also a natural framework for evaluating PFS as a surrogate for OS. A treatment's effect on OS is decomposed into its effects on progression (stable$\to$progression), on pre-progression death
(stable$\to$death) and on post-progression mortality (progression$\to$death): when the OS benefit is driven almost entirely by delayed
progression (a small or null $\gamma^{PD}_k$, as for lenalidomide in the present example, where the
overall-survival layer is semi-synthetic, the model omits a stable$\to$death effect, and the value is illustrative only), PFS carries most of
the survival signal, whereas a substantial direct $\gamma^{PD}_k$ signals an OS benefit that PFS does
not capture. Because it separates a treatment's effect on progression from its effect on post-progression
mortality, ML-NMR-MS gives a mechanistic handle on how much of an OS effect is routed through
progression: with the fitted transition effects and baselines one can recompute OS with
$\gamma^{PD}_k$ set to zero and read off the share routed through the stable state (through progression alone if
$\gamma^{SD}_k$ is also zero), under the model's Markov and no-frailty
assumptions. Extending this to formal individual- and trial-level surrogacy in the sense of Buyse et
al.~\citep{buyse2000} would require, respectively, relaxing the no-frailty assumption so that the
residual progression/death association is estimated rather than assumed, and a random-effects specification yielding study-specific transition effects, which
Section~\ref{sec:model} provides but which is not exercised for surrogacy here. A population-adjusted
surrogacy analysis, in which the surrogacy relationship itself transports to a target population, is
left to future work.

A further practical consequence of the transition-level decomposition is that a study need not report
both endpoints to be included in the analysis. A study reporting PFS but not OS (for whatever reason) still contributes coherently: its
conditional-survival counts constrain the total stable-state exit intensity $h^{SP}+h^{SD}$ and,
through the between-arm contrast, the treatment effect on it, the split between progression and
pre-progression death being carried by the priors and by the rest of the network, while the progression$\to$death transition, which overall survival would inform, takes no contribution from that study; its shared transition effects remain identified from the rest
of the network, while that study's own progression$\to$death baseline is prior-driven.

The practical motivation is HTA, and the conditional/marginal distinction between treatment effect estimates maps directly onto how these should enter a cost-effectiveness model. Because HTA adopts a treatment for a whole population, the decision-relevant cost-effectiveness estimand is marginal: the population-average net health benefit, obtained by predicting individual outcomes
and averaging over the target population's covariate distribution. The ideal cost-effectiveness model is therefore an
individual-level (state-transition) simulation whose inputs are conditional building blocks: prognostic effects, conditional treatment effects and
treatment-by-covariate interactions, which transport across jurisdictions under conditional
constancy, together with a covariate-conditional baseline for the target population. The baseline is not
protected by randomization. It needs no further assumption when it is observed in the target population,
but a judgment that its source's covariate-conditional baseline applies when it is borrowed from a study in the network
(Section~\ref{sec:estimands}). All are re-marginalized for each target population. ML-NMR-MS estimates
the relative building blocks directly and supplies the baseline from either source. ML-NMR-MS is also not exposed to the estimand mismatch that arises when a conditional HR is applied
to a marginal reference-arm curve. It also respects non-collapsibility, whereby a marginal HR transports only if both effect modifiers
and prognostic factors are balanced and the baseline hazard is the same, while the conditional building blocks require only conditional constancy. Since bias in
these inputs propagates into the incremental cost-effectiveness ratio and can flip a decision near a
cost-effectiveness threshold, producing them at the right estimand and in the jurisdiction-relevant population is of
direct practical value.

A broader direction is the extension of ML-NMR to unanchored settings: disconnected networks and
single-arm studies, where no common comparator anchors the comparison. Multilevel unanchored
meta-regression (ML-UMR) \citep{chandler2026mlumr} is one such proposal, its survival variant
marginalizing the same kind of reconstructed pseudo-individual likelihood set aside above; it is
unanchored throughout, so it forgoes randomization even where randomized evidence exists, and
further extensions that retain the within-study randomization where it is available are in
development, including an extension of ML-NMR itself to disconnected networks and single-arm studies
\citep{perren2026unanchored}. The multistate construction here could in principle be carried to any of these by
replacing conditional constancy of relative effects with treatment-specific baselines and a shared-prognostic-factor
assumption, that is, conditional constancy of absolute effects; this would extend joint PFS/OS synthesis to disconnected evidence, but would inherit the
strong, largely untestable assumptions that unanchored comparison requires.

The evidence in support of ML-NMR-MS offered here is a single illustrative network with semi-synthetic overall survival and a
simulation study of moderate size. The simulation spans
effect-modification strength, the IPD/AgD mix (including an aggregate-dominated network with a single
IPD study), a misspecified covariate correlation, unobserved frailty, and non-proportional
hazards. Of these, only unobserved frailty and a misspecified time course for the treatment effect degrade recovery. Both
are assumptions of the individual-level model, not of the aggregation, so a full-IPD analysis based on the same model shares
them (Section~\ref{sec:sim}); the others leave recovery and calibration intact. The baseline form, a clock-reset process,
between-study heterogeneity and the shared effect-modifier assumption were not varied.
It nonetheless leaves broader questions open: the true prognostic,
treatment, and effect-modifier parameters are held at a single set of values chosen to be structurally analogous to the illustrative
example, with only effect-modifier magnitude and the presence of time-varying effects
varied, so recovery and calibration are demonstrated at this parameterization rather than across
a range of true effect sizes; and performance for networks with still less individual data, thin
connectivity, or studies reporting a single endpoint is not characterized. A larger simulation study across these structures and parameterizations is the natural next step.
Simulation alone, though, is not what this family of models most lacks. Multistate NMA has seen
little applied use since it was introduced, and ML-NMR-MS inherits that position: what would test it
is application to real decision problems, where the evidence is messier than any simulation makes it,
and where the consequences of a modeling choice are visible in the resulting cost-effectiveness
estimate. Three things would make that likelier. The first is accessible software, since the method
is at present a collection of Stan models rather than a packaged interface of the kind that carried
ML-NMR into practice. The second is reporting conventions for this class of model, covering the
grid, the integration points, the baseline form and the diagnostics, so that a submitted analysis
can be checked by a reviewer who did not write it. The third is published worked examples with data
and code, against which an analyst can verify their own implementation before trusting it.

In sum, ML-NMR-MS provides a single coherent framework for the population-adjusted synthesis of
individual and aggregate progression and survival data, delivering conditional relative treatment effects on the transitions in any target population and, where a baseline hazard for the
target population can be specified, conditional effects on PFS and OS, marginal relative treatment effects, and absolute outcomes
regarding state transitions, PFS, and OS. It is well suited to parameterizing health-economic state-transition models for a target population of interest.

\section*{Acknowledgments}
The author used Claude (Opus~4.8 and Opus~5.5; Anthropic) during preparation of this manuscript: to work through and check the
mathematical derivations, review the accompanying R and Stan code, and provide editorial assistance. The proposed methodology, analyses, and interpretations are the author's own. The author reviewed and verified
all AI-assisted content and takes full responsibility for the work.

\section*{Funding}
The author received no specific grant for this research from any funding agency.

\section*{Competing interests}
The author serves as Chief Scientist for the Health Economics and Outcomes Research unit of Precision Medicine Group, a company that provides research services to the pharmaceutical and biotechnology industries.  


\clearpage
\appendix
\setcounter{page}{1}\renewcommand{\thepage}{S\arabic{page}}
\setcounter{table}{0}\setcounter{figure}{0}\setcounter{equation}{0}
\renewcommand{\thetable}{S\arabic{table}}
\renewcommand{\thefigure}{S\arabic{figure}}
\renewcommand{\theequation}{S\arabic{equation}}
\renewcommand{\theHtable}{S\arabic{table}}
\renewcommand{\theHfigure}{S\arabic{figure}}
\renewcommand{\theHequation}{S\arabic{equation}}
\begin{center}\Large\bfseries Supplemental Information\end{center}
\vspace{1em}
\begin{center}\large\bfseries Contents\end{center}
\vspace{0.4em}
{\small\setlength{\parskip}{3pt}\setlength{\parindent}{0pt}
\newcommand{\tocentry}[2]{\hyperref[#1]{\ref*{#1}\quad #2}\dotfill\pageref{#1}\par}
\newcommand{\tocsubentry}[2]{\hspace*{1.8em}\hyperref[#1]{\ref*{#1}\quad #2}\dotfill\pageref{#1}\par}
\tocentry{app:mlnmr-compare}{Relation to prior multilevel network meta-regression}
\tocentry{app:ode}{State-occupancy solution for the illness--death model}
\tocentry{app:qmc}{Quasi-Monte Carlo integration and covariate reconstruction}
\tocentry{app:data}{Constructing the aggregate-data life-table}
\tocentry{sec:ndmm-data}{Semi-synthetic overall survival for the NDMM example}
\tocentry{app:trials}{Trial characteristics and Kaplan--Meier curves}
\tocentry{app:datastruct}{Structure of the analysis data}
\tocentry{app:suppfig}{Model checking and MCMC diagnostics}
\tocsubentry{app:fit}{Fit to the observed data}
\tocsubentry{app:mcmc}{MCMC diagnostics and run time}
\tocsubentry{app:priorpost}{Prior and posterior distributions}
\tocsubentry{app:grid}{Fineness of the occupancy grid}
\tocentry{app:baselines}{Alternative stable--progression baseline hazard specifications}
\tocsubentry{app:fp2}{Fractional-polynomial stable--progression baseline}
\tocsubentry{app:mspline}{M-spline stable--progression baseline}
\tocsubentry{app:summtab}{Numerical summary across baseline forms}
\tocentry{app:stan}{Stan code for the illustrative example}
\tocentry{app:re}{Random treatment effects on stable--progression in Stan code}
\tocentry{app:calibration}{Simulation sub-study: composite versus multinomial aggregate likelihood}
\tocentry{app:simdgm}{Data-generating settings of the simulation study}
}
\clearpage

\section{Relation to prior multilevel network meta-regression}\label{app:mlnmr-compare}
ML-NMR-MS and the two ML-NMR formulations share the covariate integral \eqref{eq:mlnmr},
$\int(\cdot)\,f_{jk}(\xv)\,d\xv$, and differ in what they integrate and in the resulting aggregate
likelihood (Table~\ref{tab:mlnmr-compare}). They are nested rather than rival: the general-likelihood
formulation \citep{phillippo2024general} is the general case; it recovers the one-parameter binomial likelihood of the original
ML-NMR \citep{phillippo2020mlnmr} for binary outcomes, while the original's other aggregate likelihoods, its two-parameter
binomial included, approximate the general one in closed form; and ML-NMR-MS
is the instance in which the integrated likelihood contribution is the state occupancy of an
illness--death process, so that one transition model ties PFS to OS.

\begin{table}[htbp]\centering\footnotesize
\renewcommand{\arraystretch}{1.25}
\caption{Relation of ML-NMR-MS to the two ML-NMR formulations: what each integrates over the covariate
distribution $f_{jk}$ in the shared ML-NMR integral \eqref{eq:mlnmr}, the resulting aggregate
likelihood, the aggregate data used, and the outcomes covered; $\eta(\xv)$ is each framework's individual-level linear predictor and $g^{-1}$ the inverse
link.}
\label{tab:mlnmr-compare}
\begin{tabular}{@{}>{\raggedright\arraybackslash}p{1.9cm}>{\raggedright\arraybackslash}p{3.5cm}>{\raggedright\arraybackslash}p{3.5cm}>{\raggedright\arraybackslash}p{4.1cm}@{}}
\toprule
& Original ML-NMR \citep{phillippo2020mlnmr} & General-likelihood ML-NMR \citep{phillippo2024general} & ML-NMR-MS (this paper)\\
\midrule
Integrated over $f_{jk}$ & conditional mean outcome, $\int g^{-1}(\eta(\xv))f\,d\xv$ & individual likelihood, $\int L^{\mathrm{Con}}(\bm\xi;y,\xv)f\,d\xv$ & covariate-conditional state occupancy, $\int (S,P)(u\mid\xv)f\,d\xv$\\
Aggregate likelihood & closed-form family (Normal, Poisson, binomial approx.) & product of individual marginal likelihoods & grouped conditional-survival counts: binomial (composite) or multinomial\\
Aggregate data used & covariate $+$ outcome summaries & individual outcomes, or outcome summaries where the likelihood can be written in them (e.g., counts), $+$ covariate summaries; for survival, event and censoring times reconstructed from Kaplan--Meier curves & grouped life-table counts $+$ covariate summaries\\
Outcomes & GLM / exponential family & any likelihood; applied to single-endpoint survival & joint PFS/OS (illness--death multistate)\\
\bottomrule
\end{tabular}
\end{table}

\section{State-occupancy solution for the illness--death model}\label{app:ode}
This appendix gives the closed-form state-occupancy solution quoted as
\eqref{eq:occS}--\eqref{eq:occD} in the main text. The solution has the same form for every
arm, so the arm index $jk$ carried there is suppressed throughout. On an interval $U_m$ with intensities held constant at
$h^{SP}_m,h^{SD}_m,h^{PD}_m$, the
occupancy vector $\bm p(u)=(S,P,D)(u)$ satisfies $\dot{\bm p}=\bm p\,\bm Q_m$ with generator
\[
\bm Q_m=\begin{pmatrix}-(h^{SP}_m+h^{SD}_m) & h^{SP}_m & h^{SD}_m\\ 0 & -h^{PD}_m & h^{PD}_m\\
0 & 0 & 0\end{pmatrix}.
\]
The eigenvalues of $\bm Q_m^{\!\top}$ are $0$, $-(h^{SP}_m+h^{SD}_m)$ and $-h^{PD}_m$. From the initial condition
$(S(u_m),P(u_m),D(u_m))$, $S$ solves $\dot S=-(h^{SP}_m+h^{SD}_m)S$, giving \eqref{eq:occS}; $P$ solves
$\dot P=h^{SP}_mS-h^{PD}_mP$, and variation of constants gives \eqref{eq:occP}; \eqref{eq:occD} follows from $S+P+D=1$. Writing $\Delta=u-u_m$ and $x=(h^{PD}_m-h^{SP}_m-h^{SD}_m)\Delta$, the expression for $P$ has a
removable singularity at $x=0$, with limit $P(u)=P(u_m)e^{-h^{PD}_m\Delta}+S(u_m)h^{SP}_m\Delta\,
e^{-h^{PD}_m\Delta}$. Near it, the Stan implementation replaces the difference quotient by its series,
$P(u)=P(u_m)e^{-h^{PD}_m\Delta}+S(u_m)h^{SP}_m\Delta\,e^{-(h^{SP}_m+h^{SD}_m)\Delta}\,(1-x/2+x^2/6)$,
whenever $|x|<10^{-4}$, which also gives the sampler correct gradients there. Occupancies are propagated by
using the end-of-interval values as the next interval's initial condition; the general
finite-state model uses the matrix exponential $\bm p(u_{m+1})=\bm p(u_m)\exp(\bm Q_m\Delta_m)$, with $\Delta_m=u_{m+1}-u_m$.

\section{Quasi-Monte Carlo integration and covariate reconstruction}\label{app:qmc}
For each aggregate-data (AgD) arm the covariate-marginal occupancies \eqref{eq:marg} are evaluated by
quasi-Monte Carlo integration. A Sobol' sequence of $\tilde N$ points in $[0,1]^Q$ is transformed
into draws from the arm's baseline joint covariate distribution. When only marginal covariate
summaries are reported, the joint distribution is reconstructed with a Gaussian copula: each
marginal is mapped through its inverse distribution function (a Normal for continuous covariates
matched to the reported mean and standard deviation; a Bernoulli threshold for binary covariates
matched to the reported proportion), and the dependence is induced by a correlation matrix borrowed from the individual participant data (IPD) studies (in the example,
a single correlation between age and ASCT response, $0.03$ in the pooled IPD; ISS-III is drawn independently, and ASCT
response, although binary, is given a Normal margin with its reported mean and standard deviation, a simplification that
places $28$ to $31\%$ of its integration points outside $[0,1]$). For each integration point the illness--death occupancy is propagated
(Appendix~\ref{app:ode}) and the results are averaged; because the occupancy map is nonlinear in
the covariates, the average is taken over the trajectories, not over the covariates. The number of
integration points is chosen by the doubling rule of Phillippo et al.~\citep{phillippo2024general}: half the chains
are run at $\tilde N$ and half at $\lceil \tilde N/2\rceil$, and, once the $\hat R$ within each half
indicates convergence, $\tilde N$ is doubled until no $\hat R$ computed across all chains exceeds
$1.05$. For the worked example the model was fitted at $\tilde N=16$, $32$ and $64$, each time with
the data and sampler settings of the main fit (four chains of $3000$ iterations, $1500$ warmup, dense
mass matrix), with chains started at the main fit's posterior means. With well-mixed chains (within-resolution
$\hat R\le1.002$ for the treatment effects and effect modifiers; one and two divergent transitions at $\tilde N=16$
and $64$), the $\hat R$ computed across resolutions was at most $1.005$, far below $1.05$, and matched
the within-resolution $\hat R$ to within $0.0028$ for the check at $\tilde N=32$ (chains at $32$ and $16$) and to
within $0.0009$ for that at $\tilde N=64$, so by this rule $\tilde N=32$ suffices. $\hat R$ responds only weakly to a small shift
between resolutions, however, and comparing the posterior medians directly is more sensitive. Doubling
$\tilde N$ from $32$ to $64$ moves the medians of the treatment effects and effect modifiers by at most
$0.13$ posterior standard deviations, for thalidomide's
time-variation coefficient $\gamma^{SPt}_{\mathrm{Thal}}$ ($0.011$ per unit of $\log u$), which only the aggregate Morgan2012 study
informs. That move exceeds Monte Carlo error but is small relative to posterior uncertainty, so
$\tilde N=32$ is adequate for the example; an analysis whose conclusions rested on such a parameter
would use more points.

\section{Constructing the aggregate-data life-table}\label{app:data}
The aggregate contribution requires, for each AgD arm, a set of conditional-survival counts for
PFS and OS. These are constructed from the published Kaplan--Meier curves following the curve-digitization
approach of Jansen et al.~\citep{jansen2023multistate}, though reduced to one count per interval
rather than three nested evaluation points. Digitized survival coordinates and reported numbers at risk are used
to form the Kaplan--Meier estimators $S^{\mathrm{PFS}}_{\mathrm{KM}}(u)$ and
$S^{\mathrm{OS}}_{\mathrm{KM}}(u)$ (alternatively, event and censoring times can be reconstructed by
the algorithm of Guyot et al.~\citep{guyot2012}, which gives somewhat different numbers at risk). In the example the counts
are computed directly from the \texttt{multinma} pseudo-patients and their simulated OS (Appendix~\ref{sec:ndmm-data}). Follow-up is partitioned into short (here monthly)
intervals. For a generic interval $[u_m,u_{m'})$ and endpoint, two quantities are recorded: its own number at risk at the interval start, $n^c=n(u_m)$, where
$n(u)$ is the number at risk at time $u$ implied by the reconstructed data (reported numbers at risk
anchor that reconstruction but are published only at coarser times), and the expected number of
event-free survivors implied by the curve at the interval end,
\[
r^c = \Big\lfloor n^c\,\frac{S_{\mathrm{KM}}(u_{m'})}{S_{\mathrm{KM}}(u_m)}\Big\rceil , \] where $\lfloor\cdot\rceil$ denotes rounding to the nearest integer.
This yields one conditional-survival count per interval, each conditioned on the survivors entering that
interval; consecutive intervals are disjoint, so the counts form a grouped-survival (life-table)
likelihood rather than several nested evaluation points within an interval sharing one denominator.
Because the ratio is formed from the Kaplan--Meier estimator, the conditional-survival probability is
estimated correctly under independent censoring and never collapses at the administrative follow-up
horizon; the binomial denominator $n^c$ does not adjust for censoring within the interval. Intervals
whose risk set falls below $10$ patients are dropped. The counts $(n^c,r^c)$ enter the binomial
likelihood \eqref{eq:agdbinom} with the model probabilities \eqref{eq:cond} computed from the covariate-marginal
occupancies.

\paragraph{Ratio of marginals and marginal of the ratio.} Section~\ref{sec:agd} forms each
conditional-survival probability \eqref{eq:cond} as a ratio of marginal occupancies, whereas a direct
reading would average each patient's interval survival over the baseline distribution $f_{jk}(\xv)$. The
two differ, and the ratio is the correct one, because the patients still at risk at $u_m$ are the
survivors, whose covariate distribution is no longer the baseline $f_{jk}(\xv)$ but a depleted one
(for PFS $\propto S_{jk}(u_m\mid\xv)f_{jk}(\xv)$, for OS
$\propto\big(S_{jk}+P_{jk}\big)(u_m\mid\xv)f_{jk}(\xv)$). Averaging the individual interval survival over that at-risk distribution gives,
for the stable (PFS) and alive (OS) occupancies respectively,
\begin{equation}
\begin{aligned}
\frac{\Sbar_{jk}(u_{m'})}{\Sbar_{jk}(u_m)} &=\E_{f(\xv\,\mid\,\text{alive in }S\text{ at }u_m)}\!\left[\frac{S_{jk}(u_{m'}\mid\xv)}{S_{jk}(u_m\mid\xv)}\right],\\ \frac{\Sbar_{jk}(u_{m'})+\Pbar_{jk}(u_{m'})}{\Sbar_{jk}(u_m)+\Pbar_{jk}(u_m)} &=\E_{f(\xv\,\mid\,\text{alive at }u_m)}\!\left[\frac{S_{jk}(u_{m'}\mid\xv)+P_{jk}(u_{m'}\mid\xv)}{S_{jk}(u_m\mid\xv)+P_{jk}(u_m\mid\xv)}\right],
\end{aligned}
\label{eq:mlnmr-cond}
\end{equation}
so each ratio of marginals is exactly the average interval survival over the \emph{at-risk}
covariate distribution, not an approximation.
\section{Semi-synthetic overall survival for the NDMM example}\label{sec:ndmm-data}
The illustrative-example network and aggregate covariate summaries are real, but the PFS event times
and patient covariates distributed with \texttt{multinma} are not: the individual participant data are
simulated to resemble the original trials, and the aggregate event times are reconstructed from
digitized published Kaplan--Meier curves \citep{phillippo2024general}. Because the public data report
only PFS, they are augmented here with an OS layer to obtain a complete illness--death process; OS is
therefore simulated in addition, and the results illustrate the method rather than report
empirical survival findings. For the aggregate trials, each reconstructed pseudo-patient is first
assigned covariates drawn independently from its arm's published summaries; the covariate summaries used
in the analysis are computed from these draws. The augmentation keeps the distributed PFS as the
stable-state exit and adds two ingredients.
First, each observed PFS event is split into a progression (probability
$\pi(\xv)=\mathrm{logit}^{-1}\big(2.0-0.15\,(\text{age}-60)/10-0.30\,\text{ISS-III}\big)$, $0.88$ at age $60$ without ISS-III and
between $0.82$ and $0.90$ for $90\%$ of patients)
or a pre-progression death. The split does not depend on treatment, so pre-progression deaths share the
treatment effect on PFS. Second, progressors are assigned a clock-forward Weibull
progression$\to$death time with log-intensity
$\log h^{PD}(u)=-4.5+0.2\log u+\xv^{\!\top}\bm\beta^{PD}_1+\gamma^{PD}_k$ ($u$ in months). The true prognostic coefficients are $\bm\beta^{PD}_1=(0.20,0.40,-0.15)$ (age, ISS-III,
response). The true treatment effects,
$(\gamma^{PD}_{\mathrm{Pbo}},\gamma^{PD}_{\mathrm{Len}},\gamma^{PD}_{\mathrm{Thal}})=(0,-0.12,-0.08)$, are modest, so the OS benefit is largely
inherited from the progression delay. The fitted model's exponential progression$\to$death baseline is therefore mildly
misspecified (Weibull shape $1.2$ in the data). The baseline is set to give a
median post-progression survival of roughly $2.5$--$3$ years. Each patient has one censoring time for
both endpoints, so that censoring does not depend on the state: patients censored while stable keep
their PFS censoring time, and a patient with a PFS event is given one drawn from the arm's PFS
censoring distribution, conditional on exceeding the event time.

\section{Trial characteristics and Kaplan--Meier curves}\label{app:trials}
Table~\ref{tab:trials} summarizes the baseline patient characteristics of the illustrative-example network by
trial and arm, and Figure~\ref{fig:km} shows the trial-specific Kaplan--Meier curves. Progression-free
survival is the \texttt{multinma} data (simulated for the IPD studies, reconstructed from digitized
curves for the aggregate studies), as are the IPD covariates; the aggregate studies' covariates are drawn
from their published summaries, and overall survival is the semi-synthetic augmentation of
Appendix~\ref{sec:ndmm-data}.

\begin{table}[htbp]\centering\footnotesize
\caption{Baseline patient characteristics by trial and arm: sample size, age in years as mean (SD),
proportion with ISS-III, and proportion achieving a complete or very-good-partial response after ASCT.
For the aggregate trials, the values summarize the pseudo-patient covariates drawn from the published
summaries (Appendix~\ref{sec:ndmm-data}); the analysis uses their summaries pooled over arms (Table~\ref{tab:agd}a).}
\label{tab:trials}
\begin{tabular}{llrccc}
\toprule
Trial & Arm & $n$ & Age & ISS-III & ASCT resp.\\
\midrule
Attal2012    & Len  & 307  & 54 (6)  & 24\% & 55\%\\
Attal2012    & Pbo  & 307  & 54 (5)  & 16\% & 54\%\\
McCarthy2012 & Len  & 231  & 58 (6)  & 27\% & 62\%\\
McCarthy2012 & Pbo  & 229  & 57 (6)  & 18\% & 71\%\\
Palumbo2014  & Len  & 126  & 54 (10) & 10\% & 42\%\\
Palumbo2014  & Pbo  & 125  & 54 (9)  & 12\% & 38\%\\
Jackson2019  & Len  & 1137 & 65 (9)  & 25\% & 83\%\\
Jackson2019  & Pbo  & 864  & 65 (10) & 20\% & 83\%\\
Morgan2012   & Pbo  & 410  & 64 (9)  & 40\% & 73\%\\
Morgan2012   & Thal & 408  & 66 (8)  & 33\% & 72\%\\
\bottomrule
\end{tabular}
\end{table}
\FloatBarrier   

\IfFileExists{figures/fig_ndmm_km.png}{%
\begin{sidewaysfigure}\centering
\includegraphics[width=0.95\textheight]{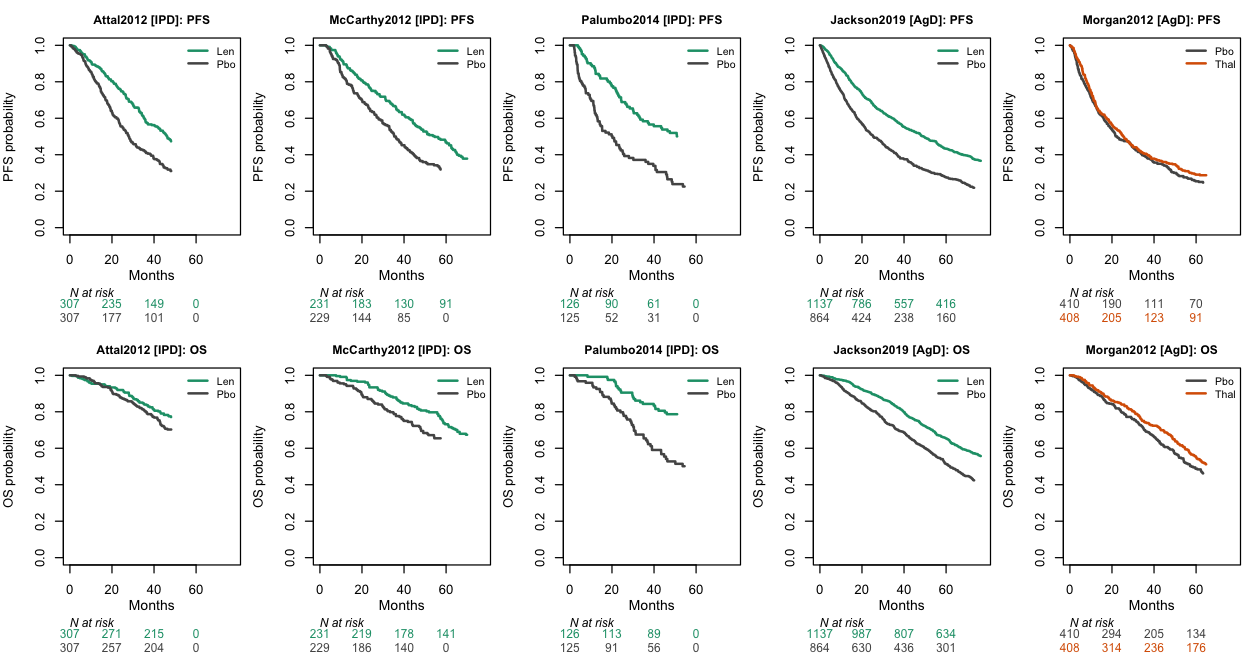}
\caption{Trial-specific Kaplan--Meier curves by arm with numbers at risk: PFS (top) and OS (bottom).
PFS is the \texttt{multinma} data (simulated for IPD studies, digitized-curve reconstructions for
aggregate studies); OS is the semi-synthetic augmentation.}
\label{fig:km}
\end{sidewaysfigure}}{}

\FloatBarrier   

\section{Structure of the analysis data}\label{app:datastruct}
This appendix illustrates the two data modalities that enter the joint likelihood. The individual
participant data (IPD) trials contribute one record per patient, with covariates and the realized
transition times, including whether the stable-state exit was a progression or a pre-progression death
(Table~\ref{tab:ipd}). The aggregate-data (AgD) trials contribute covariate summaries
(Table~\ref{tab:agd}a), from which a joint covariate distribution is reconstructed by Gaussian copula,
together with the reconstructed conditional-survival life-table counts that each AgD arm contributes for
PFS and OS over the full follow-up (Table~\ref{tab:agd}b; Appendix~\ref{app:data}).
\begin{table}[htbp]\centering\footnotesize\setlength{\tabcolsep}{4pt}
\caption{Individual participant data (IPD), illustrative excerpt (McCarthy2012): one record per patient,
with covariates, event/censoring times, and whether the stable-state exit was progression or
pre-progression death.}
\label{tab:ipd}
\begin{tabular}{llcccccccc}\toprule
Study & Arm & Age & ISS-III & ASCT resp. & PFS (mo) & PFS event & OS (mo) & OS event & Stable exit\\\midrule
McCarthy2012 & Len & 56 & no & yes & 70.0 & 0 & 70.0 & 0 & --\\
McCarthy2012 & Len & 65 & no & yes & 13.0 & 1 & 28.4 & 1 & progression\\
McCarthy2012 & Len & 61 & no & no & 70.0 & 0 & 70.0 & 0 & --\\
McCarthy2012 & Pbo & 51 & no & yes & 31.1 & 1 & 57.4 & 0 & progression\\
McCarthy2012 & Pbo & 62 & no & no & 3.3 & 0 & 3.3 & 0 & --\\
McCarthy2012 & Pbo & 57 & no & yes & 18.2 & 1 & 18.2 & 1 & death\\
$\vdots$ & $\vdots$ & $\vdots$ & $\vdots$ & $\vdots$ & $\vdots$ & $\vdots$ & $\vdots$ & $\vdots$ & $\vdots$\\
\bottomrule\end{tabular}\end{table}
\begin{table}[htbp]\centering\footnotesize
\caption{Aggregate data (AgD) as used in the analysis. \textbf{(a)} Baseline covariate summaries,
pooled over arms and computed from the pseudo-patient covariates (Appendix~\ref{sec:ndmm-data}), from
which a joint covariate distribution is reconstructed by Gaussian copula.
\textbf{(b)} Reconstructed conditional-survival life-table for an AgD arm (Jackson2019
lenalidomide), for PFS and OS: per interval $[u_m,u_{m'})$, the at-risk count $n^{c}$ and expected
event-free survivors $r^{c}$ that enter the binomial likelihood \eqref{eq:agdbinom}.}
\label{tab:agd}
\textbf{(a) Covariate summaries used in the analysis}\\[2pt]
\begin{tabular}{lccc}\toprule
Study & Mean age (SD) & \%\ ISS-III & \%\ response\\\midrule
Jackson2019 & 65 (9.4) & 23\% & 83\%\\
Morgan2012 & 65 (8.7) & 36\% & 72\%\\
\bottomrule\end{tabular}\\[10pt]
\textbf{(b) Reconstructed conditional-survival life-table}\\[2pt]
\begin{tabular}{ll c cc cc}\toprule
& & & \multicolumn{2}{c}{PFS} & \multicolumn{2}{c}{OS}\\\cmidrule(lr){4-5}\cmidrule(lr){6-7}
Study & Arm & Interval $[u_m,u_{m'})$ (mo) & $n^{c}$ & $r^{c}$ & $n^{c}$ & $r^{c}$\\\midrule
\multirow{15}{*}{Jackson2019} & \multirow{15}{*}{Len} & $[0,1)$ & 1137 & 1131 & 1137 & 1135\\
 &  & $[1,2)$ & 1131 & 1121 & 1135 & 1134\\
 &  & $[2,3)$ & 1119 & 1105 & 1132 & 1128\\
 &  & $[3,4)$ & 1105 & 1093 & 1128 & 1127\\
 &  & $[4,5)$ & 1091 & 1072 & 1125 & 1122\\
 &  & $[5,6)$ & 1068 & 1048 & 1118 & 1116\\
 &  & $[6,7)$ & 1044 & 1020 & 1112 & 1106\\
 &  & $[7,8)$ & 1014 & 999 & 1099 & 1094\\
 &  & $[8,9)$ & 997 & 983 & 1092 & 1091\\
 &  & $[9,10)$ & 979 & 969 & 1087 & 1084\\
 &  & $[10,11)$ & 966 & 951 & 1081 & 1078\\
 &  & $[11,12)$ & 946 & 934 & 1073 & 1071\\
 &  & $[12,13)$ & 929 & 914 & 1066 & 1062\\
 &  & $[13,14)$ & 910 & 886 & 1058 & 1053\\
 &  & $[14,15)$ & 883 & 865 & 1049 & 1041\\
$\vdots$ & $\vdots$ & $\vdots$ & $\vdots$ & $\vdots$ & $\vdots$ & $\vdots$\\
\bottomrule\end{tabular}\end{table}
\FloatBarrier

\section{Model checking and MCMC diagnostics}\label{app:suppfig}
This appendix reports the checks behind Section~\ref{sec:ex-model}: the fit of the illustrative
example to the observed data, its MCMC diagnostics and run time, the contraction from prior to
posterior, and the fineness of the occupancy grid. Leave-one-out cross-validation by study, which
compares the baseline forms, is reported with the alternative baselines in Appendix~\ref{app:baselines}
(Table~\ref{tab:baseline-loo}).

\subsection{Fit to the observed data}\label{app:fit}
Figure~\ref{fig:ppc} sets the model-implied population-average PFS and OS of every arm against the
observed Kaplan--Meier curve. Each arm is averaged over the covariate distribution that enters its own
likelihood contribution: the recorded covariates of the patients in an IPD arm, and the reconstructed
distribution (the $\tilde N=32$ quasi-Monte Carlo points) in an AgD arm. Because the $95\%$ bands
reflect parameter uncertainty only, an observed curve can leave them even where the model fits.

PFS is reproduced in every arm: over the follow-up with at least ten patients at risk, the observed
curve departs from the posterior median by at most $0.08$ in survival probability, the largest
departures being in the small Palumbo2014 trial. OS is reproduced in both aggregate trials (largest
departure $0.04$) and in Attal2012 ($0.02$). In McCarthy2012 and Palumbo2014, by contrast, the model
over-predicts OS on placebo and under-predicts it on lenalidomide, with the observed curve outside the
$95\%$ band in $46$--$69\%$ of months, against at most $17\%$ in the other arms (largest departures $0.04$
and $0.11$ on placebo, $0.05$ and $0.07$ on lenalidomide), so it understates the separation between the
arms. On the difference between the arms the misfit is clearer. The observed OS separation lies above
the model's $95\%$ band in $98\%$ (McCarthy2012) and $96\%$ (Palumbo2014) of months. At $48$ months the
model gives $0.05$ against an observed $0.10$ in McCarthy2012, and $0.11$ against $0.26$ in Palumbo2014.
In the other three trials the two differ by at most $0.03$. For PFS the separation is reproduced to
within $0.05$ except in Palumbo2014, the smallest trial, where the model's separation is narrower by up to $0.10$.
Part of that gap traces to the stable$\to$death transition (below), since pre-progression deaths are also PFS events.

Because the IPD record which transition each patient made, the misfit can be traced to a transition.
Table~\ref{tab:ppc-oe} compares, for each IPD arm, the observed number of transitions with the number
the model expects given each patient's time at risk. Progressions agree closely with expectation in
every arm, and post-progression deaths depart from it in both directions without a pattern by arm.
Pre-progression deaths depart systematically. In McCarthy2012 and Palumbo2014 the model expects too
few on placebo ($14.4$ and $7.4$, against $24$ and $12$ observed) and too many on lenalidomide ($23.2$
and $12.0$, against $13$ and $7$). Attal2012 departs in the same direction, by less ($18.1$ on
placebo and $25.5$ on lenalidomide, against $21$ and $22$ observed). Because the intervals reflect parameter uncertainty only and a
count also varies about its expectation, the evidence lies in this common direction across all six arms rather than in any one arm. The cause is a specification choice. Model \eqref{eq:ex-lp} gives
stable$\to$death no treatment effect, so the stable$\to$death intensity is common to both arms of a
study and the whole PFS benefit is attributed to progression. The semi-synthetic OS layer instead
splits each PFS event into a progression or a pre-progression death with a probability that does not
depend on treatment (Appendix~\ref{sec:ndmm-data}), so pre-progression deaths share the treatment
effect on PFS. Adding a term $\mathds{1}(k\neq1)\,\gamma^{SD}_k$ to the stable$\to$death log-intensity
in \eqref{eq:ex-lp} would represent this route. The example keeps the simpler specification. Sections~\ref{sec:ex-relative} and~\ref{sec:ex-absolute} use the
McCarthy2012 baseline, where this misfit is evident (an OS separation of $0.05$ against $0.10$ observed at $48$ months), so the
overall-survival benefit of lenalidomide they report is likely understated.

\IfFileExists{figures/fig_ndmm_ppc.png}{%
\begin{figure}[htbp]\centering
\includegraphics[width=0.85\linewidth]{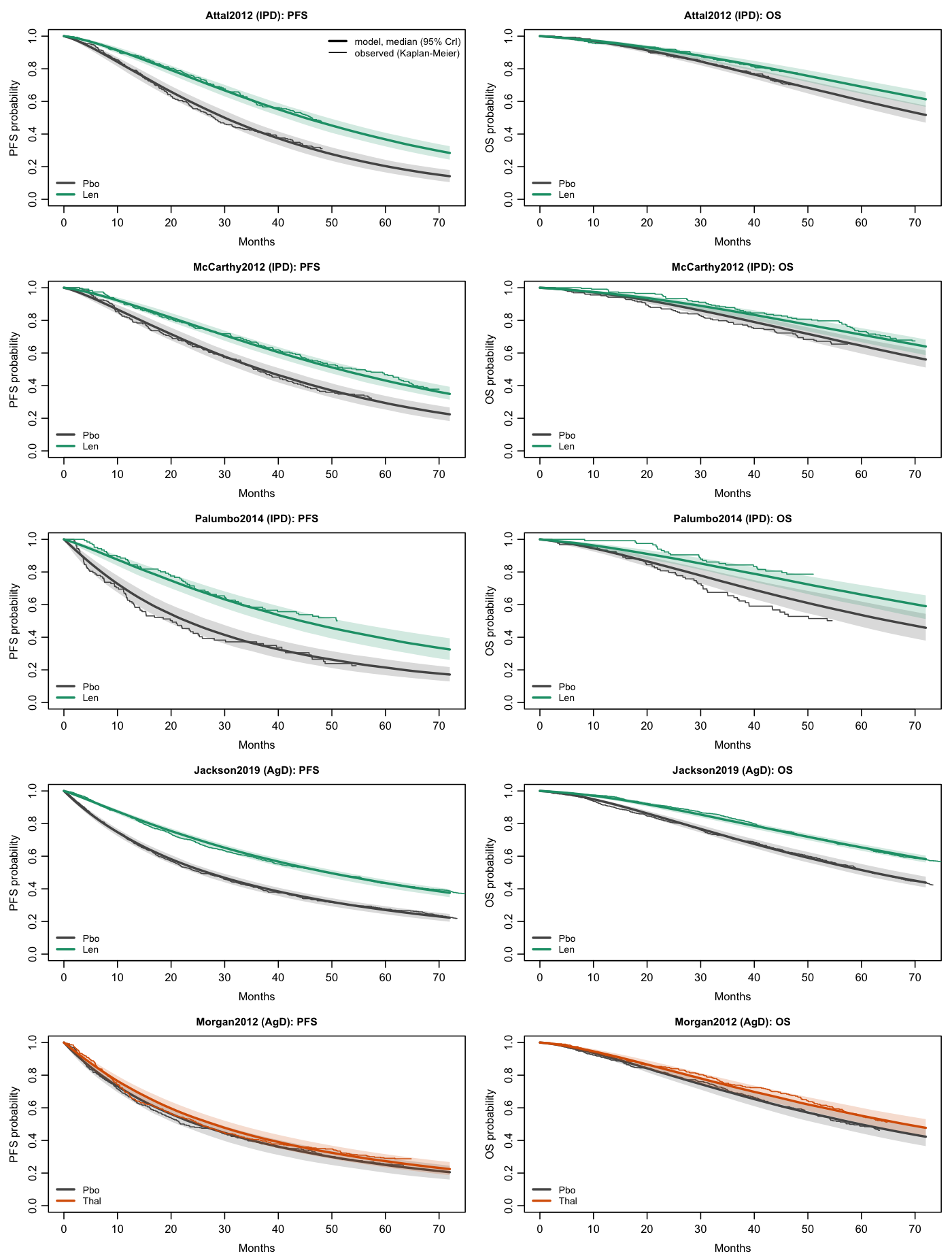}
\caption{Fit to the observed survival curves: model-implied population-average PFS (left) and OS
(right) by trial and arm (posterior median, 95\% CrI shaded, from $200$ equally spaced posterior draws; parameter uncertainty
only) against the observed Kaplan--Meier curves
(steps).}
\label{fig:ppc}
\end{figure}}{}

\FloatBarrier   
\begin{table}[htbp]\centering\footnotesize
\caption{Observed (O) and model-expected (E) numbers of transitions in each IPD arm. E is the posterior
median [95\% CrI], over 400 equally spaced posterior draws, of the cumulative transition intensity summed
over each patient's observed time at
risk in the origin state, which under the model has the same expectation as the observed count; the intervals reflect
parameter uncertainty only.}
\label{tab:ppc-oe}
\begin{tabular}{llrlrlrl}
\toprule
 & & \multicolumn{2}{c}{stable$\to$progression} & \multicolumn{2}{c}{stable$\to$death} & \multicolumn{2}{c}{progression$\to$death}\\
\cmidrule(lr){3-4}\cmidrule(lr){5-6}\cmidrule(lr){7-8}
Trial & Arm & O & E & O & E & O & E\\
\midrule
Attal2012 & Pbo & 175 & 175.0 [155.9, 196.8] & 21 & 18.1 [13.1, 23.7] & 61 & 66.3 [52.5, 78.1] \\
Attal2012 & Len & 127 & 125.7 [108.9, 144.3] & 22 & 25.5 [18.9, 33.3] & 42 & 35.7 [28.2, 44.3] \\
McCarthy2012 & Pbo & 114 & 119.3 [103.9, 136.0] & 24 & 14.4 [10.3, 19.1] & 44 & 44.9 [36.4, 55.3] \\
McCarthy2012 & Len & 120 & 114.2 [100.4, 132.7] & 13 & 23.2 [16.4, 30.5] & 55 & 52.5 [41.6, 65.6] \\
Palumbo2014 & Pbo & 72 & 69.5 [56.8, 82.5] & 12 & 7.4 [4.6, 10.9] & 38 & 33.3 [26.1, 44.3] \\
Palumbo2014 & Len & 51 & 54.0 [44.2, 66.2] & 7 & 12.0 [7.6, 18.4] & 17 & 20.8 [15.9, 27.9] \\
\bottomrule
\end{tabular}
\end{table}
\FloatBarrier

\subsection{MCMC diagnostics and run time}\label{app:mcmc}
Table~\ref{tab:mcmc-diag} reports the sampler diagnostics for the presented Weibull fit and the two
alternative baselines of Appendix~\ref{app:baselines}, by parameter group, using the rank-normalized
$\hat R$ of Vehtari et al.~\citep{vehtari2021rhat}. In all three fits every $\hat R$ is at most $1.004$, and the
Monte Carlo standard error of every posterior median at most $0.023$ posterior standard deviations. The
second-order fractional-polynomial (FP2) fit had four divergent transitions in $6000$ draws, one in each chain and not concentrated in any
parameter. It also had the lowest tail effective sample size, $604$ for its shape parameters.

The chains were initialized at the posterior means of a preliminary fit with small random
perturbations (standard deviation $0.05$ for the Weibull fit, $0.04$ for the alternative baselines and
$0.02$ for the M-spline weights), because the model initializes poorly from Stan's default random starting values.
Because these starting values are not overdispersed, $\hat R$ is weaker evidence of convergence than it
would be from overdispersed starts.

\begin{table}[htbp]\centering\footnotesize
\caption{MCMC diagnostics for the illustrative example, by S$\to$P baseline form (FP2: second-order
fractional polynomial). ESS: minimum bulk / tail
effective sample size over the parameters in the group; $\hat R$: maximum rank-normalized split-$\hat R$;
MCSE/SD: maximum Monte Carlo standard error of the posterior median relative to the posterior SD. Run
times are per chain and exclude model compilation; the chains ran in parallel on a laptop (Apple M5 Pro).}
\label{tab:mcmc-diag}
\begin{tabular}{lccc}
\toprule
 & Weibull & FP2 & M-spline\\
\midrule
Chains $\times$ post-warmup draws & 4 $\times$ 1500 & 4 $\times$ 1500 & 4 $\times$ 1500\\
Divergent transitions & 0 & 4 & 0\\
Warmup per chain (min) & 3.4--3.7 & 16.8--18.7 & 5.3--6.4\\
Sampling per chain (min) & 1.4--2.1 & 1.8--3.3 & 2.3--3.1\\
\midrule
\multicolumn{4}{l}{\textit{Treatment effects ($\gamma$)}}\\
\quad ESS bulk / tail & 8449 / 3901 & 5035 / 2912 & 7372 / 4176\\
\quad $\hat R$ & 1.002 & 1.002 & 1.001\\
\quad MCSE/SD & 0.015 & 0.018 & 0.017\\
\midrule
\multicolumn{4}{l}{\textit{Effect modifiers ($\bet_2$)}}\\
\quad ESS bulk / tail & 8347 / 4564 & 7008 / 3457 & 7668 / 4339\\
\quad $\hat R$ & 1.002 & 1.001 & 1.002\\
\quad MCSE/SD & 0.014 & 0.017 & 0.015\\
\midrule
\multicolumn{4}{l}{\textit{Prognostic effects ($\bet_1$)}}\\
\quad ESS bulk / tail & 8499 / 4108 & 6597 / 3073 & 6573 / 4281\\
\quad $\hat R$ & 1.002 & 1.002 & 1.002\\
\quad MCSE/SD & 0.017 & 0.017 & 0.018\\
\midrule
\multicolumn{4}{l}{\textit{Baseline intercepts ($c^r_j$)}}\\
\quad ESS bulk / tail & 6286 / 3614 & 1846 / 1405 & 6202 / 4002\\
\quad $\hat R$ & 1.003 & 1.003 & 1.002\\
\quad MCSE/SD & 0.017 & 0.019 & 0.017\\
\midrule
\multicolumn{4}{l}{\textit{Baseline shape parameters}}\\
\quad ESS bulk / tail & 2717 / 1993 & 1448 / 604 & 3337 / 2524\\
\quad $\hat R$ & 1.002 & 1.004 & 1.002\\
\quad MCSE/SD & 0.018 & 0.023 & 0.021\\
\bottomrule
\end{tabular}
\end{table}
\FloatBarrier   

\subsection{Prior and posterior distributions}\label{app:priorpost}
Figure~\ref{fig:priorpost} overlays prior and posterior densities for eight representative parameters of
the presented fit, each labeled with its contraction, one minus the ratio of posterior to prior variance.
Over all $43$ parameters the posterior variance is at most $6.6\%$ of the prior variance (the Morgan2012
stable$\to$death intercept), except for four of the five stable$\to$death log-shapes, which reach $22.4\%$ in Jackson2019,
$14.7\%$ in Morgan2012, $11.8\%$ in Palumbo2014 and $7.2\%$ in McCarthy2012.

\IfFileExists{figures/fig_ndmm_prior_post.png}{%
\begin{figure}[htbp]\centering
\includegraphics[width=\linewidth]{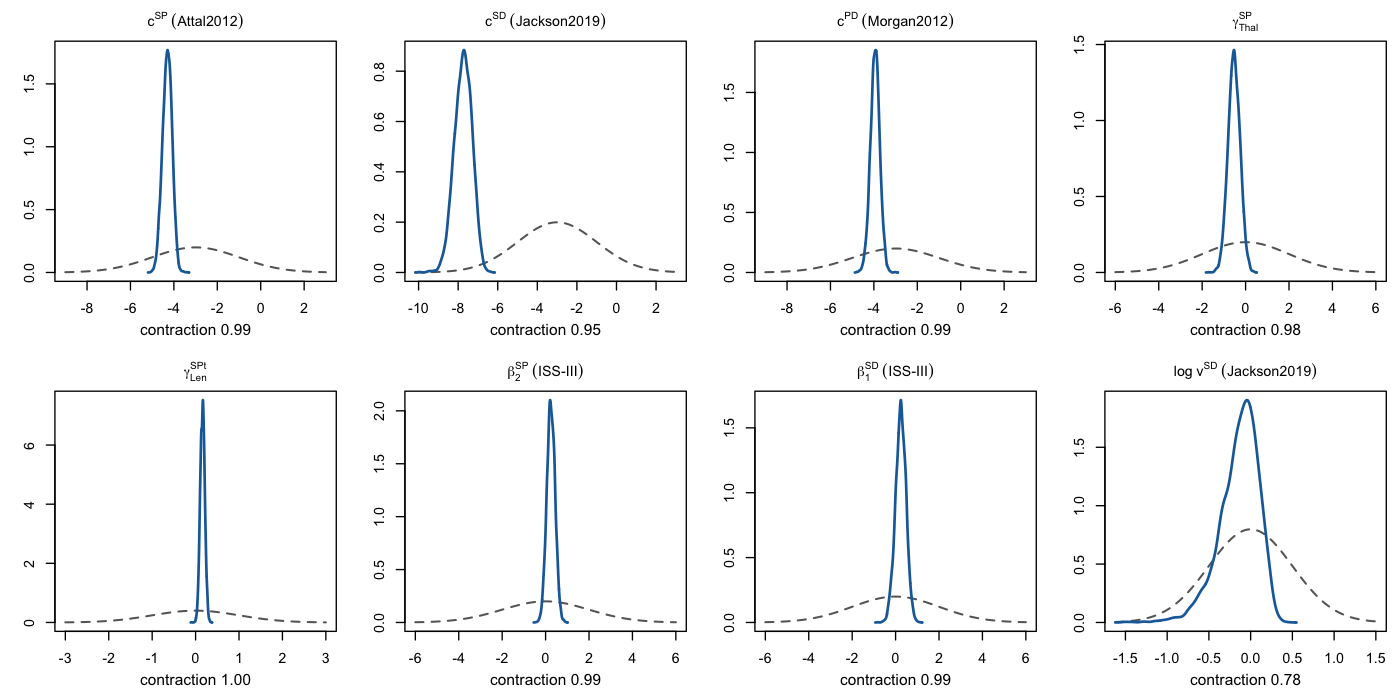}
\caption{Prior (dashed) and posterior (solid) densities for representative parameters of the
illustrative example (Weibull stable$\to$progression baseline), with the contraction
$1-\mathrm{var}_{\mathrm{post}}/\mathrm{var}_{\mathrm{prior}}$ of each.}
\label{fig:priorpost}
\end{figure}}{}
\FloatBarrier

\subsection{Fineness of the occupancy grid}\label{app:grid}
The effect of halving the monthly occupancy grid was checked by reweighting the posterior draws with Pareto-smoothed
importance sampling (Section~\ref{sec:computation}; $\hat k=-0.06$, far below the threshold of $0.7$, so the reweighting is
reliable). It changes the covariate-marginal occupancies by at most $0.003$, most of it arising in the
first month. It moves the posterior medians of the treatment effects and effect modifiers by at most
$0.02$ posterior standard deviations, and no posterior median by more than $0.09$ posterior standard
deviations. Eleven of the $43$ medians move by more than their Monte Carlo error, the largest in the aggregate studies'
baseline parameters and the age effect on stable$\to$progression, so an analysis whose conclusions rested on those
parameters should use the finer grid.

\section{Alternative stable--progression baseline hazard specifications}\label{app:baselines}
The illustrative example models the stable$\to$progression (S$\to$P) baseline hazard as Weibull
(Section~\ref{sec:example}), i.e., $\log h^{SP}_{jk}(u\mid\xv)=c^{SP}_j+(v^{SP}_j+\gamma^{SPt}_k-1)\log u
+\xv^{\!\top}(\bet^{SP}_1+\bet^{SP}_{2,k})+\gamma^{SP}_k$, with $\bet^{SP}_{2,k}=\bet^{SP}_2$ for the active treatments
and $\bm 0$ for placebo. To probe sensitivity to that functional
form (and to demonstrate that the framework accommodates the flexible baselines of its parent
methods), the model was refitted twice, replacing only the S$\to$P baseline while keeping the
stable$\to$death (Weibull) and progression$\to$death (exponential) transitions, the covariate effects,
and the time-varying S$\to$P treatment effect
$\mathrm{HR}^{SP}_k(u)=\exp(\gamma^{SP}_k+\xv^{\!\top}\bet^{SP}_2)\,u^{\gamma^{SPt}_k}$
unchanged.

\emph{(i) Second-order fractional polynomial (FP2)} \citep{jansen2023multistate}, with powers $(0,1)$ ($\log u$ and $u$) and
study-specific coefficients $a_{1j},a_{2j}$ with independent $\mathrm N(0,1)$ priors:
\begin{equation*}
\log h^{SP}_{jk}(u\mid\xv)=c^{SP}_j+a_{1j}\log u+a_{2j}\,u+\gamma^{SPt}_k\log u
 +\xv^{\!\top}(\bet^{SP}_1+\bet^{SP}_{2,k})+\gamma^{SP}_k .
\end{equation*}

\emph{(ii) Degree-3 M-spline baseline with a common shape}: one weight simplex $\bm w=(w_1,\dots,w_L)$,
$\sum_l w_l=1$, shared by all studies, on $L=6$ M-spline basis functions $M_l$ \citep{phillippo2024general,phillippo2026mspline},
with internal knots at the $1/3$ and $2/3$ quantiles of the IPD progression times ($14.5$ and $29.1$
months), boundary knots at $0$ and $77$ months, and a $\mathrm{Dirichlet}(2,\dots,2)$ prior on $\bm w$,
which pulls the weights mildly toward equality; with these knots, equal weights give a hazard that is
higher near the start and end of follow-up than in between, not a constant one. Only the log-scale $c^{SP}_j$ differs between studies:
\begin{equation*}
\log h^{SP}_{jk}(u\mid\xv)=\log\!\Big(\textstyle\sum_{l=1}^{L} w_l\,M_l(u)\Big)+c^{SP}_j
 +\gamma^{SPt}_k\log u+\xv^{\!\top}(\bet^{SP}_1+\bet^{SP}_{2,k})+\gamma^{SP}_k .
\end{equation*}

In both, the retained stable$\to$death intensity is Weibull, $\log h^{SD}_{jk}(u\mid\xv)=c^{SD}_j
+(v^{SD}_j-1)\log u+\xv^{\!\top}\bet^{SD}_1$, and progression$\to$death is exponential,
$\log h^{PD}_{jk}(u\mid\xv)=c^{PD}_j+\xv^{\!\top}\bet^{PD}_1+\gamma^{PD}_k$, as in the main model. Because the
time-varying factor $u^{\gamma^{SPt}_k}$ multiplies the (non-Weibull) baseline, the individual-level
S$\to$P cumulative hazard has no convenient closed form and is integrated by $15$-node Gauss--Legendre quadrature. All three
fits used four chains of $3000$ iterations and a dense mass matrix; their diagnostics are reported in
Table~\ref{tab:mcmc-diag} (Appendix~\ref{app:mcmc}).

Within the observed follow-up an M-spline is the more
adaptable form over time, with one weight per basis function against the two coefficients of the FP2, and
so the safer default when the hazard may change direction more than once. Here, however, its shape is
shared by all studies, whereas the Weibull and FP2 shapes are study-specific, so the comparison below
sets a form that is more flexible over time but common across studies against forms that are less
flexible over time but free to differ between them. Sharing the shape is a modeling choice in its own
right: it borrows strength across studies but makes their baselines proportional, and study-specific
weights relax that at the cost of more parameters. Beyond the observed follow-up neither
form is informed by the data. A fractional polynomial continues along its parametric form: with powers
$(0,1)$ the hazard eventually rises or falls exponentially, depending on the sign of the linear term. An
extrapolated spline follows whatever boundary behavior is imposed, such as a constant hazard beyond the
last knot or a continuation of the terminal polynomial pieces. Which extrapolation is more plausible
must therefore be judged against clinical knowledge or external data, not inferred from the form. This
matters whenever the synthesis feeds an extrapolated economic model. Sparse networks, in which each
study contributes few events, favor forms with fewer shape parameters per study: a low-order fractional
polynomial, few knots, or a shared shape. Pending a formal simulation comparison, a practical approach is to fit both, compare them by leave-one-out
cross-validation on
the IPD contribution and, separately, on the aggregate PFS and OS contributions, overall and by study
(Section~\ref{sec:computation}), and inspect
the implied hazards over the extrapolation horizon as well as the observed one.

Table~\ref{tab:baseline-loo} makes this comparison, with the aggregate PFS and OS as separate criteria
(Section~\ref{sec:com_multinomial}). On the IPD, the second-order fractional polynomial
has the highest estimated predictive accuracy, ahead of the Weibull ($\Delta$ELPD $-6.1$, SE $3.8$) and, more clearly, the M-spline ($-13.4$, SE
$6.2$). None of the per-study differences exceeds
$1.7$ standard errors, the Weibull matches the fractional polynomial in Attal2012, and the M-spline matches it
in Palumbo2014, so no single study drives the ranking. The aggregate PFS also favors the fractional
polynomial but, unlike the IPD, ranks the M-spline ($-7.7$, SE $4.8$) ahead of the Weibull ($-17.4$, SE
$7.7$). Two of its $280$ intervals, the first month of each Morgan2012 arm, have $\hat k$ above $0.7$ under
the Weibull, one of them also under the M-spline, so their estimates are unreliable; without them the
ranking is unchanged (M-spline $-5.8$, SE $4.3$; Weibull $-9.8$, SE $5.3$). On the aggregate OS, which is
semi-synthetic (Appendix~\ref{sec:ndmm-data}), the three forms are close (Weibull $0.0$, SE $1.3$; M-spline
$-2.8$, SE $1.9$). Leaving out both counts of each interval together (Section~\ref{sec:com_multinomial})
changes none of these totals by more than $0.2$. The SEs treat the intervals as independent, whereas adjacent
intervals of a curve are correlated, so they may understate the uncertainty.

\begin{table}[htbp]\centering\footnotesize
\caption{Leave-one-out comparison of the S$\to$P baseline-hazard forms on the IPD contribution and, as
separate criteria, on the aggregate PFS and aggregate OS contributions, by study and in total; the aggregate
PFS and OS criteria are not added to each other (Section~\ref{sec:com_multinomial}). LOOIC $=-2\,$ELPD (lower
is better). $\Delta$ELPD: difference in expected log pointwise predictive density from the best model on the IPD
(FP2; negative: worse than FP2), with the SE of the paired differences. $\hat k$: Pareto-smoothed
importance-sampling diagnostic; above $0.7$ the estimate for that patient or interval is unreliable. In the
aggregate panels the SEs treat the intervals as independent; adjacent intervals of a curve are correlated,
so they may understate the uncertainty. FP2: second-order fractional polynomial; M-spline: degree-3 M-spline.}
\label{tab:baseline-loo}
\begin{tabular}{lrrrcc}
\toprule
 & \multicolumn{3}{c}{LOOIC} & \multicolumn{2}{c}{$\Delta$ELPD (SE) vs FP2}\\
\cmidrule(lr){2-4}\cmidrule(lr){5-6}
Study & Weibull & FP2 & M-spline & Weibull & M-spline\\
\midrule
\multicolumn{6}{l}{\textit{IPD ($n$: patients)}}\\
Attal2012 ($n=614$) & $4625$ & $4626$ & $4641$ & $0.7\;(1.7)$ & $-7.2\;(4.4)$ \\
McCarthy2012 ($n=460$) & $3931$ & $3925$ & $3937$ & $-3.0\;(2.2)$ & $-6.2\;(3.9)$ \\
Palumbo2014 ($n=251$) & $1988$ & $1981$ & $1980$ & $-3.9\;(2.6)$ & $0.0\;(2.0)$ \\
Total & $10544$ & $10532$ & $10559$ & $-6.1\;(3.8)$ & $-13.4\;(6.2)$ \\
Largest $\hat k$ & $0.31$ & $0.25$ & $0.23$ & \\
\midrule
\multicolumn{6}{l}{\textit{Aggregate PFS ($n$: life-table intervals)}}\\
Jackson2019 ($n=151$) & $726$ & $718$ & $729$ & $-3.8\;(3.5)$ & $-5.1\;(2.6)$ \\
Morgan2012 ($n=129$) & $512$ & $484$ & $490$ & $-13.6\;(6.9)$ & $-2.6\;(4.0)$ \\
Total & $1238$ & $1203$ & $1218$ & $-17.4\;(7.7)$ & $-7.7\;(4.8)$ \\
Largest $\hat k$ & $0.82$ & $0.61$ & $0.72$ & \\
\midrule
\multicolumn{6}{l}{\textit{Aggregate OS ($n$: life-table intervals)}}\\
Jackson2019 ($n=151$) & $697$ & $698$ & $700$ & $0.3\;(0.6)$ & $-1.0\;(1.2)$ \\
Morgan2012 ($n=129$) & $480$ & $479$ & $483$ & $-0.3\;(1.1)$ & $-1.7\;(1.4)$ \\
Total & $1177$ & $1177$ & $1183$ & $0.0\;(1.3)$ & $-2.8\;(1.9)$ \\
Largest $\hat k$ & $0.29$ & $0.35$ & $0.41$ & \\
\bottomrule
\end{tabular}
\end{table}

The fitted S$\to$P baseline hazards under the three forms are overlaid in
Figure~\ref{fig:baseline-forms}: in the study shown the Weibull is monotonically increasing, whereas
the M-spline peaks near $10$ months and the FP2 near $30$, after which both decline, a shape a Weibull
cannot represent.

\IfFileExists{figures/fig_ndmm_baseline_forms.png}{%
\begin{figure}[htbp]\centering
\includegraphics[width=.8\linewidth]{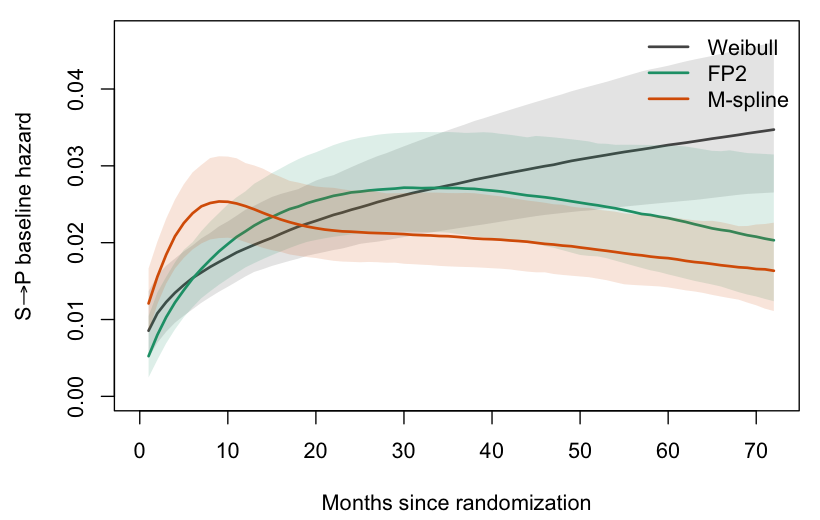}
\caption{Fitted stable$\to$progression baseline hazard for study~2 (McCarthy2012; placebo, reference
covariates) under the three baseline forms, with 95\% credible bands. The Weibull and FP2 baselines
are study-specific in scale and shape; the M-spline baselines share one shape and differ only in scale.}
\label{fig:baseline-forms}
\end{figure}}{}

The population-adjusted results under each alternative baseline (hazard ratios, baseline transition hazards, state-occupancy probabilities, and survival, as in
Sections~\ref{sec:ex-relative}--\ref{sec:ex-absolute}) are shown in the figures of Appendices~\ref{app:fp2} (fractional polynomial) and~\ref{app:mspline} (M-spline) and
summarized in Tables~\ref{tab:summ-sphr}--\ref{tab:summ-abs}; the fitted parameters are in Tables~\ref{tab:fp2-params}
and~\ref{tab:mspline-params}.
The conditional transition HRs \eqref{eq:condHR} involve no study baseline coefficients $\bm\mu^r_j$,
and all three fits retain the same $\gamma^{SP}_k+\gamma^{SPt}_k\log u$ treatment term, so the conditional
stable$\to$progression and progression$\to$death HRs (Tables~\ref{tab:summ-sphr} and~\ref{tab:summ-pdhr})
differ between the fits only through their estimated treatment and effect-modifier coefficients. These
are re-estimated under each baseline, and thalidomide's change materially (Appendix~\ref{app:summtab}).
The conditional OS HRs (Table~\ref{tab:summ-oshr}), like the absolute baseline hazards, state occupancy
and survival, also depend on the baseline.

\FloatBarrier
\subsection{Fractional-polynomial stable--progression baseline}\label{app:fp2}
\begin{table}[htbp]\centering\footnotesize
\caption{Posterior estimates (median, 95\% CrI) under the second-order fractional-polynomial S$\to$P
baseline (study~2 shown). $\gamma^{SP}_k,\gamma^{SPt}_k$: intercept and log-time components of the
S$\to$P log-HR, $\mathrm{HR}^{SP}_k(u)=\exp(\gamma^{SP}_k+\xv^{\!\top}\bet^{SP}_2)u^{\gamma^{SPt}_k}$;
$\gamma^{PD}_k$: P$\to$D log-HR; $\bet^{SP}_2$: S$\to$P effect modifiers; $a_1,a_2$: FP2 baseline coefficients on $\log u$ and
$u$.}
\label{tab:fp2-params}
\begin{tabular}{llc}
\toprule
Parameter & & Estimate (95\% CrI)\\
\midrule
\multirow{2}{*}{S$\to$P log-HR $\gamma^{SP}_k$} & Len & $-1.46\ (-1.87,-1.08)$\\
                                                       & Thal & $-0.77\ (-1.47,-0.09)$\\
\multirow{2}{*}{S$\to$P time-varying $\gamma^{SPt}_k$} & Len & $0.22\ (0.11,0.33)$\\
                                                       & Thal & $0.22\ (-0.08,0.55)$\\
\multirow{2}{*}{P$\to$D log-HR $\gamma^{PD}_k$} & Len & $-0.07\ (-0.26,0.11)$\\
                                                       & Thal & $-0.44\ (-1.30,0.14)$\\
\multirow{3}{*}{S$\to$P effect modifiers $\bet^{SP}_2$} & age & $-0.02\ (-0.21,0.17)$\\
                                                       & ISS-III & $0.23\ (-0.14,0.61)$\\
                                                       & ASCT response & $0.21\ (-0.10,0.51)$\\
\multirow{2}{*}{FP2 baseline shape (study~2)} & $a_1$ ($\log u$) & $0.65\ (0.31,1.03)$\\
                                                       & $a_2$ ($u$) & $-0.020\ (-0.038,-0.003)$\\
\bottomrule
\end{tabular}
\end{table}
Figures~\ref{fig:fp2-hrpops} to~\ref{fig:fp2-surv} repeat Figures~\ref{fig:hrpops} to~\ref{fig:survpops}
for the FP2 fit.
  \IfFileExists{figures/fig_ndmm_hr_pops_fp2.png}{\begin{figure}[htbp]\centering
    \includegraphics[width=.92\linewidth]{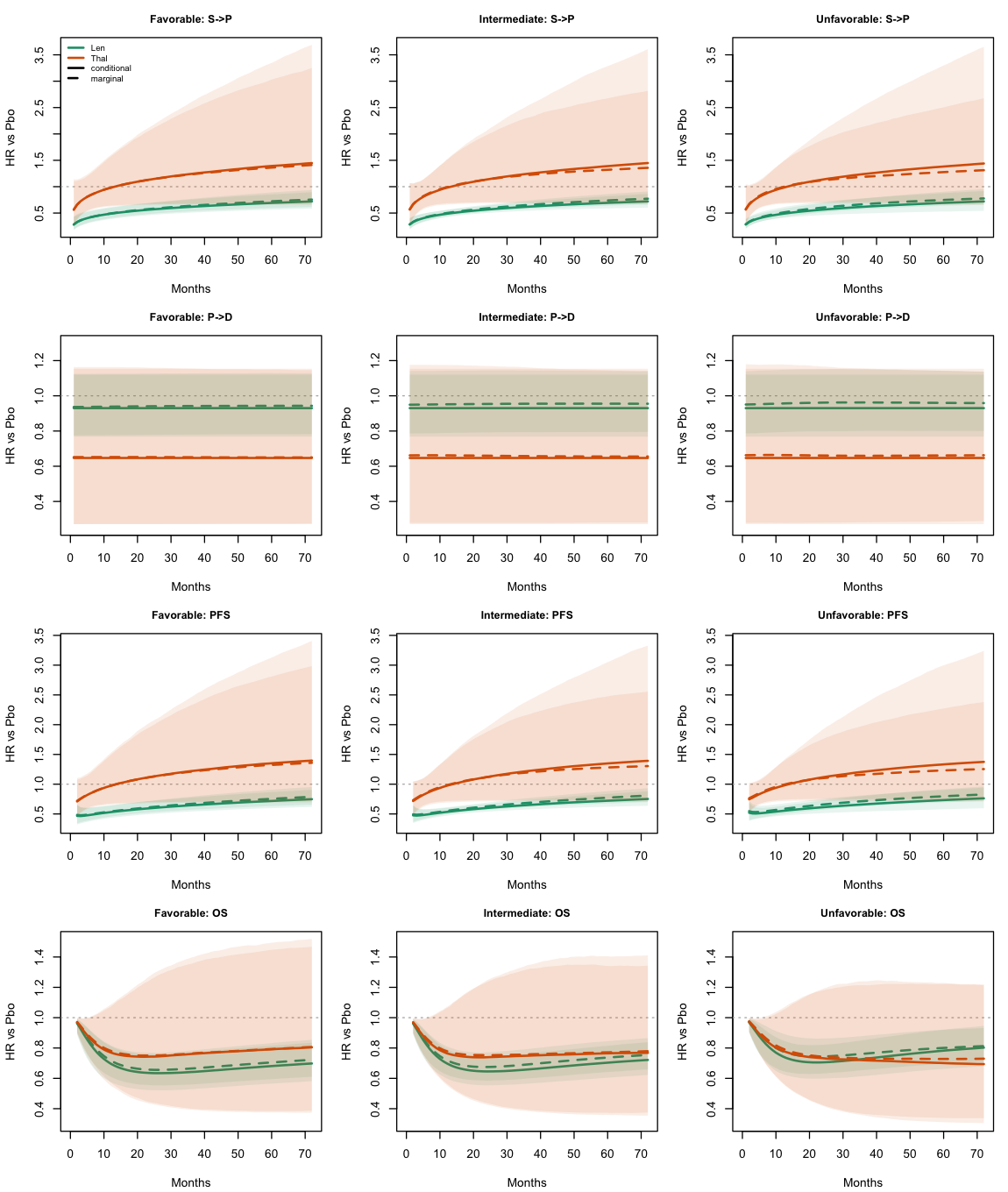}
    \caption{FP2 stable--progression baseline: hazard ratios versus placebo across the target populations,
    conditional (solid, at the population-mean covariates) and marginal (dashed, population-average).}\label{fig:fp2-hrpops}\end{figure}}{}%
  \IfFileExists{figures/fig_ndmm_baseline_pops_fp2.png}{\begin{figure}[htbp]\centering
    \includegraphics[width=\linewidth]{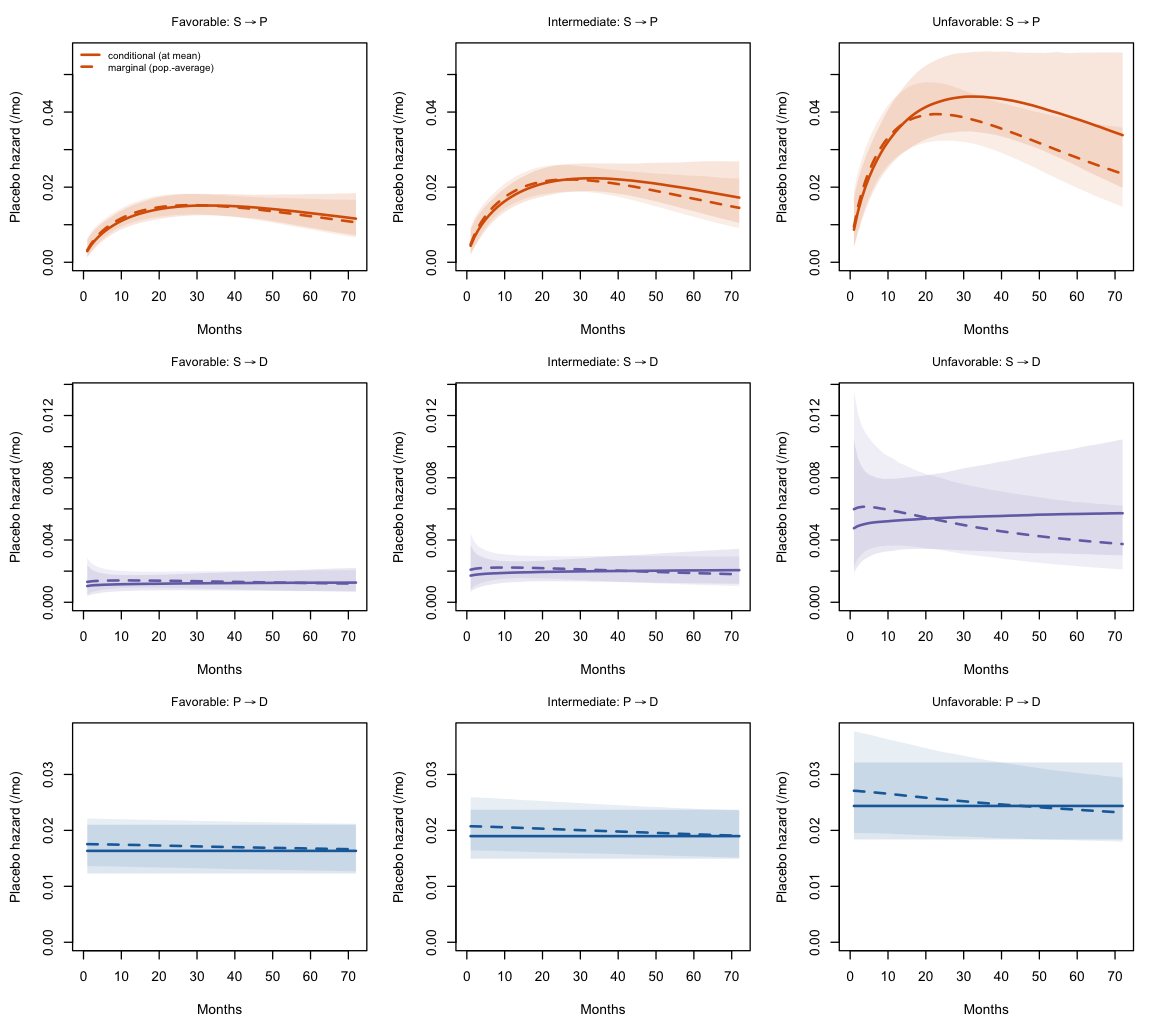}
    \caption{FP2 stable--progression baseline: placebo baseline transition hazards (rows: S$\to$P, S$\to$D,
    P$\to$D) across the three target populations (columns), shown conditionally (solid, at the
    population-mean covariates) and marginally (dashed, population-average).}\label{fig:fp2-basepops}\end{figure}}{}%
  \IfFileExists{figures/fig_ndmm_occupancy_fp2.png}{\begin{figure}[htbp]\centering
    \includegraphics[width=\linewidth]{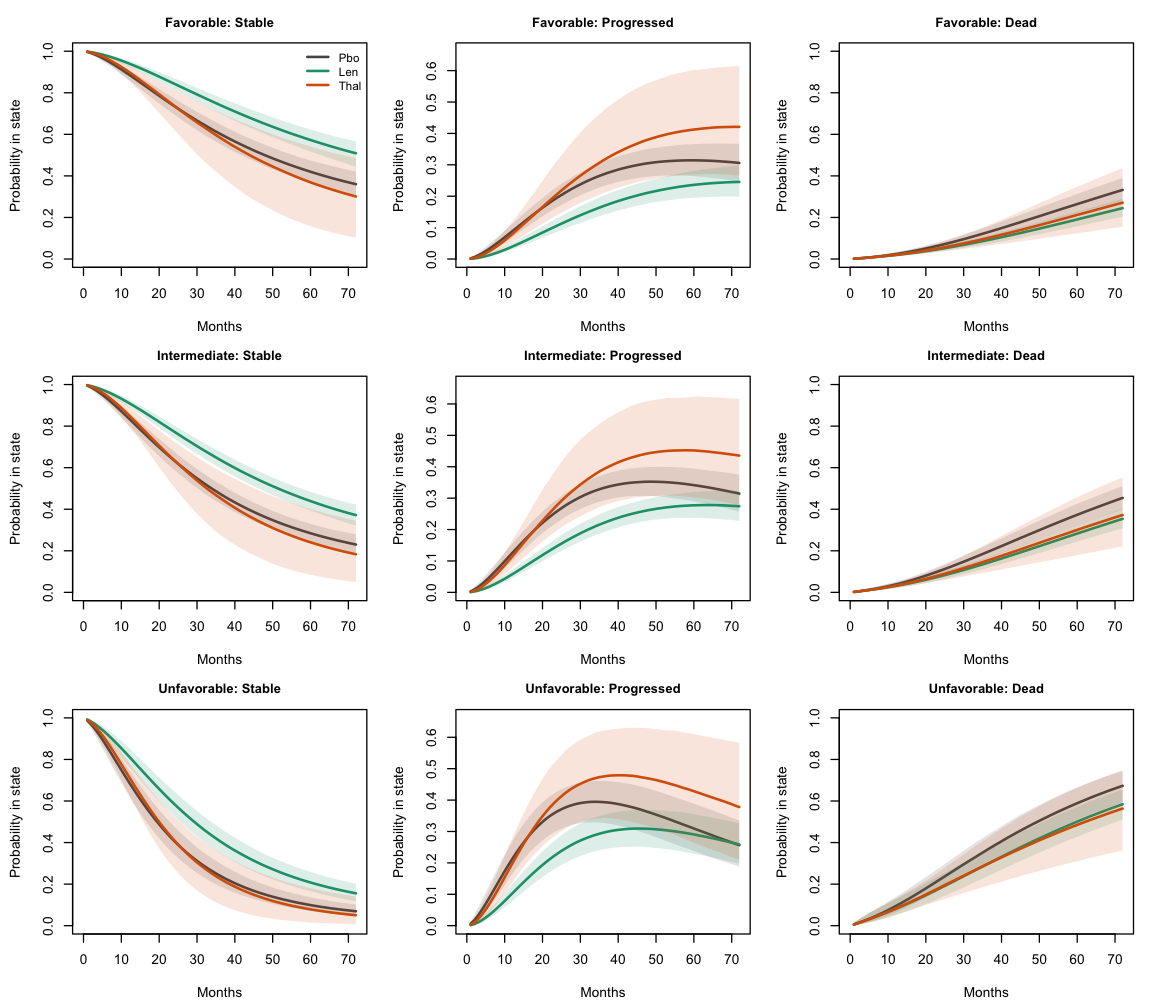}
    \caption{FP2 stable--progression baseline: marginal (population-average) state-occupancy
    probabilities (stable/progressed/dead) by treatment across the target populations.}\label{fig:fp2-occ}\end{figure}}{}%
  \IfFileExists{figures/fig_ndmm_surv_pops_fp2.png}{\begin{figure}[htbp]\centering
    \includegraphics[width=\linewidth]{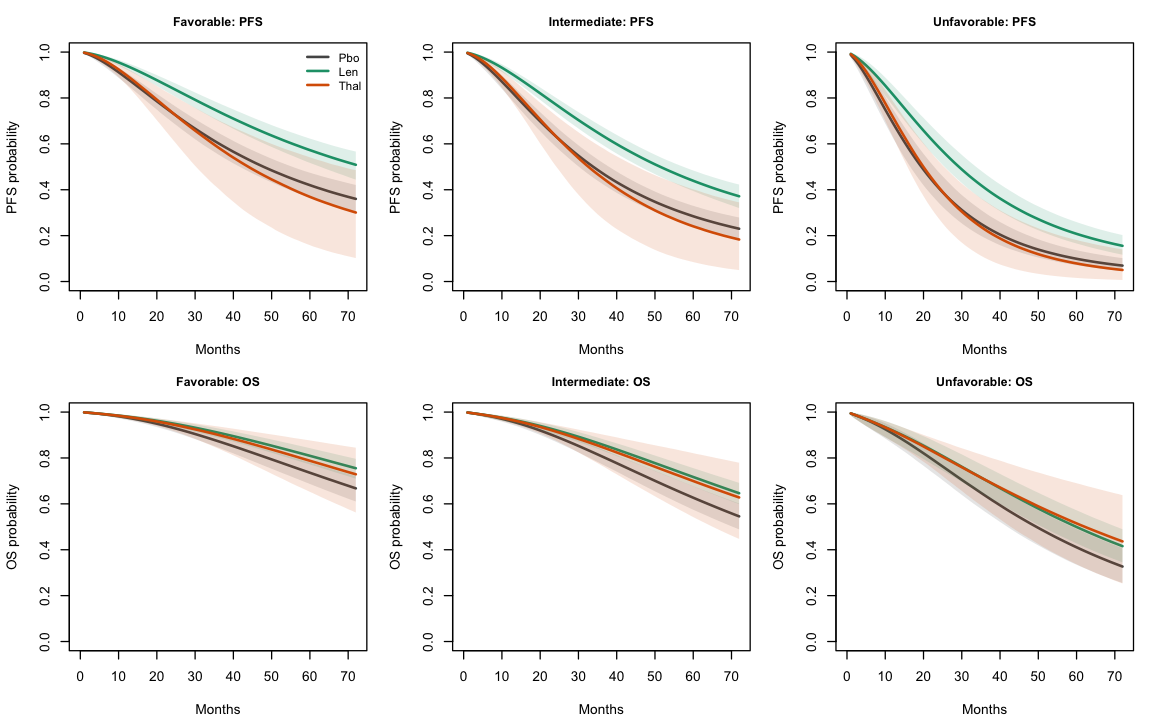}
    \caption{FP2 stable--progression baseline: marginal (population-average) PFS and OS by treatment
    across the target populations.}\label{fig:fp2-surv}\end{figure}}{}%

\FloatBarrier
\subsection{M-spline stable--progression baseline}\label{app:mspline}
\begin{table}[htbp]\centering\footnotesize
\caption{Posterior estimates (median, 95\% CrI) under the degree-3 M-spline S$\to$P baseline.}
\label{tab:mspline-params}
\begin{tabular}{llc}
\toprule
Parameter & & Estimate (95\% CrI)\\
\midrule
\multirow{2}{*}{S$\to$P log-HR $\gamma^{SP}_k$} & Len & $-1.50\ (-1.92,-1.10)$\\
                                                       & Thal & $-0.12\ (-0.66,0.42)$\\
\multirow{2}{*}{S$\to$P time-varying $\gamma^{SPt}_k$} & Len & $0.25\ (0.13,0.36)$\\
                                                       & Thal & $-0.07\ (-0.22,0.09)$\\
\multirow{2}{*}{P$\to$D log-HR $\gamma^{PD}_k$} & Len & $-0.07\ (-0.26,0.11)$\\
                                                       & Thal & $-0.20\ (-0.60,0.19)$\\
\multirow{3}{*}{S$\to$P effect modifiers $\bet^{SP}_2$} & age & $-0.08\ (-0.28,0.12)$\\
                                                       & ISS-III & $0.23\ (-0.15,0.61)$\\
                                                       & ASCT response & $0.15\ (-0.16,0.46)$\\
\multirow{6}{*}{M-spline weights $w_l$} & $w_1$ & $0.02\ (0.01,0.03)$\\
                                                       & $w_2$ & $0.14\ (0.11,0.17)$\\
                                                       & $w_3$ & $0.26\ (0.21,0.32)$\\
                                                       & $w_4$ & $0.28\ (0.19,0.36)$\\
                                                       & $w_5$ & $0.18\ (0.09,0.26)$\\
                                                       & $w_6$ & $0.12\ (0.07,0.19)$\\
\bottomrule
\end{tabular}
\end{table}
Figures~\ref{fig:mspline-hrpops} to~\ref{fig:mspline-surv} repeat Figures~\ref{fig:hrpops} to~\ref{fig:survpops}
for the M-spline fit.
  \IfFileExists{figures/fig_ndmm_hr_pops_mspline.png}{\begin{figure}[htbp]\centering
    \includegraphics[width=.92\linewidth]{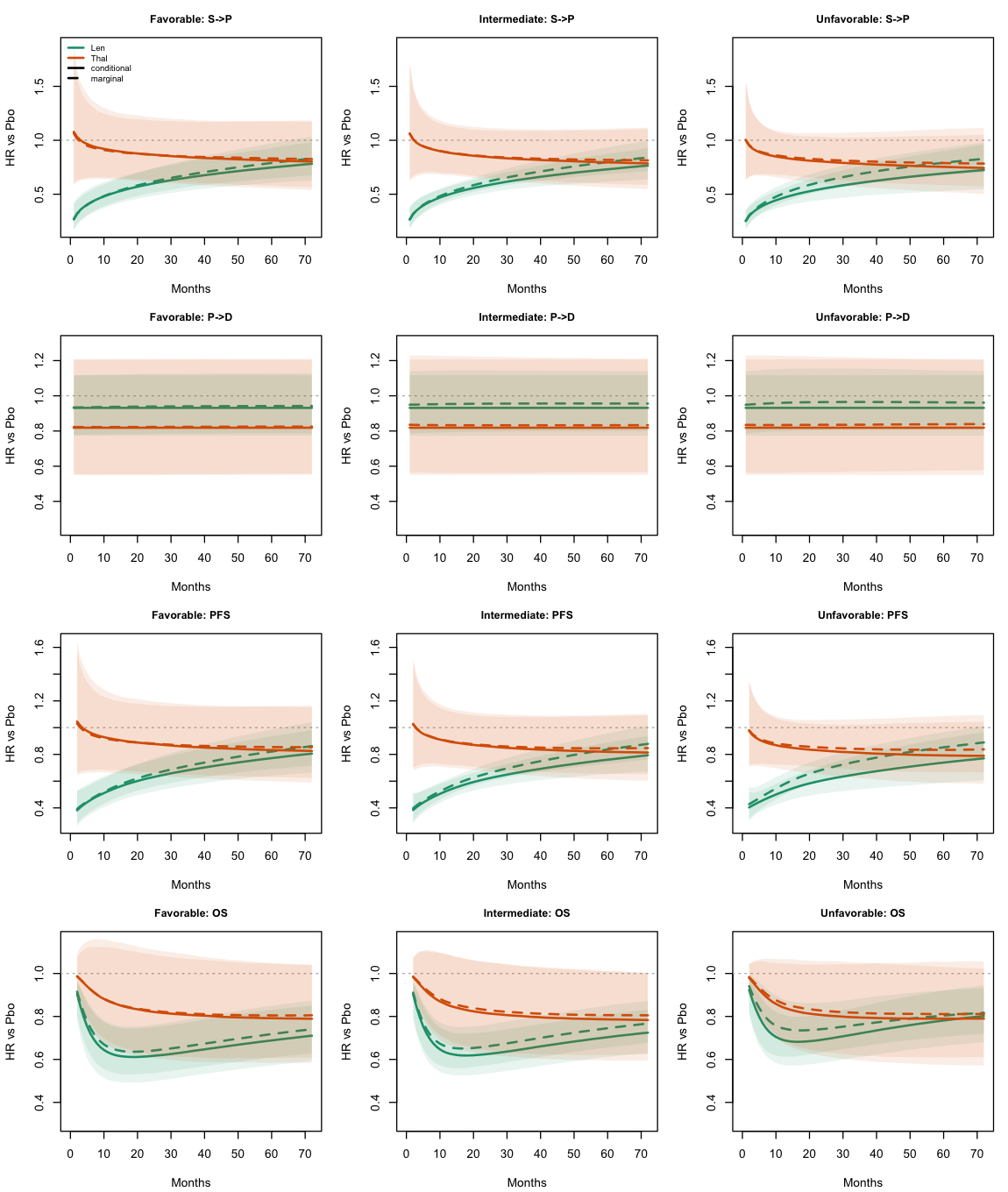}
    \caption{M-spline stable--progression baseline: hazard ratios versus placebo across the target populations,
    conditional (solid, at the population-mean covariates) and marginal (dashed, population-average).}\label{fig:mspline-hrpops}\end{figure}}{}%
  \IfFileExists{figures/fig_ndmm_baseline_pops_mspline.png}{\begin{figure}[htbp]\centering
    \includegraphics[width=\linewidth]{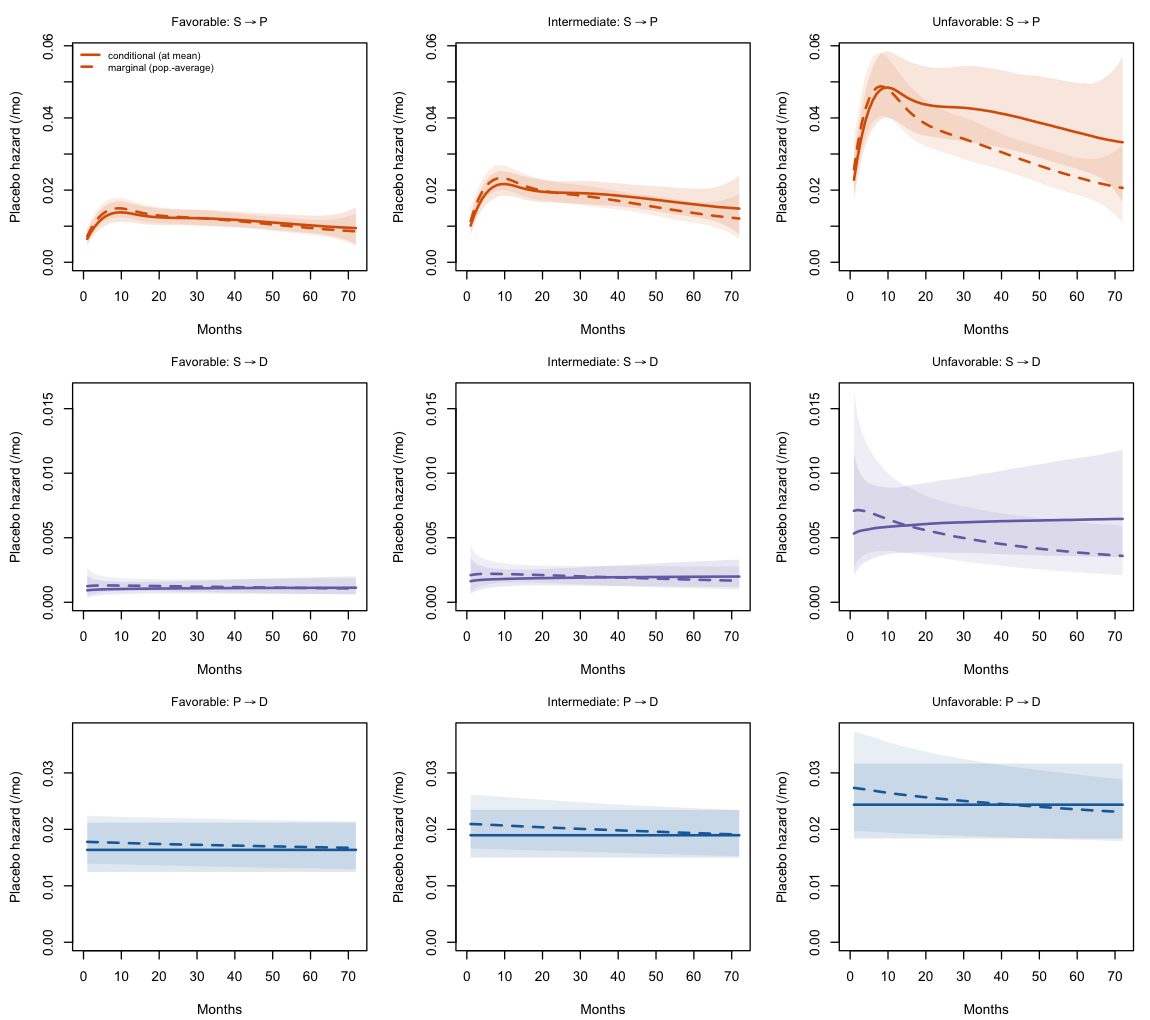}
    \caption{M-spline stable--progression baseline: placebo baseline transition hazards (rows: S$\to$P, S$\to$D,
    P$\to$D) across the three target populations (columns), shown conditionally (solid, at the
    population-mean covariates) and marginally (dashed, population-average).}\label{fig:mspline-basepops}\end{figure}}{}%
  \IfFileExists{figures/fig_ndmm_occupancy_mspline.png}{\begin{figure}[htbp]\centering
    \includegraphics[width=\linewidth]{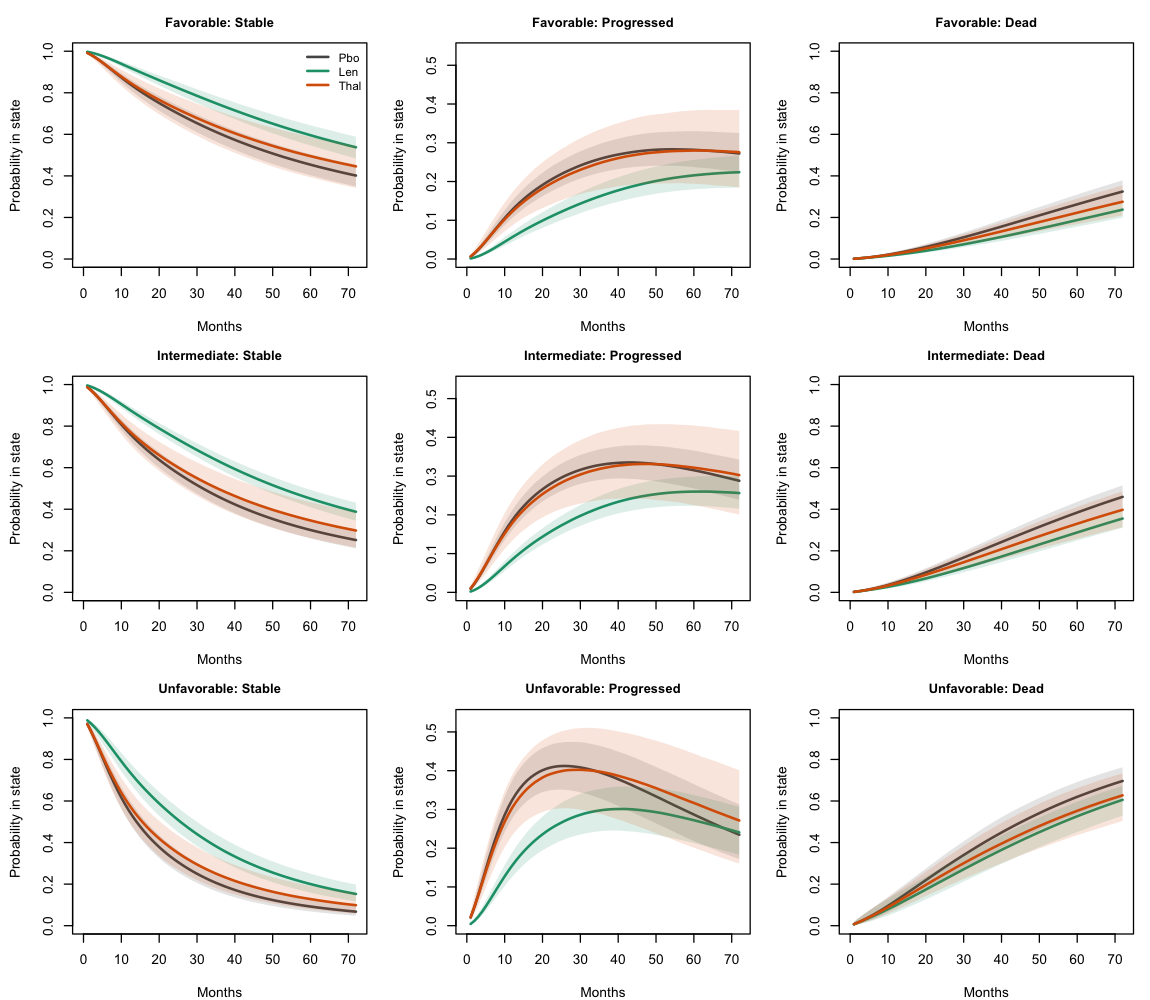}
    \caption{M-spline stable--progression baseline: marginal (population-average) state-occupancy
    probabilities (stable/progressed/dead) by treatment across the target populations.}\label{fig:mspline-occ}\end{figure}}{}%
  \IfFileExists{figures/fig_ndmm_surv_pops_mspline.png}{\begin{figure}[htbp]\centering
    \includegraphics[width=\linewidth]{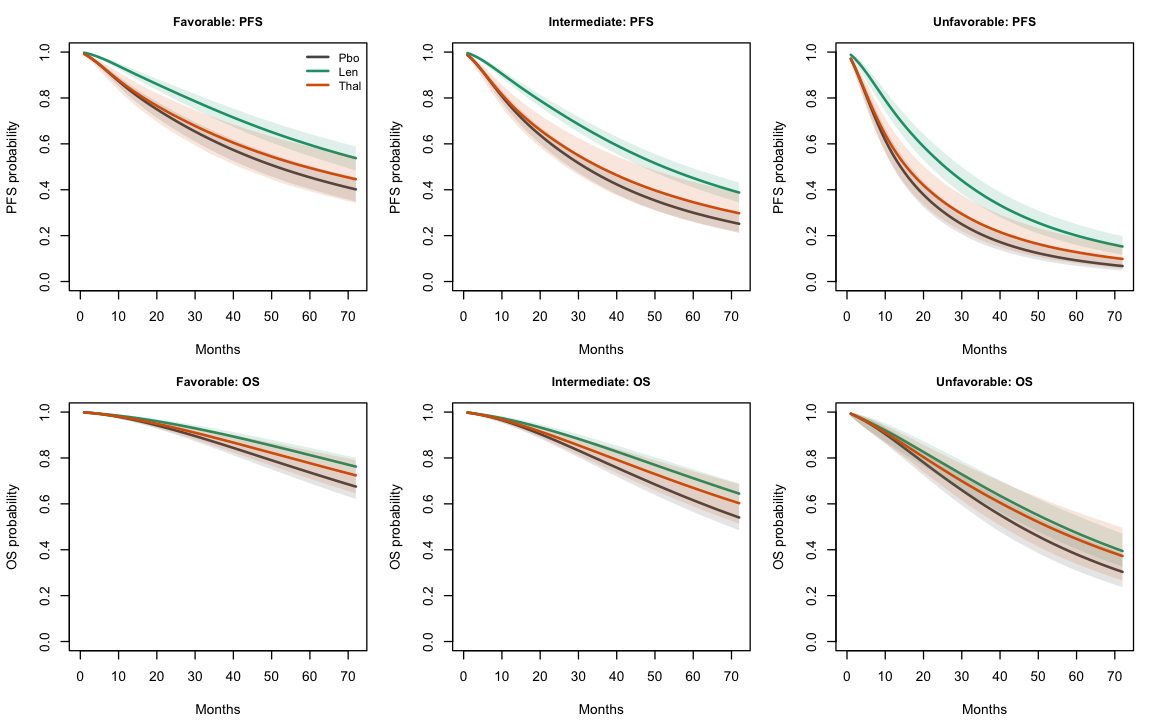}
    \caption{M-spline stable--progression baseline: marginal (population-average) PFS and OS by treatment
    across the target populations.}\label{fig:mspline-surv}\end{figure}}{}%

\FloatBarrier
\subsection{Numerical summary across baseline forms}\label{app:summtab}
Tables~\ref{tab:summ-sphr}--\ref{tab:summ-abs} give the numbers behind the curves (the
overall-survival layer is semi-synthetic, Appendix~\ref{sec:ndmm-data}): the conditional and
marginal HRs versus placebo for the two treatment-affected transitions, stable$\to$progression
(Table~\ref{tab:summ-sphr}) and progression$\to$death (Table~\ref{tab:summ-pdhr}), and for overall
survival (Table~\ref{tab:summ-oshr}); and the marginal (population-average) PFS, OS and probability
progressed (Table~\ref{tab:summ-abs}); all at 12, 24 and 48 months, for the three S$\to$P baseline forms
across the three target populations. The conditional progression$\to$death HR is constant
across populations and over time (no effect modifier, and proportional), so only its marginal
counterpart varies. The lenalidomide conditional stable$\to$progression HRs are close across the baseline forms early in
follow-up and diverge modestly later (intermediate population: $0.48$--$0.49$ at $12$ months, $0.59$--$0.69$
at $48$), while the thalidomide ones, identified from a single aggregate study, vary materially ($0.93$,
$1.32$ and $0.81$ at $48$ months under the Weibull, FP2 and M-spline). The IPD comparison of
Table~\ref{tab:baseline-loo}, which favors the FP2, cannot arbitrate between these, because no IPD study
includes thalidomide. Nor can the aggregate PFS of Morgan2012, the only thalidomide study: it does not
separate the FP2 from the M-spline ($\Delta$ELPD $-2.6$, SE $4.0$), whose thalidomide HRs differ most, and
more than half of its deficit for the Weibull ($-13.6$, SE $6.9$) comes from the two first-month intervals
with unreliable estimates ($-6.0$, SE $4.0$, without them). For thalidomide the OS HRs vary less: they agree at $48$ months ($0.76$--$0.79$),
because the FP2's higher stable$\to$progression HR is offset by its lower progression$\to$death HR
($0.65$, against $0.81$ and $0.82$). The absolute outcomes also differ, most in the unfavorable
population, where the PFS differences for placebo and lenalidomide are largest early in follow-up, whereas
for thalidomide they persist to $48$ months ($0.33$ under the FP2 against $0.41$ under the other two forms
in the intermediate population); the differences in OS grow with time. Beyond the observed follow-up the
forms would part further: the linear FP2 coefficient has a negative posterior median in every study, so
the extrapolated FP2 stable$\to$progression hazard would decay toward zero.
\begin{sidewaystable}\centering\scriptsize\setlength{\tabcolsep}{5pt}
\caption{Stable$\to$progression hazard ratio versus placebo, conditional and marginal, by treatment, target
population, and S$\to$P baseline form, at 12, 24 and 48 months. Posterior median [95\% CrI] from
all 6000 posterior draws.}
\label{tab:summ-sphr}
\begin{tabular}{llr cccc}\toprule
 & & & \multicolumn{2}{c}{Lenalidomide} & \multicolumn{2}{c}{Thalidomide}\\
\cmidrule(lr){4-5}\cmidrule(lr){6-7}
Model & Pop.\ & $u$ & conditional & marginal & conditional & marginal\\\midrule
\multirow{9}{*}{Weibull} & \multirow{3}{*}{Favorable} & 12 & $0.47$ [$0.37$, $0.60$] & $0.48$ [$0.39$, $0.59$] & $0.84$ [$0.59$, $1.19$] & $0.85$ [$0.62$, $1.15$]\\
 &  & 24 & $0.53$ [$0.43$, $0.65$] & $0.55$ [$0.46$, $0.65$] & $0.88$ [$0.62$, $1.24$] & $0.89$ [$0.64$, $1.22$]\\
 &  & 48 & $0.59$ [$0.48$, $0.72$] & $0.62$ [$0.52$, $0.75$] & $0.92$ [$0.62$, $1.34$] & $0.93$ [$0.65$, $1.32$]\\
\cmidrule(l){2-7}
 & \multirow{3}{*}{Intermediate} & 12 & $0.48$ [$0.41$, $0.55$] & $0.49$ [$0.44$, $0.56$] & $0.85$ [$0.64$, $1.11$] & $0.86$ [$0.67$, $1.10$]\\
 &  & 24 & $0.53$ [$0.47$, $0.60$] & $0.56$ [$0.50$, $0.63$] & $0.89$ [$0.66$, $1.18$] & $0.90$ [$0.68$, $1.17$]\\
 &  & 48 & $0.59$ [$0.51$, $0.68$] & $0.65$ [$0.57$, $0.73$] & $0.93$ [$0.65$, $1.30$] & $0.93$ [$0.69$, $1.25$]\\
\cmidrule(l){2-7}
 & \multirow{3}{*}{Unfavorable} & 12 & $0.48$ [$0.40$, $0.57$] & $0.51$ [$0.43$, $0.61$] & $0.85$ [$0.66$, $1.10$] & $0.87$ [$0.69$, $1.10$]\\
 &  & 24 & $0.53$ [$0.44$, $0.65$] & $0.59$ [$0.50$, $0.68$] & $0.89$ [$0.66$, $1.21$] & $0.90$ [$0.70$, $1.15$]\\
 &  & 48 & $0.60$ [$0.47$, $0.76$] & $0.67$ [$0.58$, $0.77$] & $0.93$ [$0.64$, $1.37$] & $0.92$ [$0.69$, $1.21$]\\
\midrule
\multirow{9}{*}{FP2} & \multirow{3}{*}{Favorable} & 12 & $0.49$ [$0.39$, $0.61$] & $0.49$ [$0.40$, $0.60$] & $0.98$ [$0.63$, $1.60$] & $0.98$ [$0.64$, $1.57$]\\
 &  & 24 & $0.57$ [$0.46$, $0.69$] & $0.58$ [$0.49$, $0.69$] & $1.14$ [$0.66$, $2.14$] & $1.14$ [$0.67$, $2.08$]\\
 &  & 48 & $0.66$ [$0.54$, $0.81$] & $0.68$ [$0.57$, $0.83$] & $1.32$ [$0.66$, $2.99$] & $1.31$ [$0.68$, $2.78$]\\
\cmidrule(l){2-7}
 & \multirow{3}{*}{Intermediate} & 12 & $0.49$ [$0.42$, $0.56$] & $0.50$ [$0.44$, $0.56$] & $0.98$ [$0.66$, $1.51$] & $0.99$ [$0.68$, $1.50$]\\
 &  & 24 & $0.57$ [$0.50$, $0.64$] & $0.59$ [$0.53$, $0.66$] & $1.14$ [$0.69$, $2.06$] & $1.14$ [$0.71$, $1.98$]\\
 &  & 48 & $0.66$ [$0.57$, $0.77$] & $0.70$ [$0.61$, $0.80$] & $1.32$ [$0.68$, $2.93$] & $1.28$ [$0.70$, $2.51$]\\
\cmidrule(l){2-7}
 & \multirow{3}{*}{Unfavorable} & 12 & $0.49$ [$0.41$, $0.58$] & $0.51$ [$0.42$, $0.61$] & $0.98$ [$0.67$, $1.47$] & $0.99$ [$0.69$, $1.47$]\\
 &  & 24 & $0.57$ [$0.46$, $0.69$] & $0.60$ [$0.51$, $0.71$] & $1.14$ [$0.68$, $2.07$] & $1.12$ [$0.71$, $1.89$]\\
 &  & 48 & $0.66$ [$0.51$, $0.84$] & $0.71$ [$0.61$, $0.83$] & $1.32$ [$0.67$, $2.93$] & $1.23$ [$0.70$, $2.34$]\\
\midrule
\multirow{9}{*}{M-spline} & \multirow{3}{*}{Favorable} & 12 & $0.50$ [$0.39$, $0.63$] & $0.51$ [$0.41$, $0.62$] & $0.91$ [$0.64$, $1.28$] & $0.90$ [$0.65$, $1.23$]\\
 &  & 24 & $0.60$ [$0.48$, $0.73$] & $0.61$ [$0.51$, $0.73$] & $0.87$ [$0.61$, $1.22$] & $0.87$ [$0.63$, $1.18$]\\
 &  & 48 & $0.71$ [$0.58$, $0.87$] & $0.74$ [$0.61$, $0.90$] & $0.83$ [$0.57$, $1.18$] & $0.84$ [$0.59$, $1.17$]\\
\cmidrule(l){2-7}
 & \multirow{3}{*}{Intermediate} & 12 & $0.49$ [$0.42$, $0.56$] & $0.51$ [$0.45$, $0.57$] & $0.89$ [$0.68$, $1.16$] & $0.89$ [$0.70$, $1.14$]\\
 &  & 24 & $0.58$ [$0.51$, $0.66$] & $0.61$ [$0.55$, $0.69$] & $0.85$ [$0.64$, $1.11$] & $0.85$ [$0.66$, $1.09$]\\
 &  & 48 & $0.69$ [$0.59$, $0.81$] & $0.75$ [$0.65$, $0.86$] & $0.81$ [$0.58$, $1.10$] & $0.82$ [$0.61$, $1.09$]\\
\cmidrule(l){2-7}
 & \multirow{3}{*}{Unfavorable} & 12 & $0.46$ [$0.39$, $0.55$] & $0.50$ [$0.42$, $0.59$] & $0.84$ [$0.65$, $1.08$] & $0.85$ [$0.68$, $1.06$]\\
 &  & 24 & $0.55$ [$0.45$, $0.67$] & $0.62$ [$0.53$, $0.72$] & $0.80$ [$0.60$, $1.07$] & $0.82$ [$0.64$, $1.03$]\\
 &  & 48 & $0.65$ [$0.51$, $0.84$] & $0.75$ [$0.64$, $0.87$] & $0.77$ [$0.54$, $1.09$] & $0.80$ [$0.60$, $1.03$]\\
\bottomrule\end{tabular}\end{sidewaystable}
\begin{sidewaystable}\centering\scriptsize\setlength{\tabcolsep}{5pt}
\caption{Progression$\to$death hazard ratio versus placebo, conditional and marginal, by treatment, target
population, and S$\to$P baseline form, at 12, 24 and 48 months. Posterior median [95\% CrI] from
all 6000 posterior draws.}
\label{tab:summ-pdhr}
\begin{tabular}{llr cccc}\toprule
 & & & \multicolumn{2}{c}{Lenalidomide} & \multicolumn{2}{c}{Thalidomide}\\
\cmidrule(lr){4-5}\cmidrule(lr){6-7}
Model & Pop.\ & $u$ & conditional & marginal & conditional & marginal\\\midrule
\multirow{9}{*}{Weibull} & \multirow{3}{*}{Favorable} & 12 & $0.93$ [$0.77$, $1.11$] & $0.94$ [$0.78$, $1.12$] & $0.81$ [$0.54$, $1.20$] & $0.81$ [$0.54$, $1.21$]\\
 &  & 24 & $0.93$ [$0.77$, $1.11$] & $0.94$ [$0.78$, $1.12$] & $0.81$ [$0.54$, $1.20$] & $0.81$ [$0.54$, $1.21$]\\
 &  & 48 & $0.93$ [$0.77$, $1.11$] & $0.94$ [$0.78$, $1.12$] & $0.81$ [$0.54$, $1.20$] & $0.81$ [$0.54$, $1.20$]\\
\cmidrule(l){2-7}
 & \multirow{3}{*}{Intermediate} & 12 & $0.93$ [$0.77$, $1.11$] & $0.95$ [$0.79$, $1.14$] & $0.81$ [$0.54$, $1.20$] & $0.82$ [$0.55$, $1.23$]\\
 &  & 24 & $0.93$ [$0.77$, $1.11$] & $0.95$ [$0.79$, $1.14$] & $0.81$ [$0.54$, $1.20$] & $0.82$ [$0.55$, $1.22$]\\
 &  & 48 & $0.93$ [$0.77$, $1.11$] & $0.96$ [$0.80$, $1.14$] & $0.81$ [$0.54$, $1.20$] & $0.82$ [$0.55$, $1.21$]\\
\cmidrule(l){2-7}
 & \multirow{3}{*}{Unfavorable} & 12 & $0.93$ [$0.77$, $1.11$] & $0.96$ [$0.80$, $1.14$] & $0.81$ [$0.54$, $1.20$] & $0.83$ [$0.55$, $1.22$]\\
 &  & 24 & $0.93$ [$0.77$, $1.11$] & $0.96$ [$0.80$, $1.15$] & $0.81$ [$0.54$, $1.20$] & $0.82$ [$0.55$, $1.22$]\\
 &  & 48 & $0.93$ [$0.77$, $1.11$] & $0.96$ [$0.80$, $1.14$] & $0.81$ [$0.54$, $1.20$] & $0.82$ [$0.55$, $1.20$]\\
\midrule
\multirow{9}{*}{FP2} & \multirow{3}{*}{Favorable} & 12 & $0.93$ [$0.77$, $1.12$] & $0.94$ [$0.78$, $1.13$] & $0.65$ [$0.27$, $1.15$] & $0.65$ [$0.27$, $1.16$]\\
 &  & 24 & $0.93$ [$0.77$, $1.12$] & $0.94$ [$0.78$, $1.13$] & $0.65$ [$0.27$, $1.15$] & $0.65$ [$0.27$, $1.16$]\\
 &  & 48 & $0.93$ [$0.77$, $1.12$] & $0.94$ [$0.78$, $1.13$] & $0.65$ [$0.27$, $1.15$] & $0.65$ [$0.27$, $1.15$]\\
\cmidrule(l){2-7}
 & \multirow{3}{*}{Intermediate} & 12 & $0.93$ [$0.77$, $1.12$] & $0.95$ [$0.79$, $1.14$] & $0.65$ [$0.27$, $1.15$] & $0.66$ [$0.28$, $1.18$]\\
 &  & 24 & $0.93$ [$0.77$, $1.12$] & $0.95$ [$0.79$, $1.14$] & $0.65$ [$0.27$, $1.15$] & $0.66$ [$0.28$, $1.17$]\\
 &  & 48 & $0.93$ [$0.77$, $1.12$] & $0.96$ [$0.79$, $1.14$] & $0.65$ [$0.27$, $1.15$] & $0.66$ [$0.28$, $1.15$]\\
\cmidrule(l){2-7}
 & \multirow{3}{*}{Unfavorable} & 12 & $0.93$ [$0.77$, $1.12$] & $0.96$ [$0.79$, $1.15$] & $0.65$ [$0.27$, $1.15$] & $0.66$ [$0.28$, $1.18$]\\
 &  & 24 & $0.93$ [$0.77$, $1.12$] & $0.96$ [$0.80$, $1.15$] & $0.65$ [$0.27$, $1.15$] & $0.66$ [$0.28$, $1.17$]\\
 &  & 48 & $0.93$ [$0.77$, $1.12$] & $0.96$ [$0.80$, $1.15$] & $0.65$ [$0.27$, $1.15$] & $0.66$ [$0.28$, $1.15$]\\
\midrule
\multirow{9}{*}{M-spline} & \multirow{3}{*}{Favorable} & 12 & $0.93$ [$0.77$, $1.12$] & $0.94$ [$0.78$, $1.12$] & $0.82$ [$0.55$, $1.21$] & $0.82$ [$0.55$, $1.21$]\\
 &  & 24 & $0.93$ [$0.77$, $1.12$] & $0.94$ [$0.78$, $1.12$] & $0.82$ [$0.55$, $1.21$] & $0.82$ [$0.55$, $1.21$]\\
 &  & 48 & $0.93$ [$0.77$, $1.12$] & $0.94$ [$0.79$, $1.13$] & $0.82$ [$0.55$, $1.21$] & $0.82$ [$0.56$, $1.21$]\\
\cmidrule(l){2-7}
 & \multirow{3}{*}{Intermediate} & 12 & $0.93$ [$0.77$, $1.12$] & $0.95$ [$0.79$, $1.14$] & $0.82$ [$0.55$, $1.21$] & $0.83$ [$0.56$, $1.23$]\\
 &  & 24 & $0.93$ [$0.77$, $1.12$] & $0.96$ [$0.80$, $1.14$] & $0.82$ [$0.55$, $1.21$] & $0.83$ [$0.56$, $1.22$]\\
 &  & 48 & $0.93$ [$0.77$, $1.12$] & $0.96$ [$0.80$, $1.14$] & $0.82$ [$0.55$, $1.21$] & $0.83$ [$0.56$, $1.21$]\\
\cmidrule(l){2-7}
 & \multirow{3}{*}{Unfavorable} & 12 & $0.93$ [$0.77$, $1.12$] & $0.96$ [$0.80$, $1.15$] & $0.82$ [$0.55$, $1.21$] & $0.83$ [$0.56$, $1.23$]\\
 &  & 24 & $0.93$ [$0.77$, $1.12$] & $0.96$ [$0.80$, $1.15$] & $0.82$ [$0.55$, $1.21$] & $0.84$ [$0.57$, $1.22$]\\
 &  & 48 & $0.93$ [$0.77$, $1.12$] & $0.96$ [$0.81$, $1.15$] & $0.82$ [$0.55$, $1.21$] & $0.84$ [$0.57$, $1.21$]\\
\bottomrule\end{tabular}\end{sidewaystable}
\begin{sidewaystable}\centering\scriptsize\setlength{\tabcolsep}{5pt}
\caption{Overall-survival hazard ratio versus placebo, conditional and marginal, by treatment, target
population, and S$\to$P baseline form, at 12, 24 and 48 months. Posterior median [95\% CrI] from
all 6000 posterior draws.}
\label{tab:summ-oshr}
\begin{tabular}{llr cccc}\toprule
 & & & \multicolumn{2}{c}{Lenalidomide} & \multicolumn{2}{c}{Thalidomide}\\
\cmidrule(lr){4-5}\cmidrule(lr){6-7}
Model & Pop.\ & $u$ & conditional & marginal & conditional & marginal\\\midrule
\multirow{9}{*}{Weibull} & \multirow{3}{*}{Favorable} & 12 & $0.68$ [$0.55$, $0.81$] & $0.71$ [$0.60$, $0.82$] & $0.82$ [$0.65$, $1.03$] & $0.83$ [$0.68$, $1.02$]\\
 &  & 24 & $0.62$ [$0.50$, $0.75$] & $0.65$ [$0.54$, $0.76$] & $0.78$ [$0.59$, $1.04$] & $0.79$ [$0.62$, $1.03$]\\
 &  & 48 & $0.62$ [$0.51$, $0.75$] & $0.66$ [$0.55$, $0.77$] & $0.79$ [$0.57$, $1.07$] & $0.80$ [$0.59$, $1.07$]\\
\cmidrule(l){2-7}
 & \multirow{3}{*}{Intermediate} & 12 & $0.69$ [$0.59$, $0.79$] & $0.72$ [$0.62$, $0.81$] & $0.82$ [$0.68$, $1.00$] & $0.84$ [$0.70$, $1.01$]\\
 &  & 24 & $0.63$ [$0.54$, $0.73$] & $0.67$ [$0.58$, $0.76$] & $0.79$ [$0.61$, $1.01$] & $0.81$ [$0.64$, $1.03$]\\
 &  & 48 & $0.65$ [$0.56$, $0.75$] & $0.69$ [$0.60$, $0.80$] & $0.79$ [$0.59$, $1.05$] & $0.81$ [$0.61$, $1.06$]\\
\cmidrule(l){2-7}
 & \multirow{3}{*}{Unfavorable} & 12 & $0.75$ [$0.63$, $0.86$] & $0.79$ [$0.67$, $0.92$] & $0.84$ [$0.70$, $1.00$] & $0.86$ [$0.71$, $1.03$]\\
 &  & 24 & $0.71$ [$0.60$, $0.83$] & $0.76$ [$0.65$, $0.88$] & $0.81$ [$0.63$, $1.03$] & $0.83$ [$0.64$, $1.05$]\\
 &  & 48 & $0.75$ [$0.64$, $0.89$] & $0.78$ [$0.68$, $0.90$] & $0.80$ [$0.57$, $1.09$] & $0.82$ [$0.60$, $1.07$]\\
\midrule
\multirow{9}{*}{FP2} & \multirow{3}{*}{Favorable} & 12 & $0.70$ [$0.58$, $0.81$] & $0.72$ [$0.61$, $0.82$] & $0.77$ [$0.55$, $1.09$] & $0.78$ [$0.57$, $1.07$]\\
 &  & 24 & $0.63$ [$0.52$, $0.76$] & $0.66$ [$0.56$, $0.77$] & $0.74$ [$0.43$, $1.28$] & $0.75$ [$0.44$, $1.25$]\\
 &  & 48 & $0.66$ [$0.55$, $0.79$] & $0.69$ [$0.58$, $0.81$] & $0.78$ [$0.38$, $1.46$] & $0.78$ [$0.39$, $1.41$]\\
\cmidrule(l){2-7}
 & \multirow{3}{*}{Intermediate} & 12 & $0.70$ [$0.60$, $0.79$] & $0.72$ [$0.62$, $0.81$] & $0.77$ [$0.56$, $1.06$] & $0.78$ [$0.57$, $1.07$]\\
 &  & 24 & $0.65$ [$0.55$, $0.74$] & $0.68$ [$0.59$, $0.77$] & $0.74$ [$0.43$, $1.25$] & $0.75$ [$0.44$, $1.23$]\\
 &  & 48 & $0.68$ [$0.58$, $0.79$] & $0.72$ [$0.62$, $0.82$] & $0.76$ [$0.37$, $1.39$] & $0.76$ [$0.38$, $1.35$]\\
\cmidrule(l){2-7}
 & \multirow{3}{*}{Unfavorable} & 12 & $0.74$ [$0.63$, $0.85$] & $0.78$ [$0.65$, $0.90$] & $0.78$ [$0.57$, $1.05$] & $0.80$ [$0.57$, $1.06$]\\
 &  & 24 & $0.71$ [$0.60$, $0.82$] & $0.74$ [$0.63$, $0.86$] & $0.73$ [$0.42$, $1.19$] & $0.74$ [$0.42$, $1.18$]\\
 &  & 48 & $0.76$ [$0.64$, $0.89$] & $0.78$ [$0.68$, $0.90$] & $0.71$ [$0.32$, $1.24$] & $0.73$ [$0.35$, $1.22$]\\
\midrule
\multirow{9}{*}{M-spline} & \multirow{3}{*}{Favorable} & 12 & $0.63$ [$0.50$, $0.76$] & $0.65$ [$0.54$, $0.77$] & $0.87$ [$0.66$, $1.15$] & $0.87$ [$0.68$, $1.12$]\\
 &  & 24 & $0.62$ [$0.50$, $0.75$] & $0.64$ [$0.53$, $0.76$] & $0.82$ [$0.61$, $1.11$] & $0.83$ [$0.63$, $1.09$]\\
 &  & 48 & $0.66$ [$0.55$, $0.81$] & $0.69$ [$0.58$, $0.82$] & $0.80$ [$0.59$, $1.06$] & $0.81$ [$0.61$, $1.05$]\\
\cmidrule(l){2-7}
 & \multirow{3}{*}{Intermediate} & 12 & $0.63$ [$0.53$, $0.73$] & $0.66$ [$0.56$, $0.76$] & $0.86$ [$0.67$, $1.09$] & $0.87$ [$0.69$, $1.09$]\\
 &  & 24 & $0.62$ [$0.53$, $0.73$] & $0.66$ [$0.57$, $0.76$] & $0.81$ [$0.62$, $1.05$] & $0.83$ [$0.64$, $1.05$]\\
 &  & 48 & $0.68$ [$0.59$, $0.78$] & $0.72$ [$0.63$, $0.82$] & $0.79$ [$0.60$, $1.02$] & $0.81$ [$0.62$, $1.02$]\\
\cmidrule(l){2-7}
 & \multirow{3}{*}{Unfavorable} & 12 & $0.69$ [$0.57$, $0.81$] & $0.74$ [$0.61$, $0.88$] & $0.84$ [$0.68$, $1.05$] & $0.86$ [$0.69$, $1.07$]\\
 &  & 24 & $0.69$ [$0.58$, $0.81$] & $0.74$ [$0.63$, $0.86$] & $0.81$ [$0.62$, $1.04$] & $0.83$ [$0.64$, $1.06$]\\
 &  & 48 & $0.75$ [$0.64$, $0.89$] & $0.79$ [$0.69$, $0.91$] & $0.79$ [$0.58$, $1.05$] & $0.81$ [$0.61$, $1.04$]\\
\bottomrule\end{tabular}\end{sidewaystable}
\begin{sidewaystable}\centering\scriptsize\setlength{\tabcolsep}{2.5pt}
\caption{Marginal (population-average) progression-free survival (PFS), overall survival (OS), and
probability of being alive and progressed (Prog., OS $-$ PFS) by treatment, target population, and S$\to$P
baseline form, at 12, 24 and 48 months. Posterior median [95\% CrI] from all 6000 posterior draws.}
\label{tab:summ-abs}
\begin{tabular}{llr *{9}{c}}\toprule
 & & & \multicolumn{3}{c}{Placebo} & \multicolumn{3}{c}{Lenalidomide} & \multicolumn{3}{c}{Thalidomide}\\
\cmidrule(lr){4-6}\cmidrule(lr){7-9}\cmidrule(lr){10-12}
Model & Pop.\ & $u$ & PFS & OS & Prog. & PFS & OS & Prog. & PFS & OS & Prog.\\\midrule
\multirow{9}{*}{Weibull} & \multirow{3}{*}{Favorable} & 12 & $0.89$ [$0.86$,$0.91$] & $0.98$ [$0.97$,$0.98$] & $0.09$ [$0.07$,$0.11$] & $0.94$ [$0.93$,$0.96$] & $0.98$ [$0.97$,$0.99$] & $0.04$ [$0.03$,$0.05$] & $0.91$ [$0.87$,$0.93$] & $0.98$ [$0.97$,$0.99$] & $0.07$ [$0.05$,$0.10$]\\
 &  & 24 & $0.76$ [$0.72$,$0.79$] & $0.94$ [$0.92$,$0.95$] & $0.18$ [$0.15$,$0.21$] & $0.86$ [$0.84$,$0.88$] & $0.95$ [$0.94$,$0.96$] & $0.09$ [$0.07$,$0.11$] & $0.79$ [$0.73$,$0.84$] & $0.95$ [$0.93$,$0.96$] & $0.16$ [$0.11$,$0.21$]\\
 &  & 48 & $0.53$ [$0.48$,$0.57$] & $0.82$ [$0.78$,$0.85$] & $0.29$ [$0.25$,$0.33$] & $0.68$ [$0.64$,$0.72$] & $0.87$ [$0.85$,$0.90$] & $0.19$ [$0.16$,$0.22$] & $0.56$ [$0.46$,$0.65$] & $0.85$ [$0.80$,$0.89$] & $0.29$ [$0.20$,$0.38$]\\
\cmidrule(l){2-12}
 & \multirow{3}{*}{Intermediate} & 12 & $0.83$ [$0.80$,$0.86$] & $0.96$ [$0.95$,$0.97$] & $0.13$ [$0.10$,$0.16$] & $0.91$ [$0.89$,$0.93$] & $0.97$ [$0.95$,$0.98$] & $0.06$ [$0.05$,$0.08$] & $0.85$ [$0.81$,$0.89$] & $0.96$ [$0.95$,$0.98$] & $0.11$ [$0.08$,$0.15$]\\
 &  & 24 & $0.65$ [$0.60$,$0.69$] & $0.89$ [$0.87$,$0.91$] & $0.25$ [$0.21$,$0.28$] & $0.78$ [$0.76$,$0.81$] & $0.92$ [$0.90$,$0.94$] & $0.14$ [$0.12$,$0.16$] & $0.68$ [$0.61$,$0.74$] & $0.91$ [$0.88$,$0.93$] & $0.23$ [$0.17$,$0.29$]\\
 &  & 48 & $0.37$ [$0.33$,$0.41$] & $0.72$ [$0.68$,$0.76$] & $0.35$ [$0.31$,$0.39$] & $0.55$ [$0.51$,$0.58$] & $0.80$ [$0.76$,$0.83$] & $0.25$ [$0.22$,$0.28$] & $0.41$ [$0.31$,$0.50$] & $0.76$ [$0.70$,$0.82$] & $0.35$ [$0.26$,$0.46$]\\
\cmidrule(l){2-12}
 & \multirow{3}{*}{Unfavorable} & 12 & $0.65$ [$0.58$,$0.71$] & $0.89$ [$0.84$,$0.92$] & $0.24$ [$0.19$,$0.30$] & $0.78$ [$0.73$,$0.83$] & $0.90$ [$0.85$,$0.94$] & $0.12$ [$0.09$,$0.15$] & $0.69$ [$0.61$,$0.76$] & $0.90$ [$0.85$,$0.93$] & $0.21$ [$0.15$,$0.29$]\\
 &  & 24 & $0.38$ [$0.32$,$0.45$] & $0.75$ [$0.69$,$0.80$] & $0.37$ [$0.31$,$0.43$] & $0.56$ [$0.50$,$0.62$] & $0.80$ [$0.73$,$0.85$] & $0.23$ [$0.19$,$0.28$] & $0.43$ [$0.34$,$0.51$] & $0.78$ [$0.71$,$0.84$] & $0.35$ [$0.26$,$0.45$]\\
 &  & 48 & $0.13$ [$0.10$,$0.17$] & $0.49$ [$0.42$,$0.57$] & $0.36$ [$0.29$,$0.43$] & $0.27$ [$0.23$,$0.32$] & $0.58$ [$0.50$,$0.64$] & $0.30$ [$0.25$,$0.37$] & $0.16$ [$0.10$,$0.23$] & $0.55$ [$0.45$,$0.66$] & $0.39$ [$0.27$,$0.51$]\\
\midrule
\multirow{9}{*}{FP2} & \multirow{3}{*}{Favorable} & 12 & $0.89$ [$0.86$,$0.91$] & $0.98$ [$0.97$,$0.98$] & $0.09$ [$0.07$,$0.11$] & $0.94$ [$0.92$,$0.95$] & $0.98$ [$0.97$,$0.99$] & $0.04$ [$0.03$,$0.05$] & $0.90$ [$0.86$,$0.93$] & $0.98$ [$0.97$,$0.99$] & $0.08$ [$0.05$,$0.12$]\\
 &  & 24 & $0.74$ [$0.70$,$0.77$] & $0.93$ [$0.92$,$0.95$] & $0.20$ [$0.17$,$0.23$] & $0.84$ [$0.81$,$0.87$] & $0.95$ [$0.94$,$0.96$] & $0.11$ [$0.09$,$0.13$] & $0.74$ [$0.62$,$0.81$] & $0.95$ [$0.92$,$0.96$] & $0.21$ [$0.14$,$0.32$]\\
 &  & 48 & $0.50$ [$0.45$,$0.55$] & $0.81$ [$0.77$,$0.84$] & $0.31$ [$0.26$,$0.35$] & $0.65$ [$0.60$,$0.69$] & $0.86$ [$0.83$,$0.89$] & $0.21$ [$0.18$,$0.25$] & $0.46$ [$0.25$,$0.61$] & $0.85$ [$0.75$,$0.91$] & $0.38$ [$0.25$,$0.56$]\\
\cmidrule(l){2-12}
 & \multirow{3}{*}{Intermediate} & 12 & $0.84$ [$0.80$,$0.87$] & $0.96$ [$0.95$,$0.97$] & $0.13$ [$0.10$,$0.15$] & $0.91$ [$0.89$,$0.93$] & $0.97$ [$0.96$,$0.98$] & $0.06$ [$0.05$,$0.07$] & $0.85$ [$0.80$,$0.89$] & $0.97$ [$0.95$,$0.98$] & $0.11$ [$0.08$,$0.16$]\\
 &  & 24 & $0.63$ [$0.59$,$0.68$] & $0.89$ [$0.87$,$0.91$] & $0.26$ [$0.22$,$0.30$] & $0.77$ [$0.74$,$0.80$] & $0.92$ [$0.90$,$0.94$] & $0.15$ [$0.13$,$0.17$] & $0.63$ [$0.51$,$0.72$] & $0.92$ [$0.88$,$0.94$] & $0.28$ [$0.20$,$0.40$]\\
 &  & 48 & $0.36$ [$0.32$,$0.41$] & $0.72$ [$0.67$,$0.76$] & $0.35$ [$0.31$,$0.40$] & $0.53$ [$0.48$,$0.57$] & $0.79$ [$0.76$,$0.82$] & $0.26$ [$0.23$,$0.30$] & $0.33$ [$0.15$,$0.48$] & $0.77$ [$0.65$,$0.86$] & $0.44$ [$0.31$,$0.61$]\\
\cmidrule(l){2-12}
 & \multirow{3}{*}{Unfavorable} & 12 & $0.69$ [$0.63$,$0.75$] & $0.91$ [$0.86$,$0.94$] & $0.21$ [$0.16$,$0.27$] & $0.81$ [$0.77$,$0.86$] & $0.92$ [$0.87$,$0.95$] & $0.10$ [$0.08$,$0.13$] & $0.72$ [$0.63$,$0.79$] & $0.92$ [$0.87$,$0.95$] & $0.20$ [$0.14$,$0.28$]\\
 &  & 24 & $0.41$ [$0.34$,$0.48$] & $0.77$ [$0.72$,$0.82$] & $0.36$ [$0.30$,$0.43$] & $0.59$ [$0.53$,$0.65$] & $0.82$ [$0.76$,$0.86$] & $0.23$ [$0.19$,$0.28$] & $0.41$ [$0.27$,$0.52$] & $0.81$ [$0.73$,$0.88$] & $0.40$ [$0.29$,$0.53$]\\
 &  & 48 & $0.15$ [$0.11$,$0.20$] & $0.51$ [$0.44$,$0.59$] & $0.36$ [$0.29$,$0.44$] & $0.29$ [$0.24$,$0.34$] & $0.60$ [$0.53$,$0.66$] & $0.31$ [$0.25$,$0.37$] & $0.13$ [$0.04$,$0.25$] & $0.61$ [$0.45$,$0.75$] & $0.47$ [$0.32$,$0.63$]\\
\midrule
\multirow{9}{*}{M-spline} & \multirow{3}{*}{Favorable} & 12 & $0.85$ [$0.82$,$0.87$] & $0.97$ [$0.96$,$0.98$] & $0.13$ [$0.11$,$0.15$] & $0.92$ [$0.91$,$0.94$] & $0.98$ [$0.97$,$0.99$] & $0.06$ [$0.04$,$0.07$] & $0.85$ [$0.80$,$0.89$] & $0.98$ [$0.97$,$0.98$] & $0.12$ [$0.09$,$0.17$]\\
 &  & 24 & $0.71$ [$0.67$,$0.75$] & $0.92$ [$0.91$,$0.94$] & $0.21$ [$0.18$,$0.25$] & $0.83$ [$0.80$,$0.86$] & $0.95$ [$0.93$,$0.96$] & $0.12$ [$0.10$,$0.14$] & $0.73$ [$0.65$,$0.79$] & $0.93$ [$0.91$,$0.95$] & $0.20$ [$0.15$,$0.28$]\\
 &  & 48 & $0.52$ [$0.47$,$0.57$] & $0.80$ [$0.76$,$0.83$] & $0.28$ [$0.24$,$0.32$] & $0.66$ [$0.62$,$0.71$] & $0.86$ [$0.83$,$0.89$] & $0.20$ [$0.17$,$0.23$] & $0.56$ [$0.46$,$0.64$] & $0.83$ [$0.78$,$0.87$] & $0.27$ [$0.19$,$0.37$]\\
\cmidrule(l){2-12}
 & \multirow{3}{*}{Intermediate} & 12 & $0.77$ [$0.74$,$0.80$] & $0.95$ [$0.94$,$0.96$] & $0.19$ [$0.16$,$0.21$] & $0.88$ [$0.86$,$0.90$] & $0.97$ [$0.95$,$0.98$] & $0.08$ [$0.07$,$0.10$] & $0.78$ [$0.72$,$0.83$] & $0.96$ [$0.94$,$0.97$] & $0.18$ [$0.13$,$0.24$]\\
 &  & 24 & $0.59$ [$0.54$,$0.62$] & $0.88$ [$0.85$,$0.90$] & $0.29$ [$0.26$,$0.33$] & $0.75$ [$0.72$,$0.77$] & $0.91$ [$0.89$,$0.93$] & $0.17$ [$0.14$,$0.19$] & $0.61$ [$0.53$,$0.68$] & $0.89$ [$0.86$,$0.92$] & $0.28$ [$0.21$,$0.36$]\\
 &  & 48 & $0.37$ [$0.32$,$0.41$] & $0.70$ [$0.65$,$0.74$] & $0.33$ [$0.29$,$0.38$] & $0.53$ [$0.49$,$0.57$] & $0.78$ [$0.75$,$0.81$] & $0.25$ [$0.22$,$0.29$] & $0.41$ [$0.32$,$0.50$] & $0.74$ [$0.67$,$0.80$] & $0.33$ [$0.24$,$0.43$]\\
\cmidrule(l){2-12}
 & \multirow{3}{*}{Unfavorable} & 12 & $0.55$ [$0.50$,$0.61$] & $0.88$ [$0.83$,$0.91$] & $0.32$ [$0.28$,$0.37$] & $0.74$ [$0.69$,$0.79$] & $0.90$ [$0.85$,$0.94$] & $0.16$ [$0.13$,$0.19$] & $0.58$ [$0.50$,$0.66$] & $0.89$ [$0.84$,$0.93$] & $0.30$ [$0.23$,$0.39$]\\
 &  & 24 & $0.32$ [$0.27$,$0.37$] & $0.73$ [$0.67$,$0.79$] & $0.41$ [$0.35$,$0.47$] & $0.52$ [$0.46$,$0.58$] & $0.79$ [$0.72$,$0.84$] & $0.26$ [$0.22$,$0.31$] & $0.36$ [$0.28$,$0.45$] & $0.76$ [$0.69$,$0.82$] & $0.40$ [$0.30$,$0.50$]\\
 &  & 48 & $0.13$ [$0.10$,$0.17$] & $0.48$ [$0.40$,$0.55$] & $0.34$ [$0.27$,$0.42$] & $0.27$ [$0.22$,$0.32$] & $0.57$ [$0.50$,$0.64$] & $0.30$ [$0.24$,$0.36$] & $0.17$ [$0.11$,$0.24$] & $0.54$ [$0.43$,$0.64$] & $0.36$ [$0.25$,$0.48$]\\
\bottomrule\end{tabular}\end{sidewaystable}

\FloatBarrier
\section{Stan code for the illustrative example}\label{app:stan}
The Stan model fitted in the illustrative example (time-varying stable$\to$progression effect, exponential
progression$\to$death baseline, no stable$\to$death treatment effect) is listed below; the individual
and aggregate contributions correspond to the likelihoods of Section~\ref{sec:datalik}.
\begin{lstlisting}[language=C]
// =============================================================================
// mlnmr_multistate_ndmm.stan
//
// ML-NMR-MS: multilevel network meta-regression for an ILLNESS-DEATH multistate
// model, fitted jointly to individual participant data (IPD) and aggregate data
// (AgD). This is the specification used for the NDMM illustrative example.
//
//   States       S = stable, P = progressed, D = dead
//   Transitions  S->P progression, S->D pre-progression death,
//                P->D post-progression death
//   Endpoints    PFS = time spent in S;  OS = time spent in S or P
//
// Specification fitted here:
//   S->P  Weibull baseline; TIME-VARYING treatment effect (non-PH), so that
//         HR^SP(u) = exp(gSP[k] + x'beta2SP) * u^gSPt[k]
//   S->D  Weibull baseline; NO treatment effect (HR = 1)
//   P->D  exponential baseline (constant hazard); CONSTANT treatment effect,
//         HR^PD = exp(gPD[k])
//
// The two evidence types enter ONE likelihood:
//   IPD arms  the full multistate likelihood over the linked transitions.
//   AgD arms  the SAME individual model, integrated over the study's
//             reconstructed covariate distribution (quasi-Monte-Carlo over
//             Ntilde points) to give covariate-marginal state-occupancy
//             probabilities, which enter a conditional-survival (life-table)
//             binomial likelihood.
// Both contributions share every transition parameter, so IPD and AgD inform a
// single coherent network of treatment effects.
// =============================================================================


// -----------------------------------------------------------------------------
// FUNCTIONS -- one exact step of the Kolmogorov forward system for occupancy
//
// The illness-death state occupancies solve
//     dS/du = -(h^SP + h^SD) S
//     dP/du =   h^SP S - h^PD P
// Holding the intensities constant over an interval of width dt (the grid step
// used in the model block) these have closed-form solutions:
//     S(u+dt) = S(u) exp(-(h^SP + h^SD) dt)
//     P(u+dt) = P(u) e^{-h^PD dt}
//               + h^SP S(u) [e^{-(h^SP+h^SD)dt} - e^{-h^PD dt}]/(h^PD-h^SP-h^SD)
// The ratio has a REMOVABLE SINGULARITY when h^PD = h^SP + h^SD; its l'Hopital
// limit is h^SP S(u) dt e^{-h^PD dt}. Near it, the branch below replaces the
// ratio by its series expansion, which also keeps the gradient exact (see there).
// -----------------------------------------------------------------------------
functions {
  vector pstep(vector S, vector P, vector hSP, vector hSD, vector hPD, real dt) {
    int M = num_elements(S);
    vector[M] eS = exp(-(hSP + hSD) * dt);   // survival in S over the interval
    vector[M] eP = exp(-hPD * dt);           // survival in P over the interval
    vector[M] denom = hPD - hSP - hSD;       // vanishes at the singularity
    vector[M] Pnew;
    for (n in 1:M) {
      // x = denom*dt is the DIMENSIONLESS quantity that controls both the series
      // truncation error and the cancellation error, so the branch tests x, not
      // denom: an absolute test on denom is scale-correct only when dt = 1.
      real x = denom[n]*dt;
      if (abs(x) < 1e-4)
        // Near the singularity, expand (eS-eP)/denom = dt*eS*(1 - x/2 + x^2/6 - ...).
        // Truncating after x^2/6 keeps the VALUE to ~x^3/24 and, unlike the bare
        // l'Hopital limit dt*eP, reproduces BOTH partial derivatives exactly at
        // x = 0: d/d(hPD) and d/d(hSP+hSD) are each -dt^2*eS/2 there. The bare
        // limit gives -dt^2*eP for the first (twice too large) and zero for the
        // second, so Stan's autodiff would see a wrong gradient on this branch.
        Pnew[n] = P[n]*eP[n] + S[n]*hSP[n]*dt*eS[n]*(1 - x/2 + x*x/6);
      else
        // Algebraically identical to (eS-eP)/denom, but eS-eP loses precision to
        // cancellation for small x; expm1 computes 1-exp(-x) accurately instead.
        Pnew[n] = P[n]*eP[n] - S[n]*hSP[n]*eS[n]*expm1(-x)/denom[n];
    }
    return Pnew;
  }
}


// -----------------------------------------------------------------------------
// DATA (1/2): INDIVIDUAL PARTICIPANT DATA
// One row per patient: which study and treatment, the covariates, and the
// observed multistate history (exit from S, and if progressed, exit from P).
// -----------------------------------------------------------------------------
data {
  // ---- network sizes ----
  int<lower=0> N;   // IPD patients
  int<lower=1> J;   // studies in the network
  int<lower=2> K;   // treatments in the network (k = 1 is the network reference)
  int<lower=1> Q;   // covariates

  // ---- IPD records ----
  array[N] int<lower=1,upper=J> study;    // study index j of patient i
  array[N] int<lower=1,upper=K> trt;      // treatment index k of patient i
  matrix[N,Q] X;                          // covariates (row i = patient i)
  vector<lower=0>[N] t1;                  // time of exit from S (or censoring)
  array[N] int<lower=0,upper=2> eventS;   // 0 censored, 1 progression, 2 death in S
  array[N] int<lower=0,upper=1> prog;     // 1 if the patient entered P
  vector<lower=0>[N] t2;                  // time of exit from P (or censoring)
  array[N] int<lower=0,upper=1> eventP;   // 1 if post-progression death observed

  // ---------------------------------------------------------------------------
  // DATA (2/2): AGGREGATE DATA
  // Each aggregate arm supplies (i) a cloud of Ntilde reconstructed covariate
  // points -- the ML-NMR integration cloud -- and (ii) life-table counts giving,
  // per interval, how many were at risk and how many remained event-free.
  // ---------------------------------------------------------------------------
  int<lower=0> n_agd_arms;   // aggregate arms
  int<lower=1> Ntilde;       // integration points per aggregate arm (QMC cloud)
  int<lower=2> n_grid;       // points on the shared occupancy time grid
  vector[n_grid] grid;       // the occupancy grid itself

  array[n_agd_arms] int<lower=1,upper=J> agd_study;   // study index of each arm
  array[n_agd_arms] int<lower=1,upper=K> agd_trt;     // treatment index of each arm
  array[n_agd_arms,Ntilde,Q] real Xtilde;             // the reconstructed covariates

  // ---- life-table rows (conditional-survival counts) ----
  int<lower=0> n_lt;                                   // number of life-table rows
  array[n_lt] int<lower=1,upper=n_agd_arms> lt_arm;    // which aggregate arm
  array[n_lt] int<lower=1,upper=2> lt_endpoint;        // 1 = PFS, 2 = OS
  array[n_lt] int<lower=1,upper=n_grid> lt_idx_start;  // grid index, interval start
  array[n_lt] int<lower=1,upper=n_grid> lt_idx_eval;   // grid index, interval end
  array[n_lt] int<lower=0> lt_n;                       // n^c at risk at start
  array[n_lt] int<lower=0> lt_r;                       // r^c event-free at end
}
transformed data {
  // The occupancy march hard-codes the initial condition (S,P,D)(0) = (1,0,0) at
  // grid index 1, so the grid must start at 0 and be strictly increasing. The R
  // pipeline always supplies 0:Tmax; assert it rather than trust the caller.
  if (grid[1] != 0)
    reject("occupancy grid must start at 0; got grid[1] = ", grid[1]);
  for (m in 2:n_grid)
    if (grid[m] <= grid[m-1])
      reject("occupancy grid must be strictly increasing at index ", m);
}


// -----------------------------------------------------------------------------
// PARAMETERS
// Split as in any NMA: STUDY-SPECIFIC baselines (nuisance, never pooled) plus
// NETWORK-WIDE relative effects (the quantities of interest), here estimated for
// each transition separately.
// -----------------------------------------------------------------------------
parameters {
  // ---- study effects: baseline hazards, one set per study ----
  matrix[J,3] c_base;   // log baseline intercept per study, per transition
  matrix[J,2] logv;     // log Weibull shape per study: (,1) S->P, (,2) S->D
                        // (P->D is exponential, so its shape is fixed at 1)

  // ---- treatment effects: the BASIC PARAMETERS of the network ----
  // K-1 free effects per transition, each versus the reference treatment k = 1.
  vector[K-1] gamma_SP;    // S->P log-hazard-ratio (scale)
  vector[K-1] gamma_SPt;   // S->P log-time slope: makes the HR time-varying
  vector[K-1] gamma_PD;    // P->D log-hazard-ratio

  // ---- covariate effects ----
  matrix[Q,3] beta1;   // PROGNOSTIC effects, per transition (columns SP, SD, PD)
  vector[Q] beta2SP;   // EFFECT MODIFIERS: treatment-by-covariate INTERACTIONS on
                       // S->P, shared across the active treatments
}

transformed parameters {
  matrix<lower=0>[J,2] v = exp(logv);   // Weibull shapes on the natural scale

  // ---- NMA CONSISTENCY ASSUMPTION ----
  // Prepending 0 pins the reference treatment at zero (gamma^r_1 = 0), so the K-1
  // estimated effects are the BASIC parameters: each is a contrast against that
  // reference. Every other contrast is a FUNCTIONAL parameter, obtained by
  // DIFFERENCING the basic ones rather than estimated in its own right. Writing
  // d^r_XY for the transition-r treatment coefficient of Y versus X, with A the reference:
  //     d^r_AB = gamma^r[B],    d^r_AC = gamma^r[C],
  //     d^r_BC = d^r_AC - d^r_AB = gamma^r[C] - gamma^r[B].
  // That identity IS the consistency assumption, and it holds in every study, so
  // a direct B-vs-C trial and the indirect A-anchored path are forced to agree.
  // It is what makes this a network meta-analysis rather than a set of separate
  // pairwise comparisons.
  //
  // A contrast between two non-reference treatments therefore needs no extra
  // parameters, only the difference. Because the S->P effect here is
  // time-varying, that contrast is a FUNCTION of time u:
  //     log HR^SP_{C vs B}(u) = (gSP[C] - gSP[B]) + (gSPt[C] - gSPt[B]) * log(u)
  //     log HR^PD_{C vs B}    =  gPD[C] - gPD[B]                (constant in u)
  // Here these differences are formed from the posterior draws after fitting. To
  // have Stan report them directly, append e.g. (with B = 2 and C = 3):
  //     generated quantities {
  //       vector[n_grid-1] log_HR_SP_CvsB;        // grid[1] = 0, so start at m = 2
  //       for (m in 2:n_grid)
  //         log_HR_SP_CvsB[m-1] = (gSP[3] - gSP[2])
  //                               + (gSPt[3] - gSPt[2]) * log(grid[m]);
  //     }
  // Such quantities are pure transformations of the basic parameters: they add no
  // parameters and change no inference.
  vector[K] gSP  = append_row(0, gamma_SP);
  vector[K] gSPt = append_row(0, gamma_SPt);
  vector[K] gPD  = append_row(0, gamma_PD);
}


// -----------------------------------------------------------------------------
// LOG POSTERIOR (Stan's model block): the priors, then the likelihood
// contributions. The linear predictors and the log link that turn parameters into
// transition intensities are built inside the likelihood loops below.
// -----------------------------------------------------------------------------
model {
  // ---------------------------------------------------------------------------
  // PRIORS -- weakly informative and independent
  // ---------------------------------------------------------------------------
  to_vector(c_base) ~ normal(-3,2);   // low monthly baseline hazards a priori
  to_vector(logv) ~ normal(0,0.5);    // centers the Weibull shapes at 1 (constant hazard)
  gamma_SP ~ normal(0,2);             // diffuse on the log-hazard-ratio scale
  gamma_SPt ~ normal(0,1);            // tighter: time-variation is a smaller effect
  gamma_PD ~ normal(0,2);
  to_vector(beta1) ~ normal(0,2);
  beta2SP ~ normal(0,2);

  // ---------------------------------------------------------------------------
  // LIKELIHOOD (1/2): INDIVIDUAL PARTICIPANT DATA
  // Each patient contributes: survival in S up to t1, times the intensity of the
  // exit actually observed, times (if progression occurred) the post-progression
  // survival and death intensity.
  // ---------------------------------------------------------------------------
  for (i in 1:N) {
    int j=study[i]; int k=trt[i]; row_vector[Q] xi=X[i]; real active=(k==1)?0:1;

    // LINEAR PREDICTORS. The LINK IS LOG: log hazard = linear predictor.
    // prognostic effect + effect modification (active arms only) + treatment effect
    real lpSP = dot_product(xi,beta1[,1]) + active*dot_product(xi,beta2SP) + gSP[k];
    real lpSD = dot_product(xi,beta1[,2]);                   // no treatment effect
    real lpPD = dot_product(xi,beta1[,3]) + gPD[k];

    // TIME-VARYING HAZARD RATIO on S->P: the treatment shifts the Weibull SHAPE,
    // giving HR^SP(u) = exp(gSP[k] + x'beta2SP) * u^gSPt[k] instead of a constant ratio.
    real vSP = v[j,1] + gSPt[k];
    // v[j,*] is constrained positive, the log-time slope is not; for shape <= 0 the
    // cumulative hazard below diverges, so reject rather than evaluate it
    if (vSP <= 0) reject("S->P shape v[j,1]+gSPt[k] must be positive; got ", vSP);
    real vSD = v[j,2];

    // WEIBULL cumulative hazards H(t) = exp(c + lp)/v * t^v, for both exits from S
    real logL = -(exp(c_base[j,1]+lpSP)/vSP*pow(t1[i],vSP)
                  + exp(c_base[j,2]+lpSD)/vSD*pow(t1[i],vSD));

    // log-intensity of the exit observed (a censored exit adds neither term)
    if (eventS[i]==1) logL += c_base[j,1]+(vSP-1)*log(t1[i])+lpSP;       // progression
    else if (eventS[i]==2) logL += c_base[j,2]+(vSD-1)*log(t1[i])+lpSD;  // death in S

    // post-progression: EXPONENTIAL survival over (t2 - t1), plus the death
    // intensity when the death is observed
    if (prog[i]==1) { logL += -(exp(c_base[j,3]+lpPD)*(t2[i]-t1[i]));
      if (eventP[i]==1) logL += c_base[j,3]+lpPD; }

    target += logL;
  }

  // ---------------------------------------------------------------------------
  // LIKELIHOOD (2/2): AGGREGATE DATA
  // The ML-NMR step. For each aggregate arm the individual model is evaluated at
  // all Ntilde covariate points, the occupancies are marched along the grid, and
  // only THEN averaged across points (marginalizing last). The resulting
  // covariate-marginal occupancies feed a conditional-survival binomial.
  // ---------------------------------------------------------------------------
  if (n_agd_arms > 0) {
    matrix[n_agd_arms,n_grid] SbarM;   // marginal P(stable) on the grid, per arm
    matrix[n_agd_arms,n_grid] PbarM;   // marginal P(progressed) on the grid, per arm

    for (a in 1:n_agd_arms) {
      int j=agd_study[a]; int k=agd_trt[a]; real active=(k==1)?0:1;
      matrix[Ntilde,Q] Xa=to_matrix(Xtilde[a]);   // this arm's QMC covariate cloud

      // The SAME linear predictors as the IPD block, at all Ntilde points.
      vector[Ntilde] offSP = Xa*beta1[,1] + gSP[k]; if (active==1) offSP += Xa*beta2SP;
      vector[Ntilde] offSD = Xa*beta1[,2];
      vector[Ntilde] offPD = Xa*beta1[,3] + gPD[k];
      real vSP = v[j,1] + gSPt[k];                // time-varying S->P shape
      // guard as in the IPD block: v[j,1] is constrained positive, gSPt is not
      if (vSP <= 0) reject("S->P shape v[j,1]+gSPt[k] must be positive; got ", vSP);

      // everyone starts in the stable state
      vector[Ntilde] S=rep_vector(1,Ntilde); vector[Ntilde] P=rep_vector(0,Ntilde);
      SbarM[a,1]=1; PbarM[a,1]=0;

      // KOLMOGOROV GRID: march interval by interval, holding the intensities
      // constant at the interval midpoint (lu = log midpoint).
      for (m in 2:n_grid) {
        real dt=grid[m]-grid[m-1]; real lu=log((grid[m]+grid[m-1])/2);
        vector[Ntilde] hSP = exp(c_base[j,1]+(vSP-1)*lu+offSP);     // Weibull
        vector[Ntilde] hSD = exp(c_base[j,2]+(v[j,2]-1)*lu+offSD);  // Weibull
        vector[Ntilde] hPD = exp(c_base[j,3]+offPD);                // exponential
        vector[Ntilde] Snew = S .* exp(-(hSP+hSD)*dt);              // exact S step
        P = pstep(S,P,hSP,hSD,hPD,dt);                              // exact P step
        S=Snew;
        SbarM[a,m]=mean(S); PbarM[a,m]=mean(P);   // MARGINALIZE over the cloud
      }
    }

    // Life-table (conditional-survival) likelihood: of the n^c at risk at the
    // interval start, r^c remain event-free at its end, with probability p given
    // by a ratio of marginal occupancies.
    for (l in 1:n_lt) {
      int a=lt_arm[l]; real p;
      if (lt_endpoint[l]==1)         // PFS: still stable
        p = SbarM[a,lt_idx_eval[l]]/SbarM[a,lt_idx_start[l]];
      else                           // OS: stable OR progressed, i.e. still alive
        p = (SbarM[a,lt_idx_eval[l]]+PbarM[a,lt_idx_eval[l]])
            /(SbarM[a,lt_idx_start[l]]+PbarM[a,lt_idx_start[l]]);
      p = fmin(fmax(p,1e-12),1-1e-12);   // guard the binomial away from 0 and 1
      target += binomial_lpmf(lt_r[l] | lt_n[l], p);
    }
  }
}
\end{lstlisting}

\noindent Code identifiers correspond to the notation of Sections~\ref{sec:setting}--\ref{sec:datalik} and of
\eqref{eq:ex-lp} as follows. Indices: \texttt{study}~$=j$, \texttt{trt}~$=k$, and the loop counter \texttt{i} the
individual~$i$; \texttt{J}, \texttt{K}, \texttt{Q} and \texttt{N} are $J$, $K$, $Q$ and the number of
IPD records, and row~$i$ of \texttt{X} is $\xv_{ijk}$ for the patient with \texttt{study[i]}~$=j$ and
\mbox{\texttt{trt[i]}~$=k$}. Times: \texttt{t1}~$=u_S$ and \texttt{t2}~$=u_P$, the exit times from the stable
and progressed states, with \texttt{eventS}~$=1,2$ and \texttt{eventP}~$=1$ the indicators
$\mathds{1}(S\to P)$, $\mathds{1}(S\to D)$ and $\mathds{1}(P\to D)$ of \eqref{eq:ipdlik} and
\texttt{prog} the outer $\mathds{1}(S\to P)$; \texttt{grid} is the occupancy grid $u_m$ and
\texttt{m}~$=m$ its loop counter.
Parameters: \texttt{c\_base}~$=c^r_j$; \texttt{logv}~$=\log v^r_j$, the declared parameter, with
\texttt{v}~$=v^r_j=\exp(\texttt{logv})$ transformed, for the two Weibull transitions (S$\to$P and
S$\to$D; $v^{PD}_j\equiv1$ is fixed, not a parameter); \texttt{gSP}, \texttt{gSPt} and
\texttt{gPD}~$=\gamma^{SP}_k,\gamma^{SPt}_k,\gamma^{PD}_k$ (each formed by
\texttt{append\_row(0,$\cdot$)}, so entry~1 is the network reference $\bm\gamma^r_1=\bm 0$; the
time-varying effect enters as a shape shift, \texttt{vSP}~$=v^{SP}_j+\gamma^{SPt}_k$),
\texttt{beta1}~$=\bet^r_1$ and \texttt{beta2SP}~$=\bet^{SP}_2$. Aggregate quantities:
\texttt{Ntilde}~$=\tilde N$, \texttt{Xtilde} the reconstructed covariate points
$\tilde\xv^{(l)}_{jk}$, \texttt{SbarM} and \texttt{PbarM} the marginal occupancies
$\Sbar_{jk},\Pbar_{jk}$, and \texttt{lt\_idx\_start}, \texttt{lt\_idx\_eval}, \texttt{lt\_n}, \texttt{lt\_r} the interval
endpoints $m,m'$ and counts $n^c,r^c$ of \eqref{eq:agdbinom}, with \texttt{lt\_arm} the arm $(j,k)$
and \texttt{lt\_endpoint} the endpoint label $c\in\{\mathrm{cPFS},\mathrm{cOS}\}$ selecting between
the two binomials. Two loop
counters are reused inside the aggregate block for local purposes: \texttt{a} indexes aggregate arms
and \texttt{l} life-table rows; \texttt{a} is not the contrast label $a$ of \eqref{eq:condHR}, and \texttt{l} is neither the integration point $l$ of \eqref{eq:marg} nor the M-spline basis index $l$ of Section~\ref{sec:model}; likewise \texttt{M} inside \texttt{pstep} is the number of integration points
$\tilde N$, not the M-spline basis.

\section{Random treatment effects on stable$\to$progression in Stan code}\label{app:re}
The model of Appendix~\ref{app:stan} treats the stable$\to$progression treatment effect as common to every study. To
allow between-study heterogeneity, replace that common effect by a study-specific one,
\begin{equation}
\delta^{SP}_{jk}\sim\mathrm{N}\!\big(\gamma^{SP}_{k},\ \tau^{2}\big)\quad (k\neq1),
\qquad \delta^{SP}_{j1}=0,
\label{eq:re-sp}
\end{equation}
so that $\gamma^{SP}_{k}$ becomes the mean effect of treatment $k$ versus the network
reference and $\tau$ the between-study standard deviation; $\tau=0$ recovers that model.
Consistency is untouched, because it is a constraint on the means: in every study the mean of
$\delta^{SP}_{jb}-\delta^{SP}_{ja}$ is $\gamma^{SP}_{b}-\gamma^{SP}_{a}$.

Three edits implement this. First, add the heterogeneity parameters, in the non-centered form that
samples better when $\tau$ is small:
\begin{lstlisting}[language=C]
parameters {
  // ... as in Appendix J, plus:
  real<lower=0> tau_SP;    // between-study SD of the S->P treatment effect
  matrix[J,K-1] z_SP;      // standardized study deviations (non-centered)
}
\end{lstlisting}
Second, give them priors alongside the others:
\begin{lstlisting}[language=C]
  tau_SP ~ normal(0, 0.5);          // half-normal, since tau_SP is declared lower=0
  to_vector(z_SP) ~ std_normal();
\end{lstlisting}
Third, use the study-specific effect \texttt{dSP} in place of \texttt{gSP[k]}, in the individual and
the aggregate contribution alike, so that IPD and AgD studies draw their deviations from the same distribution:
\begin{lstlisting}[language=C]
  real dSP = gSP[k] + (k == 1 ? 0 : tau_SP * z_SP[j, k-1]);   // delta^SP_{jk}
  // IPD block:  ... + active*dot_product(xi,beta2SP) + dSP;
  // AgD block:  vector[Ntilde] offSP = Xa*beta1[,1] + dSP; ...
\end{lstlisting}
The deviations are drawn independently with the network reference held
fixed, which induces the usual contrast heterogeneity $\tau^{2}$ in a two-arm study that includes the
network reference, as every study in the example network does. A study that omits the network
reference, or that has three or more arms, needs a correlated multivariate Normal across its arms so
that the induced contrasts carry $\tau^{2}$ rather than $2\tau^{2}$.

With few studies $\tau$ is only weakly identified, so the half-normal prior is influential and its
scale should be set deliberately on the log-hazard-ratio scale. The estimand also has to be stated: $\gamma^{SP}_{k}$
is the mean effect, whereas a prediction for a new study also carries the variance $\tau^{2}$, and
the population-adjusted quantities reported in the main text should be formed from whichever of the
two is intended. The same construction extends to the time-varying component $\gamma^{SPt}$, and to the
progression$\to$death effect, by adding further variance components in the same way; independent
scalar variances correspond to a diagonal $\bm\Sigma$ in Section~\ref{sec:model}. Correlated
components require the full $\bm\Sigma$.

\section{Simulation sub-study: composite versus multinomial aggregate likelihood}\label{app:calibration}
The two-binomial composite likelihood \eqref{eq:agdbinom} treats the PFS and OS grouped-survival streams as
independent, whereas the multinomial state-occupancy likelihood \eqref{eq:multinom} models each interval's
$(n_S,n_P,n_D)$ occupancy counts jointly. To quantify the practical cost of the composite's
independence assumption, both are fitted to $200$ replicate networks of the base-scenario design
(Section~\ref{sec:sim}), with $250$ rather than $200$ patients per arm, monthly life-table intervals and
$\tilde N=24$ integration points. Both use four chains of $1000$ iterations ($500$ warmup). The composite
is fitted as in the main study, from random starting values in a narrow range of plausible values, since
Stan's default random starting values can fail to initialize the model. The multinomial, whose posterior is
more strongly correlated, uses a dense mass matrix and starting values near the data-generating values of the baselines and treatment effects.
Each fit is scored on the population-adjusted conditional S$\to$P
log-hazard-ratio for the three treatment contrasts in the three target populations. Metrics are
computed over the converged fits (Table~\ref{tab:calibration}). The
simulated aggregate arms report only separate PFS and OS curves, so the multinomial's joint
occupancy counts must be reconstructed (Section~\ref{sec:com_multinomial}), as is common in
practice.

Both likelihoods have small bias (mean absolute bias $0.010$ and $0.020$), although the multinomial's is systematic,
about $-0.03$ in the favorable population and $+0.02$ in the unfavorable one. The multinomial's estimates are also about
$20\%$ more variable (empirical SE $0.115$ against $0.096$), and the two differ sharply
in interval calibration (Table~\ref{tab:calibration}). The two-binomial composite is well calibrated: its average
posterior standard deviation (ModSE, $0.099$) matches the empirical standard error ($0.096$) and its
coverage is at the nominal level ($0.955$). The multinomial, despite respecting the PFS/OS dependence,
is overconfident: its posterior is too narrow (ModSE $0.063$, little more than half the empirical SE of
$0.115$), and coverage falls to $0.696$. Two mechanisms contribute. Reconstructing the joint counts from the separate PFS and OS curves
adds information that was never observed. The second is a property of \eqref{eq:multinom} itself: each interval is conditioned only on the
number of patients alive at its start, so the stable count is referred to everyone alive and the progression information of earlier
intervals is used again. It does not depend on reconstruction: in a check with genuine counts (exponential illness--death model,
$250$ patients, $400$ replicates) the multinomial's model-based standard error for the log progression hazard is $0.044$ against an
empirical standard deviation of $0.098$, while the two-binomial's is $0.098$ against $0.092$. It was also the more fragile fit: despite the dense mass
matrix and the informed starting values, $20$ of its $200$ fits did not converge, against none of the
composite's, and $177$ of its $200$ fits ($89\%$) had at least one divergent transition after warmup, against $4$ ($2\%$). Per-fit
Monte Carlo error is similar to the main study's: the per-fit Monte Carlo standard error of the interval
limits has a median of $0.08$ posterior standard deviations for both likelihoods ($99$th percentile $0.12$
and $0.15$). These results support the two-binomial as the default. Because the second mechanism persists with genuine
counts, they give no support for the multinomial in that setting either: using genuine occupancy counts efficiently requires a
multinomial conditioned on the state occupied at the start of each interval (Section~\ref{sec:com_multinomial}).

\begin{table}[htbp]\centering\footnotesize\setlength{\tabcolsep}{5pt}
\caption{Simulation sub-study: composite (two-binomial) versus multinomial aggregate likelihood for the population-adjusted conditional S$\to$P log-hazard-ratio, on networks of the base-scenario design with 250 patients per arm (200 replicates, monthly life-table intervals, $\tilde N=24$). The five performance measures are averaged over the nine target-population\,$\times$\,contrast cells, with the mean per-cell Monte Carlo SE in parentheses. Converged fits ($\hat R\le1.05$): 100\% (composite), 90\% (multinomial).}
\label{tab:calibration}
\begin{tabular}{l ccccc}\toprule
Aggregate likelihood & $|$bias$|$\,(MCSE) & empSE\,(MCSE) & ModSE & RMSE & coverage\,(MCSE)\\\midrule
Two-binomial (composite) & $0.010$\,(0.007) & $0.096$\,(0.005) & $0.099$ & $0.097$ & $0.955$\,(0.014)\\
Multinomial (state occupancy) & $0.020$\,(0.009) & $0.115$\,(0.006) & $0.063$ & $0.116$ & $0.696$\,(0.034)\\
\bottomrule\end{tabular}\end{table}

\section{Data-generating settings of the simulation study}\label{app:simdgm}
Table~\ref{tab:sim-dgm} gives the settings of the simulation study that Section~\ref{sec:sim} does not
list: the design of each study, the target populations, and what each scenario changes, including the
true values that differ in the strong-EM scenario.

\begin{table}[htbp]\centering\footnotesize
\caption{Data-generating settings of the simulation study that Section~\ref{sec:sim} does not list.
Top: the six studies, with the studies that contribute IPD in each data configuration (the others
report aggregate data), the administrative horizon and the covariate means, and the three target
populations. Bottom: what each scenario changes from the base. In every scenario each arm has $200$
patients; $x_1$ and $x_3$ are normal with standard deviation $1$ and correlation $0.30$, and $x_2$
is Bernoulli and independent of them; patients are also censored by exponential dropout at $0.004$ per
month; and the replicates use seeds $2001$ to $2200$, the same in every scenario.}
\label{tab:sim-dgm}
\begin{tabular}{@{}llccccccc@{}}\toprule
 & & \multicolumn{3}{c}{IPD in configuration} & Horizon & \multicolumn{3}{c}{Covariate means}\\
\cmidrule(lr){3-5}\cmidrule(lr){7-9}
Study & Comparison & Base & More IPD & Agg.-dom. & (months) & $x_1$ & $\Pr(x_2=1)$ & $x_3$\\\midrule
1 & $B$ vs $A$ & $\bullet$ & $\bullet$ &  & 30 & $0.0$ & $0.40$ & $0.0$\\
2 & $C$ vs $A$ & $\bullet$ & $\bullet$ & $\bullet$ & 36 & $0.3$ & $0.60$ & $0.5$\\
3 & $C$ vs $B$ &  &  &  & 30 & $-0.2$ & $0.30$ & $-0.3$\\
4 & $D$ vs $A$ &  & $\bullet$ &  & 24 & $0.5$ & $0.70$ & $0.2$\\
5 & $D$ vs $C$ &  &  &  & 36 & $0.1$ & $0.50$ & $0.0$\\
6 & $B$ vs $A$ &  & $\bullet$ &  & 30 & $-0.4$ & $0.20$ & $-0.5$\\
\midrule
\multicolumn{6}{@{}l}{Target population} & & & \\
\multicolumn{6}{@{}l}{\quad Favorable} & $-0.5$ & $0.20$ & $0.5$\\
\multicolumn{6}{@{}l}{\quad Intermediate (as study 1)} & $0.0$ & $0.40$ & $0.0$\\
\multicolumn{6}{@{}l}{\quad Unfavorable} & $0.5$ & $0.60$ & $-0.5$\\
\bottomrule\end{tabular}\par\medskip
\begin{tabular}{@{}>{\raggedright\arraybackslash}p{0.2\linewidth}p{0.74\linewidth}@{}}\toprule
Scenario & Change from the base\\\midrule
(a) Strong EM & Effect modifiers multiplied by $2$, $\bet^{SP}_2=(0.00, 0.80, -0.60)$. Population-adjusted truths for $B$, $C$ and $D$
versus $A$: favorable $-0.64$, $-0.84$, $-0.74$; intermediate $-0.18$, $-0.38$, $-0.28$; unfavorable $0.28$, $0.08$, $0.18$.\\
(b) More IPD; Agg.-dom. & IPD from studies 1, 2, 4 and 6; from study 2 only (top panel).\\
(c) Frailty & A shared Gamma frailty with mean $1$ and variance $0.5$ multiplies all three transition intensities of each patient.\\
(d) Copula mis. & The correlation of $x_1$ and $x_3$ used to reconstruct the aggregate studies' covariates
is set to $-0.30$, instead of the IPD estimate (about $0.30$) used otherwise.\\
(e) Non-PH & $\gamma^{SPt}_k=0.15$ for $k=B,C,D$, as described in Section~\ref{sec:sim}.\\
\bottomrule\end{tabular}\end{table}

\end{document}